\documentclass[9pt,journal]{IEEEtran}
\usepackage{graphicx}
\usepackage{amssymb} 
\usepackage{epstopdf}
\usepackage[ruled,norelsize]{algorithm2e}
\makeatletter
\newcommand{\removelatexerror}{\let\@latex@error\@gobble}
\makeatother
\usepackage{pseudocode}
\usepackage{amsthm}
\usepackage{amsmath}
\usepackage{cite}
\usepackage{tikz}
\usepackage{marginnote}
\usepackage{subfigure}
\usepackage{hyperref}

\newtheorem{thm}{Theorem}

\newtheorem{lem}[thm]{Lemma}
\newtheorem{prop}[thm]{Proposition}
\newtheorem{defn}[thm]{Definition}

\begin{document}
%
\title{Bare Demo of IEEEtran.cls\\ for IEEE Journals}

\title{Sampling of Planar Curves: Theory and Fast Algorithms}
\author{Qing Zou$^{*}$ \IEEEmembership{Student Member,~IEEE}, Sunrita Poddar$^*$ \IEEEmembership{Student Member,~IEEE},  and Mathews Jacob,~\IEEEmembership{Senior Member,~IEEE}
\thanks{$^*$Both authors contributed equally to the paper. Qing Zou is with the Applied Mathematics \& Computational Sciences and Sunrita Poddar, Mathews Jacob are with the Department of Electrical and Computer Engineering at the University of Iowa, Iowa City, IA, 52242, USA (e-mail: zou-qing@uiowa.edu; sunrita-poddar@uiowa.edu; mathews-jacob@uiowa.edu)}
\thanks{This work is supported by grants NIH 1R01EB019961-01A1 and R01 EB019961-02S1.}}
\maketitle

\begin{abstract}
We introduce a continuous domain framework for the recovery of a planar curve from a few samples. We model the curve as the zero level set of a trigonometric polynomial. We show that the exponential feature maps of the points on the curve lie on a low-dimensional subspace. We show that the null-space vector of the feature matrix can be used to uniquely identify the curve, given a sufficient number of samples. The worst-case theoretical guarantees show that the number of samples required for unique recovery depends on the bandwidth of the underlying trigonometric polynomial, which is a measure of the complexity of the curve. We introduce an iterative algorithm that relies on the low-rank property of the feature maps to recover the curves when the samples are noisy or when the true bandwidth of the curve is unknown. We also demonstrate the preliminary utility of the proposed curve representation in the context of  image segmentation. 
\end{abstract}



\IEEEpeerreviewmaketitle

\section{Introduction}

The recovery of a curve from finite number of unorganized and noisy points is an important problem, with applications to computer vision \cite{Botsch2007,Jain1995} and image processing \cite{jacobtip,Uhlmann2016,Delgado-gonzalo2015}.  This problem is fundamentally ill-posed because one can find infinite number of curves that pass through the measured points. In addition, the reconstruction is challenging due to curve topology (e.g. the shape of the curve), and the variation of topology with noise. Popular approaches for shape representation with arbitrary topology include (a) explicit representations using a mesh or graphs \cite{Botsch2007,Badoual2017}, and (b) implicit level-set representations \cite{burger2005survey,Aghasi2011,Bernard2009}. In the first scheme, the shape is constructed from the noisy points as a graph, where the nodes corresponding to adjacent data points are connected. In the second approach, level set functions are constructed from the points \cite{Turk1999}. Several methods were introduced to account for noisy data, including spectral graph theory, Laplacian/curvature flow \cite{Crane2013,Nealen2006}. All of these methods suffer from the inherent parametrization of the curve, which often depends on the sampling density. 

The main focus of this paper is to introduce a unified continuous domain theory for the recovery of a planar curve. We assume that the points live on a curve, which is the zero level set of a bandlimited function $\psi$. This property enables us to express $\psi$ as a finite linear combination of complex exponentials, where the weights are specified by the vector $\mathbf c$. The bandwidth of $\psi$ denoted by $\Lambda$ is a measure of the complexity of the curve \cite{ongie2017convex}. We show that when $\psi$ is irreducible (c.f. Definition \ref{defired}), the curve consists of a single closed connected component, which we term as an irreducible curve. When $\psi$ can be factorized into multiple irreducible factors, we obtain an union of irreducible curves; each of the irreducible factors correspond to a closed connected curve. 

The function $\psi$ vanishes at all points $\mathbf x$ on the curve; i.e, $\psi(\mathbf x)=0; \forall \mathbf x \in \mathcal C$. This implies that the weighted linear combination of complex exponential features of the point $\exp\left(j2\pi \mathbf k^T \mathbf x\right); \mathbf k \in \Lambda$, weighted by $\mathbf c$, will vanish for all points on the curve. In particular, $\mathbf c$ is the normal vector to the complex exponential features of the points on the curve. We term this property as the annihilation relation, which suggests that the complex exponential maps of the points on the curve lie in a subspace, whose normal vector is $\mathbf c$. Thus, the non-linear exponential mapping transforms the non-linear curve structure to the familiar low-rank or subspace structure, which well-studied in signal processing. When we have a union of irreducible curves, the samples from each one of the irreducible components lie on a subspace; the mapping transforms the complex structure to a union of subspaces structure \cite{elhamifar2013sparse}. The dimension of the subspace spanned by the feature maps is dependent on the bandwidth $\Lambda$, and hence the complexity of the curve. 

We use the subspace structure of the feature vectors to recover the curve from a few measurements. Specifically, we identify the coefficient vector as the unit norm null space vector of the feature matrix, which is unique up to a scaling with magnitude one. Our worst-case results show that the band-limited function, and hence the curve, can be recovered uniquely, when the number of points exceed a bound that is dependent on the curve complexity. We also show that if the curve is irreducible (single connected component), then one can recover it from any arbitrary sampling pattern. When we have a union of irreducible curves, then the number of samples on each component would depend on the bandwidth of the corresponding irreducible polynomial. We also introduce efficient strategies when the bandwidth of the curve is unknown. Specifically, we show that when the support is overestimated, there exist multiple linearly independent filters that will annihilate the exponential maps; the common zeros, or equivalently the zero level set of the greatest common divisor of the filters, uniquely specifies the curve in this case. 

When the support is overestimated, the feature matrix has multiple linearly independent null-space vectors, and hence is low-rank. We note that the Gram matrix of the exponential features correspond to a kernel matrix, which connects the band-limited curve model with widely used non-linear low-rank kernel methods \cite{scholkopf}. When the curve samples are noisy, we rely on a nuclear norm minimization formulation to denoise the points. Specifically, we seek to find the denoised curve samples such that their feature vectors form a low-rank matrix. We use an iterative re-weighted algorithm to solve the above optimization problem, which alternates between the estimation of a weight matrix that approximates the null-space and a quadratic sub-problem to recover the data. We note that the iterative algorithm bears strong similarity to Laplacian/curvature flows used in graph denoising, which provides the connection between implicit level-set and explicit graph-based curve representations. One can also derive a graph Laplacian matrix from the weight matrix, which will facilitate the smoothing of signals that live on the nodes of the graph. This graph can be viewed as a discrete mesh approximation to the points that live on the curve. Our experiments show that the Laplacian matrix obtained by solving the proposed optimization algorithm is more representative of the graph structure than classical methods \cite{Belkin2006}, especially when it is estimated from noisy data. This framework reveals links between recent advances in superresolution theory \cite{candes2014towards,Ongie2016b,recht}, manifold smoothness based regularization, as well as graph signal processing \cite{spm}.

This work has connections with Logan's results \cite{Logan1977} for the recovery of 1-D band-limited functions from their zero crossings, as well as their extensions to 2-D \cite{Zakhor1990}. The main challenge of these works is the extreme sensitivity of the band-limited function to the location of the zero-crossings, when no amplitude information of the signal is used \cite{Zakhor1990}; this has prompted the use of additional information including multi-level contours \cite{Zakhor1990} and multi-scale edges \cite{Mallat1992}. By contrast, we focus on the recovery of the curve itself, rather than the band-limited function, which is considerably simpler. Specifically, we propose to recover the curve as the zero level set of the sum of squares of all band-limited functions that satisfy the sampling conditions. In addition, unlike \cite{Zakhor1990}, our results are also valid for the union of irreducible curves. The proposed work is built upon our prior work on annihilation based super-resolution image recovery \cite{Ongie2017fast,Ongie2016b,Ongie2015,Mohsin2015,gregvariety,ongie2017convex} that has similarities to algebraic shape recovery \cite{Fatemi2016} and the recent work by Ongie et al., which considered polynomial kernels \cite{gregvariety}. Our main focus is to generalize \cite{gregvariety} to shift invariant kernels, which are more widely used in applications. This approach can also be viewed as the generalization of the finite rate of innovation (FRI) theory \cite{vetterli2002sampling,blu2008sparse,dragotti2007sampling,pan2014sampling,shukla2007sampling} in terms of the non-uniform sampling and the knowledge of the samples. We also introduce sampling conditions and algorithms to determine the curve, when the dimension is low. In addition, the iterative algorithm using the kernel trick shows the connections with graph Laplacian based methods used in graph signal processing. The conference version of this work has been published in \cite{icassp2018,isbi2018}; this paper provides the proofs of these results, and more elaborate description of details.

\begin{figure}[t!]
\centering
\includegraphics[width=0.48\textwidth]{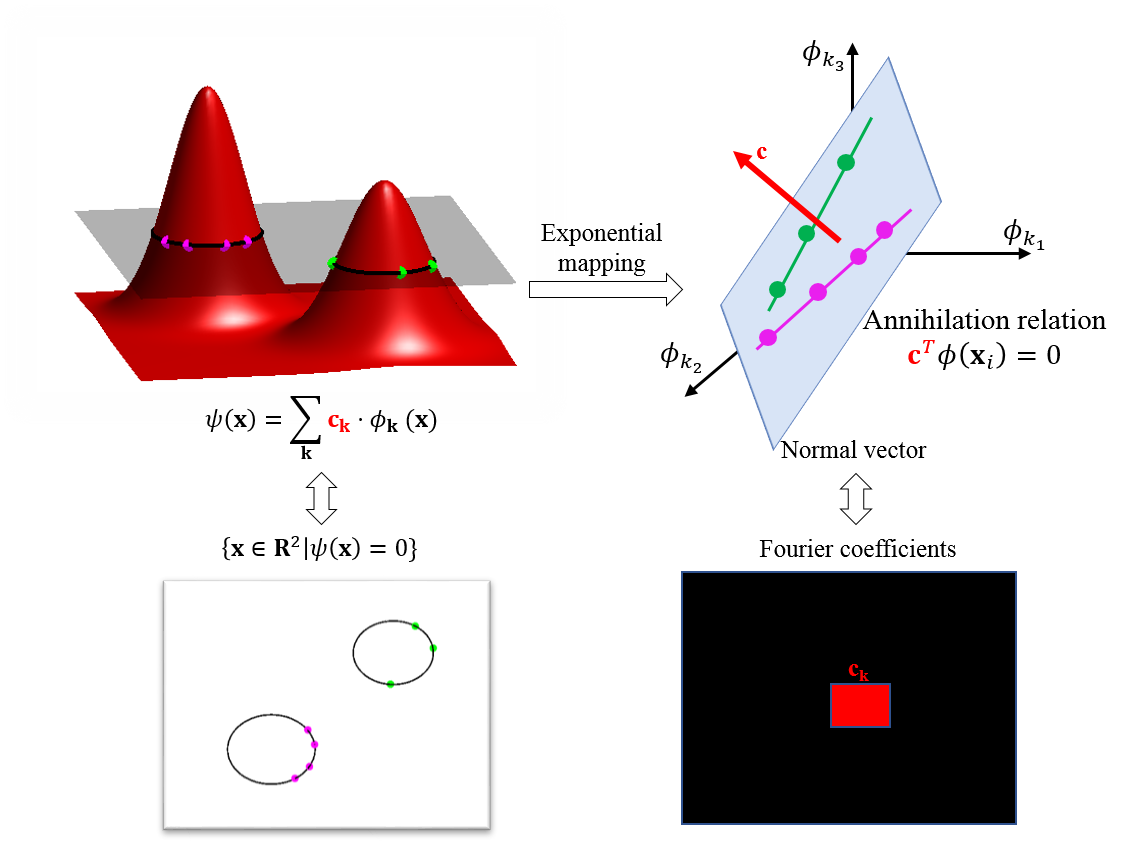}
\caption{Illustration of the annihilation relations in 2D. We assume that the curve is the zero level set of a band-limited function $\psi(\mathbf x)$, shown by the red function in the top left and the plane slicing  the function gives us the level set of the function. The Fourier coefficients of $\psi$, denoted by $\mathbf c$, are support limited in $\Lambda$, denoted by the red square on the figure in the bottom right. Each point on the curve satisfies $\psi(\mathbf x_i)=0$. Using the representation of the curve specified by \eqref{model}, we thus have  $\mathbf c^T\phi_{\Lambda}(\mathbf x_i)=0$. Note that $\phi_{\Lambda}(\mathbf x_i)$ is the exponential feature map of the point $\mathbf x_i$, whose dimension is specified by the cardinality of the set $\Lambda$. This means that the feature map will lift each point in the level set to a $\Lambda$ dimensional subspace whose normal vector is specified by $\mathbf c$, as illustrated by the plane and the red vector $\mathbf c$ in the top right. Note that if more than one closed curve are presented, each curve will be lifted to a lower dimensional subspace in the feature space, as shown by the two lines in the plane, and the lower dimensional spaces will span the $\Lambda$ dimensional subspace.}
\label{illus}
\end{figure}

\section{Bandlimited Curves}

\subsection{Parametric level set representation of curves}
We model the curve in $[0,1)^n$, as the zero level set 
\begin{equation}
\label{implicit}
\{\mathbf x \in \mathbb R^n|\psi(\mathbf x)=0\}
\end{equation}
of $\psi$. We denote the curve specified by the zero level set of $\psi$ by $\mathcal C[\psi]$. Note that the level set representation of curves is widely used in segmentation and shape representation \cite{burger2005survey}. Several authors have proposed to represent $\psi$ as a linear combination of basis functions $\varphi_{\mathbf k}(\mathbf x)$ \cite{Aghasi2011,Bernard2009}:
\begin{equation}
\label{generalmodel}
\psi(\mathbf x) = \sum_{\mathbf k \in \Lambda} \mathbf c_{\mathbf k} ~\varphi_{\mathbf k}(\mathbf x).
\end{equation}
A popular choice is the shift invariant representation \cite{Unser:2000p5230} using compactly supported basis functions such as B-splines \cite{Bernard2009,Aghasi2011} or radial basis functions \cite{Turk1999}. In this case, we have $\varphi_{\mathbf k}(\mathbf x) = \varphi(\mathbf x-\mathbf k)$, where $\varphi(\mathbf x)$ is a B-spline or Gaussian function. Here, $|\Lambda|$ denotes the number of basis functions, which is equivalent to the number of grid points in the context of shift-invariant representations. 

\subsection{Band-limited level set representation of curves}
\label{trig}
We now consider a special case of the parametric level set curve representation, in which the basis functions are chosen as $\varphi_{\mathbf k}= \exp(j~2\pi\mathbf k^T \mathbf x); \mathbf k \in \Lambda$, $j=\sqrt{-1}$. Then
\begin{equation}
\label{model}
\psi(\mathbf x) = \sum_{\mathbf k \in \Lambda} \mathbf c_{\mathbf k} \exp(j~2\pi\mathbf k^T \mathbf x); ~\mathbf x\in [0,1)^2
\end{equation}
Here, the Fourier coefficients $\mathbf c_{\mathbf k}$ are supported on a rectangular grid $\Lambda\subset \mathbb Z^{n}$. We use Hermitian symmetry property to create the curves in our experiements. But in the proofs and estimation, the Hermitian symmetry property is not utilized. Here, ${\rm BW}(\varphi) = |\Lambda|$ is the bandwidth of the curve $\mathcal C[\psi]$; $|\Lambda|$ denotes the cardinality of the set. See Fig. \ref{illus} for an illustration. Note that for each index $\mathbf k$, the basis functions satisfy $|\varphi_{\mathbf k}(\mathbf x)|=1; \forall \mathbf x$. The non-compact nature of the basis functions unlike \cite{Aghasi2011,Bernard2009} makes the analysis of the problem easy. We note that the band-limited level set model can represent closed curves with arbitrary complexity. See Fig. 2 in \cite{Ongie2016b} for examples. The complexity of the curve is dependent on the cardinality of the set $\Lambda$, which we denote by $|\Lambda|$, which is the number of free parameters in the curve representation. We now illustrate the geometric properties of bandlimited curves.

\subsection{Properties of bandlimited curves}
\label{trigBack}
We now briefly introduce the relation between  band-limited functions and complex polynomials \cite{Ongie2016b} in 2-D. We use the change of variables: $(x_1,x_2) \mapsto (z_1,z_2) = (\exp(j2\pi x_1), \exp(j2\pi x_2))$, which gives a one-to-one mapping \cite{Ongie2016b} $\mathcal P$ from $[0,1)^2$ to the unit torus $\mathbb{T}^2$. This mapping allows us to rewrite \eqref{model} as the complex polynomial:
\begin{equation}\label{key}
\mathcal{P}[\psi](\mathbf z) = \sum_{\mathbf k \in \Lambda} c_{\mathbf k}~\prod_{i=1}^{2}z_i^{k_i}
\end{equation}
Since $\mathcal{P}$ is one-to-one, the space of bandlimited functions $\psi(\mathbf{x})$ in \eqref{model} and the space of complex polynomials living on $\mathbb{T}^2$ are isomorphic. We can analyze the properties of $\mathcal{P}[\psi]$ to deduce the properties of the trigonometric polynomial \eqref{model}.



\subsubsection{\underline{Irreducible curves}} We now consider the simplest possible curve, which consists of a single connected component.
\begin{defn}[Irreducible trigonometric polynomials]\label{defired}
	A polynomial $\eta(\mathbf z)$ with complex coefficients is irreducible, if it cannot be expressed as the product of two or more non-constant polynomials with complex coefficients \cite{ency}. The trigonometric polynomial $\eta(\mathbf x)$ is termed as irreducible, if the polynomial specified by $\mathcal P[\eta]$ is irreducible.
\end{defn}
\begin{defn}[Irreducible curve] A trigonometric curve $\mathcal C$ is irreducible if it is the zero level set of an irreducible trigonometric polynomial.
\end{defn}
In most cases, the zero level sets of  irreducible trigonometric polynomials have only one component, except for some pathological cases. We consider the irreducible curves with only one connected component in the rest of the paper.

 \begin{prop}\cite[Prop.11]{Ongie2016b}
	\label{propirred}
	Trigonometric curves are closed in $[0,1)^2$.
\end{prop}
Note that by definition, the curves are periodic in  $[0,1)^2$. We observe from the segmentation examples in Fig. \ref{segment} that this restriction does not pose a  problem in practical applications; some additional boundary curves are added, which can be ignored in practice. If this poses a problem, the image may be padded by additional zero regions. While the representation cannot represent open curves in the strict sense, an open curve can be approximated with arbitrary accuracy by a bounded closed curve \cite{Kublik2016}. 
 
 \subsubsection{\underline{Union of irreducible curves}}
The above curves are often too simple to represent real-world structures. One can represent curves with multiple closed \& connected components as the zero level set of a bandlimited function with multiple irreducible factors $\psi=\eta_1\cdot\eta_2\ldots \eta_n$. In this case, the resulting curve is the union of the corresponding irreducible curves
 \begin{equation}\label{uic}
 \mathcal C[{\psi}] =  \mathcal C[\eta_1] \bigcup  \mathcal C[\eta_2] \ldots  \bigcup \mathcal C[\eta_n]
 \end{equation} 
Note that the coefficient vector of the union of irreducible curves $\psi$ is obtained as the convolution between the coefficient vectors of the irreducible components $\eta_i$:
\begin{equation}\label{key}
\mathbf c = \mathbf d_1 * \mathbf d_2\ldots * \mathbf d_n, 
\end{equation}
where $\mathbf d_i \stackrel{\mathcal F}{\leftrightarrow} \eta_i$. This implies that the bandwidth of $\psi$ is greater than that of the components.

 \subsubsection{\underline{Minimal Polynomial}}  Note that if $\psi(\mathbf{x})$ defines a curve, then $\eta(\mathbf{x})=\exp(j2\pi\mathbf{k}^T\mathbf{x})\cdot\psi(\mathbf{x})$ will also define the same curve for $\mathbf{k}\in\mathbb{Z}^2$ since $\exp(j2\pi\mathbf{k}^T\mathbf{x})$ never vanishes in $[0,1)^2$. But the bandwidth of $\eta$ is higher than that of $\psi$. We term the polynomial with the smallest bandwidth to be the minimal polynomial of a curve, which is the object we are interested in in this paper. The following result about minimal polynomial of a curve guarantees the existence of the minimal polynomial. 

 \begin{prop}\cite[Prop. 2]{Ongie2016b}
For every trigonometric curve $\mathcal C$ there is a unique trigonometric polynomial $\mu_0$ with $\mathcal C = \{\mu_0 = 0\}$ such that for any other trigonometric polynomial $\mu$ with $\mathcal C = \{\mu = 0\}$, we have ${\rm BW}(\mu_0) \leq {\rm BW}(\mu)$ componentwise.
\end{prop}
We note that in the 1-D setting ($n=1$), there is a one-to-one correspondence between the number of points on the zero set of $\psi$ and the bandwidth $|\Lambda|$ \cite{Ongie2016b,Fatemi2016}; this relation enabled the use of $|\Lambda|$ in \cite{Ongie2016b,ongie2017convex} as a surrogate for sparsity in signal recovery. In higher dimensions, $|\Lambda|$ can still serve as a complexity measure. Note that the zero-set can consist of isolated Diracs in 2D, which implies that ${\rm BW}(\psi)$ can still serve as the surrogate for sparsity. However, we emphasize that \eqref{model} can provide a significantly richer representation, even when the signal is not isolated and consists of points on a curve. 

\section{Sampling of band-limited curves} 
\label{sampling}
\subsection{Annihilation relations for points on the curve}
\label{single}
Consider an arbitrary point $\mathbf x$ on the curve specified by \eqref{implicit} and \eqref{generalmodel}. By definition, we have $\psi(\mathbf x)=0$, which translates to:
\begin{eqnarray}\label{annn}
\label{curverep}
\psi(\mathbf x) &=& \sum_{\mathbf k \in \Lambda} \mathbf c_{\mathbf k}\; \varphi_\mathbf{k}(\mathbf x)= \mathbf c^T \underbrace{ \begin{bmatrix} \varphi_{k_1}(\mathbf x)\\
 \vdots\\
 \varphi_{k_{|\Lambda|}}(\mathbf x)
\end{bmatrix}}_{\phi_{\Lambda}(\mathbf x)} = 0
\end{eqnarray}

Note that  $\phi_{\Lambda}: \mathbb R^n \rightarrow \mathbb C^{|\Lambda|}$ is a non-linear mapping or lifting of a point $\mathbf x$ to a high dimensional space, whose dimension is given by the cardinality of the set $\Lambda$, denoted by $|\Lambda|$. Note that this non-linear lifting strategy is similar to feature maps used in kernel methods. We hence term  $\phi_{\Lambda}(\mathbf x)$ as the feature map of the point $\mathbf x$. Note that every point on the curve satisfies \eqref{annn}, which we term as the annihilation relation. 

Let us now consider a set of $N$ points on the curve, denoted by $\mathbf x_1,\cdots,\mathbf x_N$. Note that  the feature maps of each one of the points satisfy the above annihilation relations, which can be compactly represented as: 
\begin{equation}
\label{ann}
 \mathbf c^T \underbrace{ \begin{bmatrix}
\phi_{\Lambda}(\mathbf x_1) & \phi_{\Lambda}(\mathbf x_2) & \ldots &\phi_{\Lambda}(\mathbf x_N) 
\end{bmatrix}}_{\Phi_{\Lambda}(\mathbf X)} =\mathbf{0}.
\end{equation}
Here, $\Phi_{\Lambda}(\mathbf X)$ is the feature matrix of the points and $\mathbf X = [\mathbf x_1~ \mathbf x_2 ~\ldots ~ \mathbf x_N]$. 

Assume that we have a union of irreducible curves as in \eqref{uic}, where the bandwidth of each of the irreducible components $\mathcal C[\eta_i]$ is $\Lambda_i$ and bandwidth of $\mathcal C[\psi]$ is $\Lambda$. In this case, the $|\Lambda_i|$ dimensional lifting $\Phi_{\Lambda_i}(\mathbf x)$ of the samples on $\mathcal C[\eta_i]$ will lie on a $|\Lambda_i|-1$ dimensional subspace. Similarly, the $|\Lambda|$ dimensional lifting $\Phi_{\Lambda}(\mathbf x)$ of the samples on the union of irreducible curve $\mathcal C[\psi]$ will lie on a $|\Lambda|-1$ dimensional subspace. 

\subsection{Curve recovery from samples}
\label{curverecoveryalgorithm}
When $\Phi_{\Lambda}$ is rank-deficient by one, the coefficient vector $\mathbf c$ can be identified as the unique non-zero null-space basis vector of $\Phi_{\Lambda}(\mathbf X)$. This implies that the features lie in an $|\Lambda|-1$ dimensional subspace, whose normal is specified by $\mathbf c$. This annihilation relation is illustrated in Fig \ref{illus}, in the context of band-limited curves considered in the next subsection. We will show that there exists a unique null-space basis vector when complex exponential basis functions are chosen as in Section \ref{trig}.

In practice, the points are often corrupted by noise. In the presence of noise, the null-space conditions are often not satisfied exactly. In this case, we can pose the least square estimation of the coefficients from the noisy data points $\{\mathbf x_i\}_{i=1}^{N}$ as the minimization of the criterion: 
\begin{equation}
\mathcal C(\mathbf c) = \sum_{i=1}^N \|\psi(\mathbf x_i)\|^2= \mathbf c^T \mathbf Q_{\Lambda} \mathbf c
\end{equation}
where $\mathbf Q_{\Lambda}=\sum_{i=1}^N \phi_{\Lambda}(\mathbf x_i)\phi_{\Lambda}(\mathbf x_i)^T$. To eliminate the trivial solution $\mathbf c=0$, we pose the recovery as the constrained optimization scheme:
\begin{equation}
\label{eigen}
\mathbf c^* = \arg \min_{\mathbf c} \mathbf c^T ~\mathbf Q_{\Lambda}~ \mathbf c ~~\mbox{such that }~~ \|\mathbf c\|^2 = 1
\end{equation}
The solution is the eigenvector corresponding to the minimum eigenvalue of $\mathbf Q_{\Lambda}$. Note that $\mathbf Q_{\Lambda}$ is nothing but $\Phi_{\Lambda}(\mathbf X)\Phi_{\Lambda}^T(\mathbf X)$. Thus we just need to use the singular value decomposition of $\Phi_{\Lambda}^T(\mathbf X)$ to obtain the desired solution.

\subsection{Irreducible band-limited planar curve: sampling theorem}

We now focus on the problem of the recovery of the curve \eqref{implicit}, given a few points $\{\mathbf x_i\in \mathbb R^2; i=1, \cdots ,N\}$ on the curve \footnote{All the experiments on curve recovery in Section \ref{sampling} are run on a desktop computer with an Intel Core i7-2600 CPU.}. Let us take the band-limited curve representation as \eqref{model} for the rest of the section to derive our sampling conditions. We now determine the sampling conditions for the perfect recovery of the curve $\psi(\mathbf x)=0$ using \eqref{eigen}, when the curve is specified by \eqref{model}.  In this case, the annihilation relation \eqref{curverep} is satisfied with the feature maps defined as 
\begin{eqnarray}
\phi_{\Lambda}(\mathbf x)&=&  \begin{bmatrix} \exp(j~2\pi\mathbf k_1^T\mathbf x)\\
 \vdots\\
  \exp(j~2\pi\mathbf k_{|\Lambda|}^T\mathbf x)\\
\end{bmatrix}
\end{eqnarray}

We also assume that $\Lambda$ is a rectangular neighborhood in $\mathbb Z^2$ of size $k_1\times k_2$. We first review some results from algebraic geometry.

There is a one-to-one correspondence between trigonometric polynomials and complex polynomials, as shown in Section \ref{trigBack}. We use the extension of B\'ezout's inequality for trigonometric polynomials, which bounds the number of solutions of the system $\mu(\mathbf x)=\eta(\mathbf x)=0$ that do not have any common factors.

\begin{lem}[B\'ezout's inequality for band-limited polynomials]  
\label{prop0}
Let $\mu(\mathbf x)$ and $\eta(\mathbf x)$ be two band-limited polynomials,  whose Fourier coefficients are support limited to $k_1\times k_2$ and $l_1\times l_2$, respectively. If $\mu$ and $\eta$ have no common factor, then the system of equations 
\begin{equation}
\label{syseq}
\mu(\mathbf x)=\eta(\mathbf x)=0
\end{equation}
has a maximum of $(k_1+k_2)(l_1+l_2) = {\rm deg}(\mu){\rm deg}(\eta)$ solutions in $[0,1)^2$. 
\end{lem}
The proof of Lemma \ref{prop0} is given in Appendix \ref{proofBez}. We use this property to derive our main results. We first focus on the case where $\psi$ is an irreducible band-limited function. 

 \begin{prop}
 \label{prop2D1}
Let $\{\mathbf x_i\}_{i=1}^{N}$ be $N$ distinct points on the zero level set of an irreducible band-limited function $\psi(\mathbf x), \mathbf x \in \mathbb{R}^2$, whose Fourier coefficients are restricted to a rectangular region $\Lambda$ with size $k_1\times k_2$. Then the curve $\psi(\mathbf x) = 0$ can be uniquely recovered by \eqref{ann}, when:
\begin{equation}
\label{nsamples1}
N> (k_1+k_2)^2 = {\rm deg}^2(\psi)
\end{equation}
\end{prop}

The proof is provided in Appendix \ref{proof2D1}.

\begin{figure}[t!]
	\centering
          \subfigure[Original curve]{\includegraphics[width=0.115\textwidth]{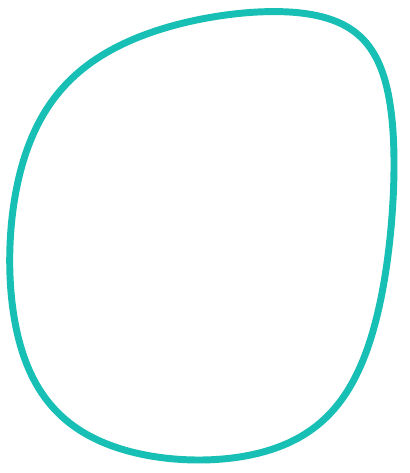}}
          \subfigure[36 random samples]{\includegraphics[width=0.115\textwidth]{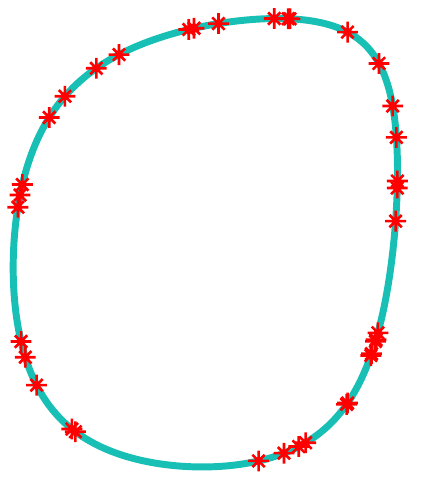}}
          \subfigure[36 samples in left half]{\includegraphics[width=0.115\textwidth]{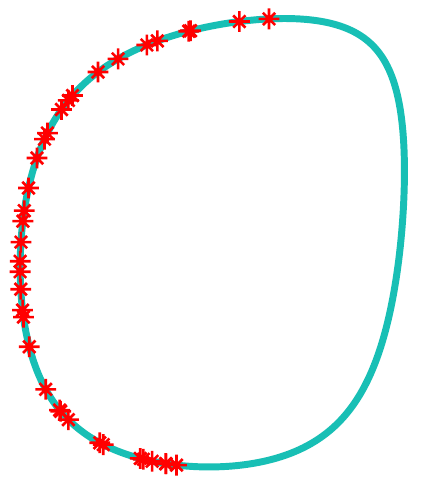}}
          \subfigure[36 samples in right half]{\includegraphics[width=0.115\textwidth]{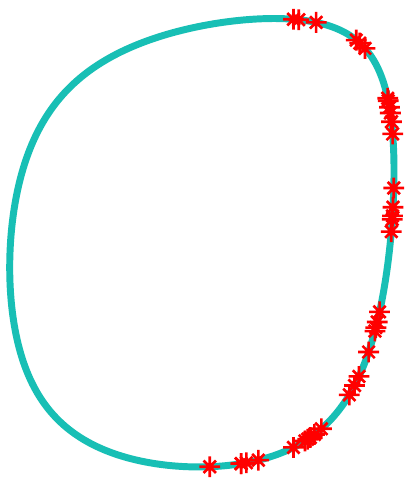}}
	\caption{Illustration of Proposition \ref{prop2D1}: We consider a curve $C[\psi]$ given by $\psi(\mathbf{x})$, where  $\mathbf c_{\psi} \stackrel{\mathcal F}{\leftrightarrow} \psi$ is support limited to a $3\times 3$ region, shown in (a). The theorem guarantees the perfect recovery will happen if we have no less than $(k_1+k_2)^2=36$ samples. We first randomly chose 36 samples on the curve. Then from these 36 randomly chosen samples, we obtained (b), which gives us perfect recovery of the original curve. Furthermore, we mentioned that we do not require any constraint on the distribution of samples on the curve. In (c), we randomly chose 36 samples from the left half part of the curve and we got perfect recovery as well. In (d), 36 samples are randomly chosen from the right half of the curve. From (d), we saw that perfect recovery of the whole curve was also obtained. For each case, the average time required for the recovery is about 1.2 second.}
	\label{prop0Sim1}
\end{figure}

Note that the sampling condition for a single irreducible curve does not specify any constraint on the distribution of points on the curve; any set of $N> (k_1+k_2)^2$ points are sufficient for the recovery of the curve. This proposition is illustrated in Fig. \ref{prop0Sim1}, which shows that the recovery is guaranteed irrespective of the distribution of samples. This property is similar to well-known results in non-uniform sampling of band-limited signals \cite{Maymon2011}, where the recovery is guaranteed under weak conditions on the nonuniform grid and  the average sampling rate exceeding Nyquist rate. 

We compare this setting with the sampling conditions for the recovery of a piecewise constant image, whose gradients vanish on the zero level set of a band-limited function \cite{Ongie2016b}. The minimum number of Fourier measurements required to recover the function there is $|3\Lambda|$. When $k_1=k_2=K$, then $9K^2$ complex Fourier samples are required, which is far more than $4K^2$ real samples required for the recovery of the curve in our setting. Note that the constant values within the regions bounded by the curves also need to be recovered in \cite{Ongie2016b}, which explains the higher sampling requirement. We note that the above bounds are looser than the ones in \cite{pan2014sampling}, which are based on the number of available equations; they assume the chances of the equations being linear dependent is unlikely \cite{pan2014sampling}. Unlike the high-probability results in \cite{pan2014sampling} that our bounds are worst-case guarantees, which will hold irrespective of the sampling geometry. We note from the experiments in Fig. \ref{prop1Sim2} that recovery succeeds in most cases whenever the number of samples exceed $|\Lambda|-1 = k_1k_2-1$, which is the number of degrees of freedom in representing the curve. 

\subsection{Union of irreducible curves: sampling theorem}
We now generalize the previous result to the setting where the composite curve is a union of multiple irreducible curves. Equivalently, the level set function is the product of multiple irreducible band-limited functions. We have the following result for this general case:

\begin{prop}
\label{prop2D2}
Let $\{\mathbf x_i\}_{i=1}^{N}$ be points on the zero level set of a band-limited function $\psi(\mathbf x), \mathbf x \in \mathbb{R}^2$, where the bandwidth of $\psi$ is specified by $|\Lambda| = k_1\times k_2$. Assume that $\psi(\mathbf x)$ has $J$ irreducible factors (i.e., $\psi = \eta_1\cdots\eta_J$), where the bandwidth of the $j^{th}$ factor is given by $k_{1,j} \times k_{2,j}$.  The curve $\psi(\mathbf x) = 0$ can be uniquely recovered by \eqref{ann}, when each of the irreducible curves are sampled with 
\begin{equation}
\label{nsamples2}
N_j~>~ (k_1+k_2)(k_{1,j}+k_{2,j}) = {\rm deg}(\psi){\rm deg}(\eta_j); ~j=1,\cdots,J.
\end{equation}
The total number of samples needed for unique recovery is specified by 
\begin{equation}
N = \sum_{j=1}^{J}N_j = {\rm deg}(\psi) \sum_{j=1}^{J}{\rm deg}(\eta_j),
\end{equation}
which is bounded above by $(k_1+k_2)(k_1+k_2+2(J-1))$.
\end{prop}
We note that the upper bound can be approximated as $(k_1+k_2)^2$ for small values of $J$, which is the upper bound in Proposition \ref{prop2D1}. The above result is proved in Appendix \ref{proof2D2}. Note that unlike the case considered in Section \ref{single}, an arbitrary set of $N$ samples cannot guarantee the perfect recovery. Each of the $J$ irreducible curves $C[\eta_j]$ need to be sampled proportional to their complexity, specified by ${\rm deg}(\eta_j)$ to guarantee perfect recovery. 
\begin{figure}[t!]
	\centering
	\subfigure[$\psi(x,y)$]{\includegraphics[width=0.135\textwidth]{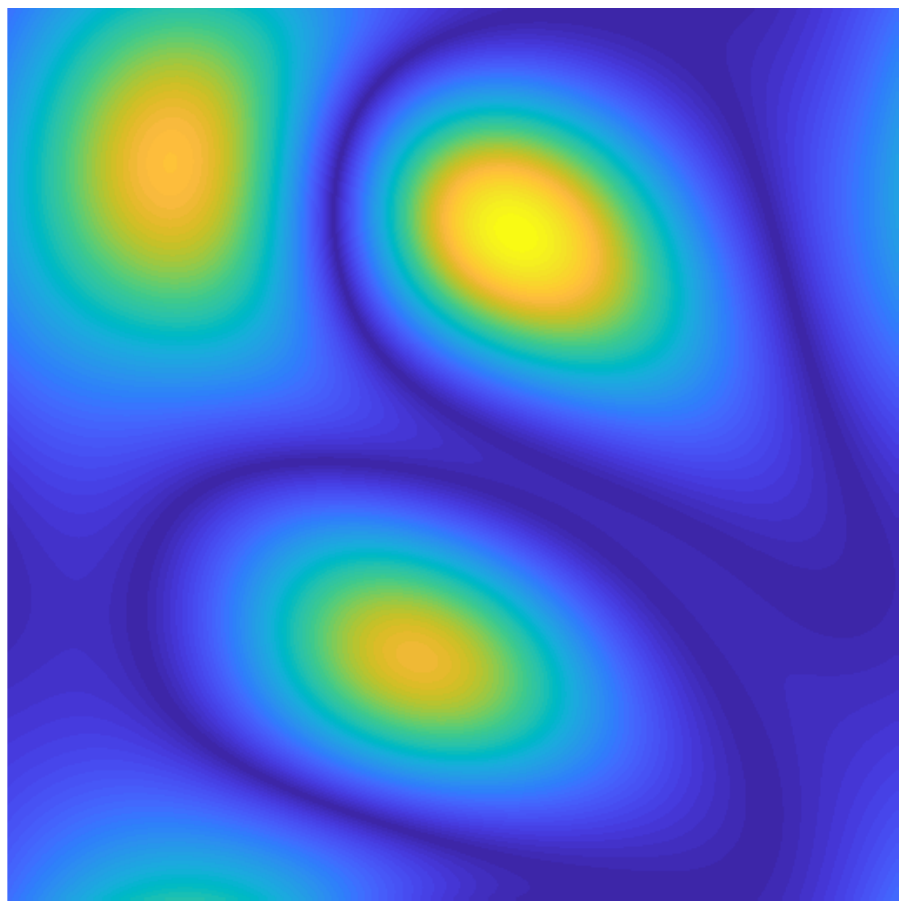}}\qquad
	\subfigure[$\psi(x,y)=0$]{\includegraphics[width=0.135\textwidth]{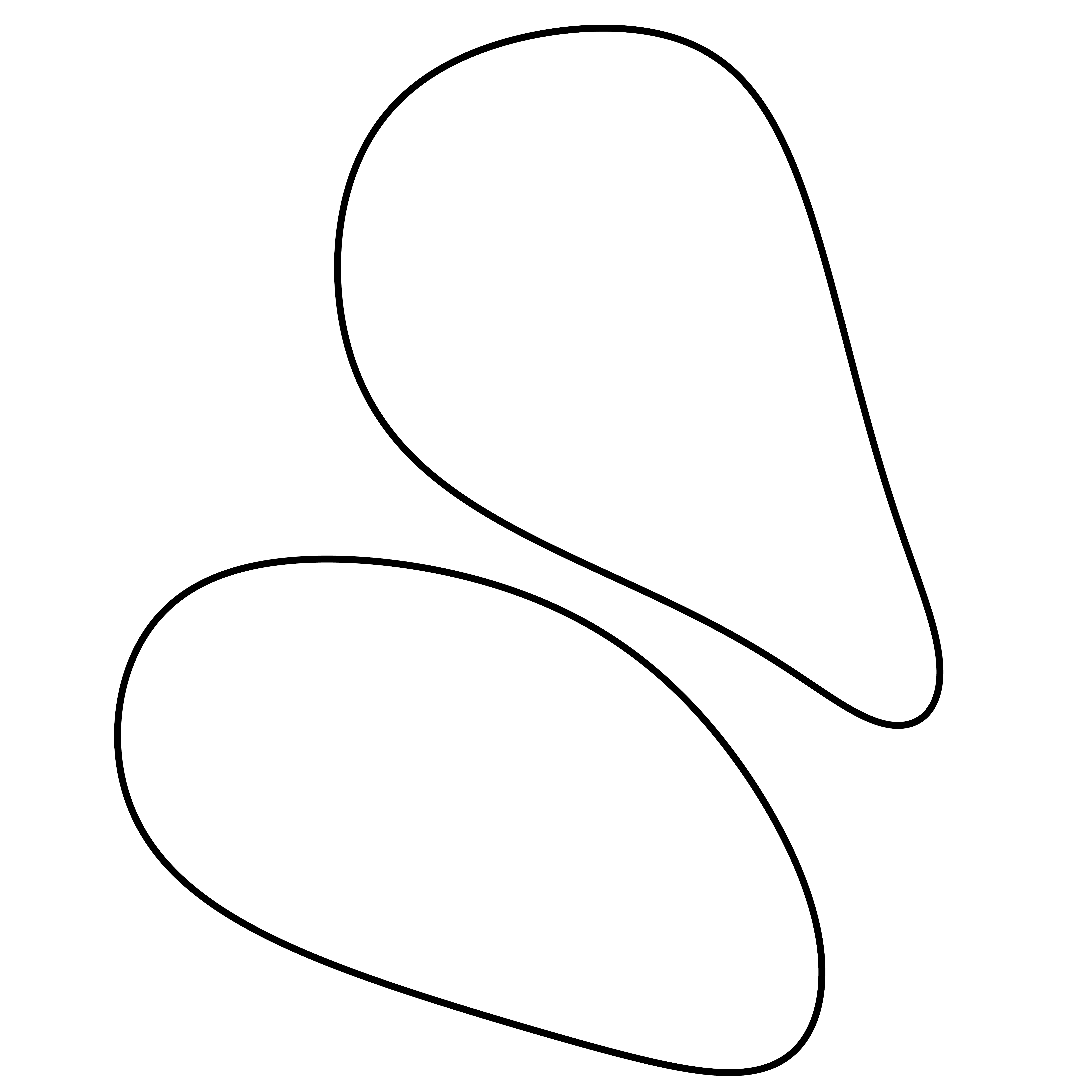}}\\
	\subfigure[10 points]{\includegraphics[width=0.135\textwidth]{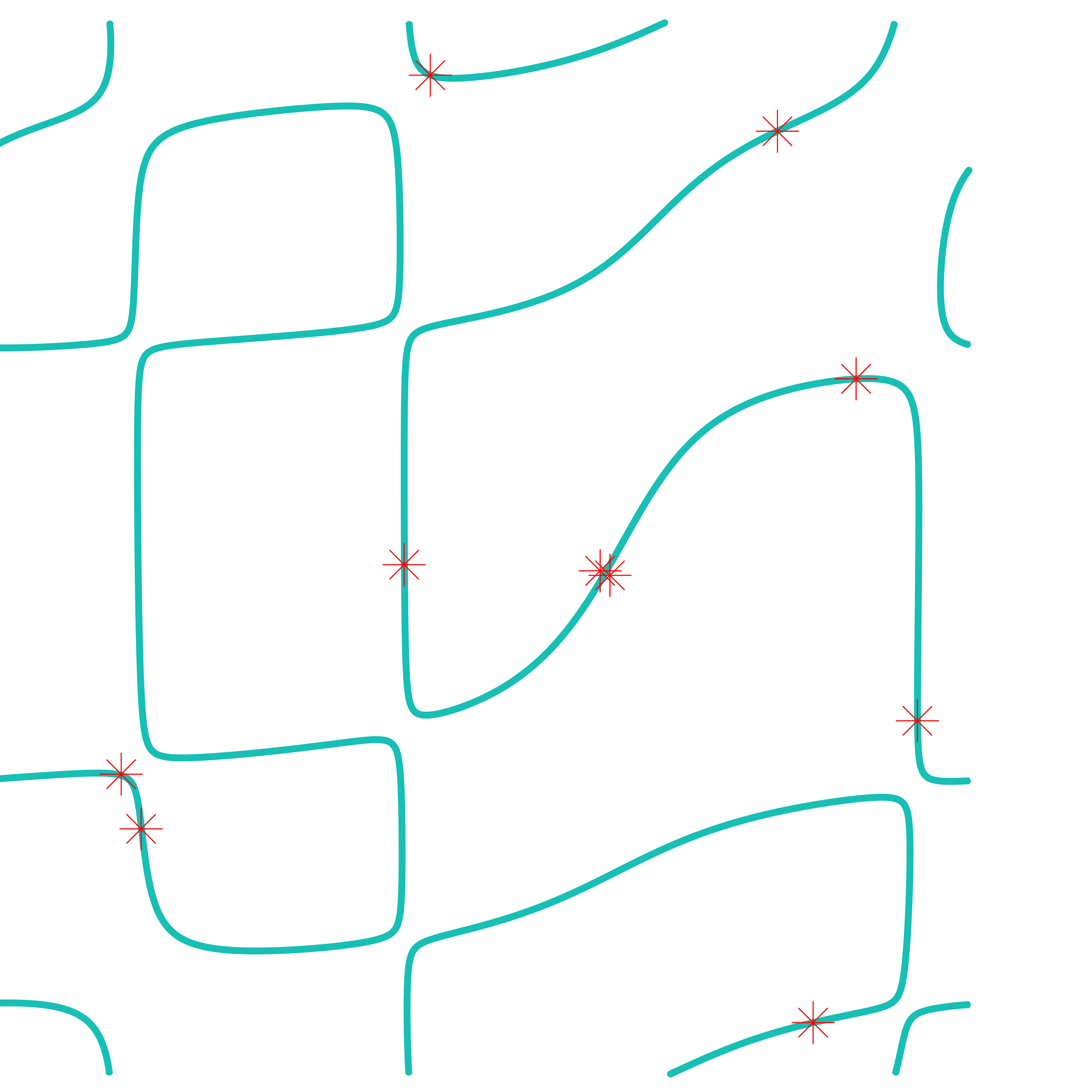}}
	\subfigure[25 points]{\includegraphics[width=0.135\textwidth]{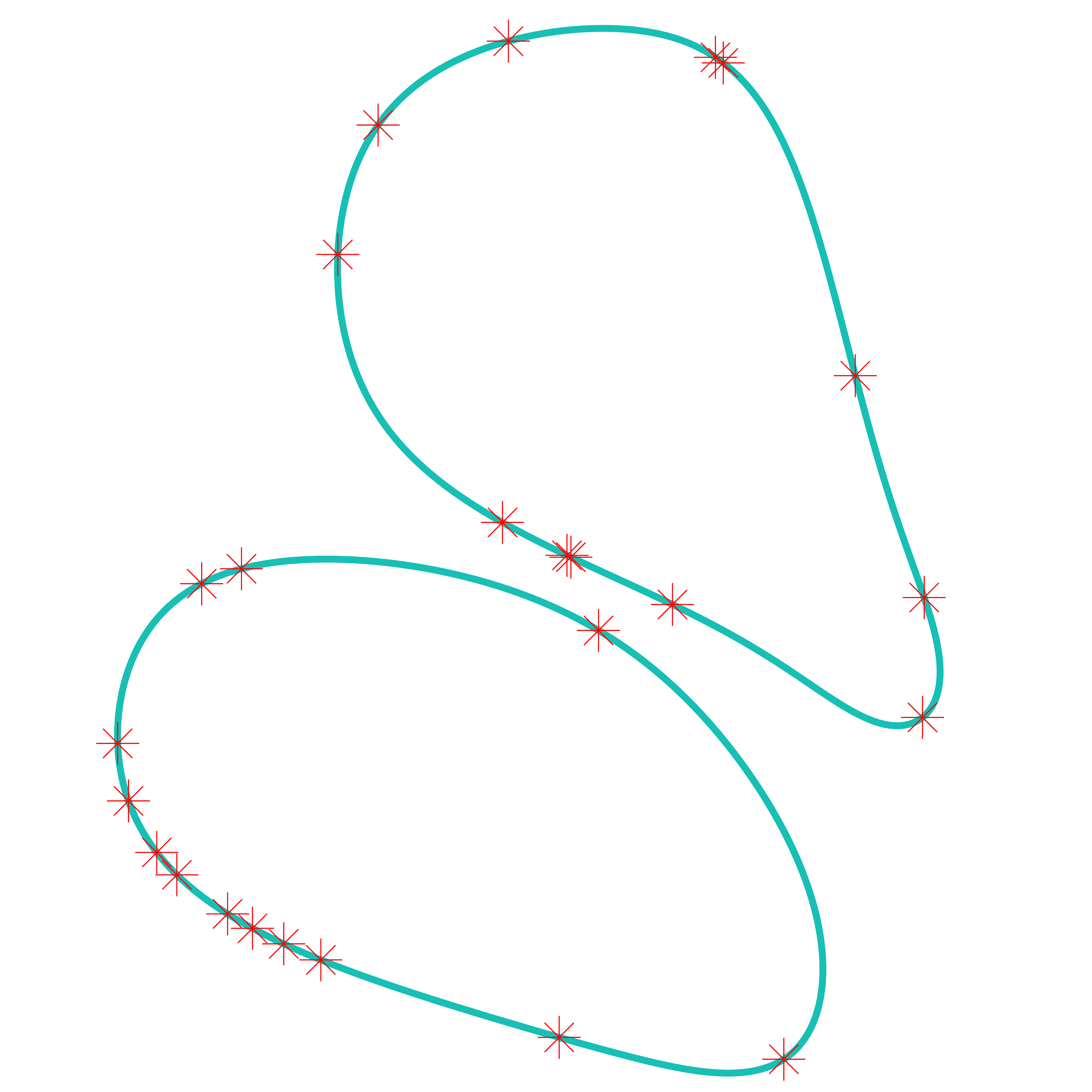}}
	\subfigure[50 points]{\includegraphics[width=0.135\textwidth]{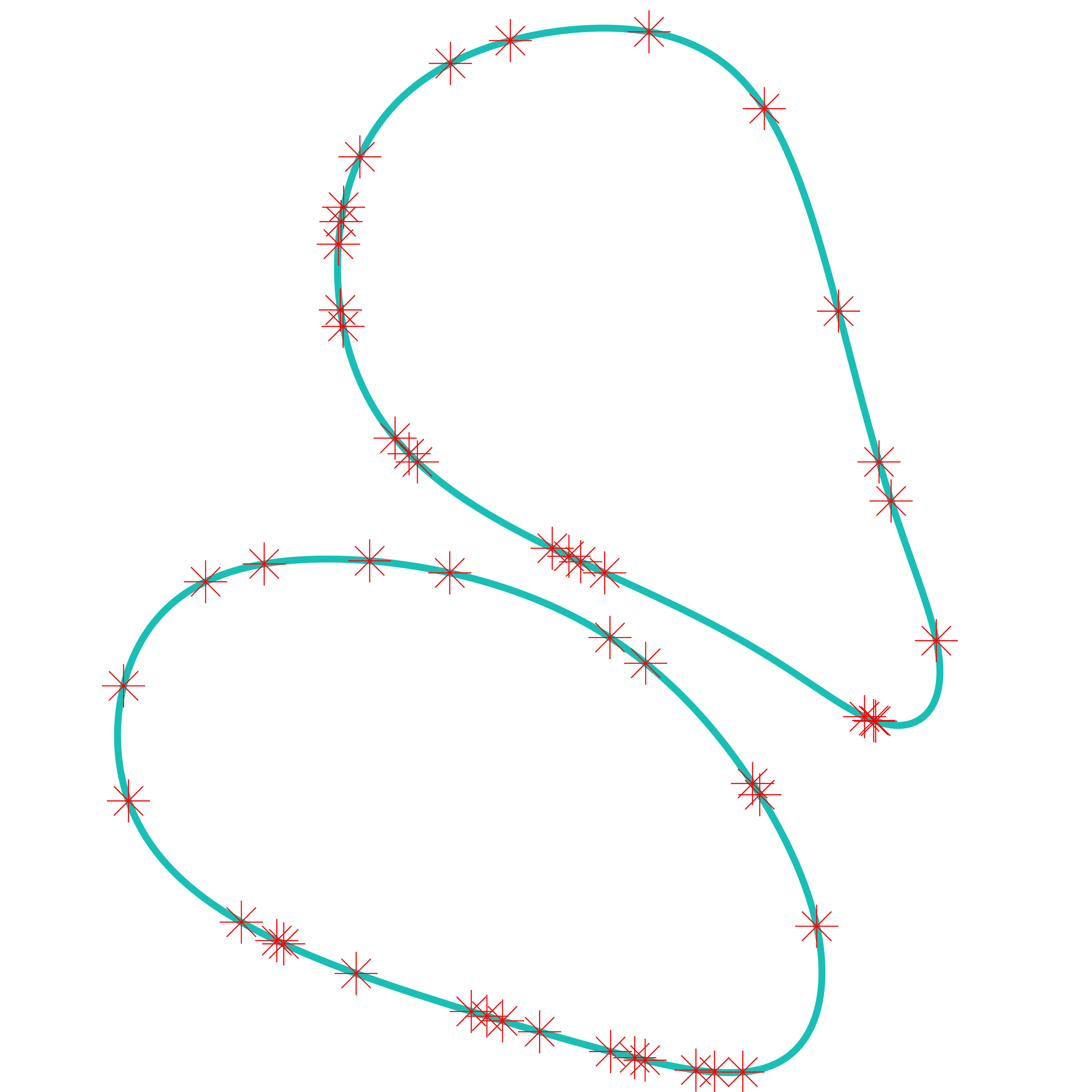}}
	\caption{Illustration of Proposition \ref{prop2D2}: We consider a curve $C[\psi]$ on the top right, where  $\mathbf c_{\psi} \stackrel{\mathcal F}{\leftrightarrow} \psi$ is support limited to a $5\times 5$ region. The level set function is shown in the top left. We consider the recovery from different number of samples of $C[\psi]$, sampled randomly. The sampling locations are marked by red crosses. Note that the theory guarantees the recovery when the number of samples exceeds $(k_1+k_2)^2=100$ samples. However, we observe good recovery of the curve around 50 samples. Note that our theoretical results are worst-case guarantees, and in practice fewer samples are sufficient for good recovery as seen from Fig, \ref{prop1Sim2}. On average, the computational time required for the recovery of the curve using 50 points is about 1.5 second.}
	\label{prop1Sim1}
\end{figure}

\begin{figure}[t!]
	\centering
	\subfigure[$25+25$]{\includegraphics[width=0.135\textwidth]{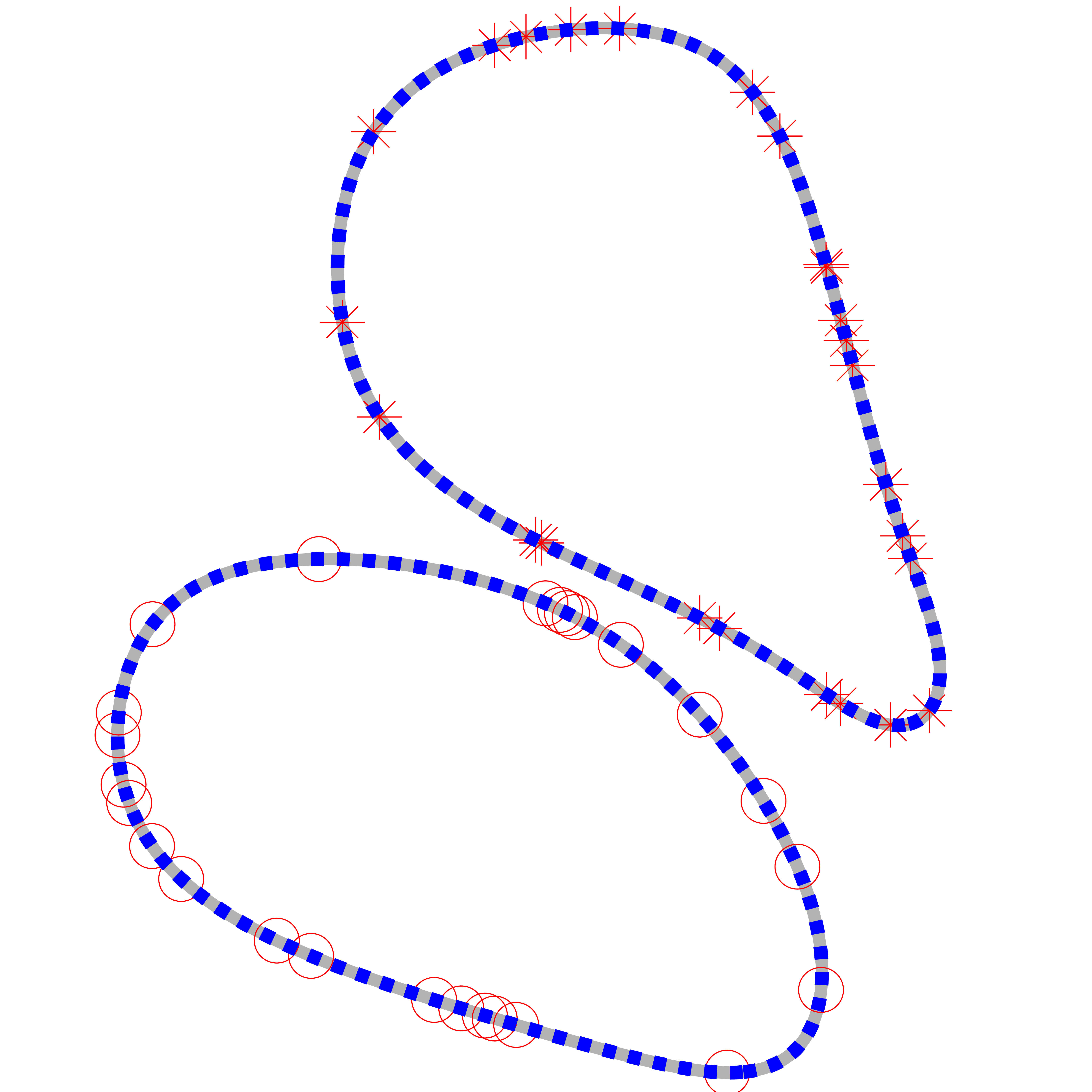}}
	\subfigure[$49+1$]{\includegraphics[width=0.135\textwidth]{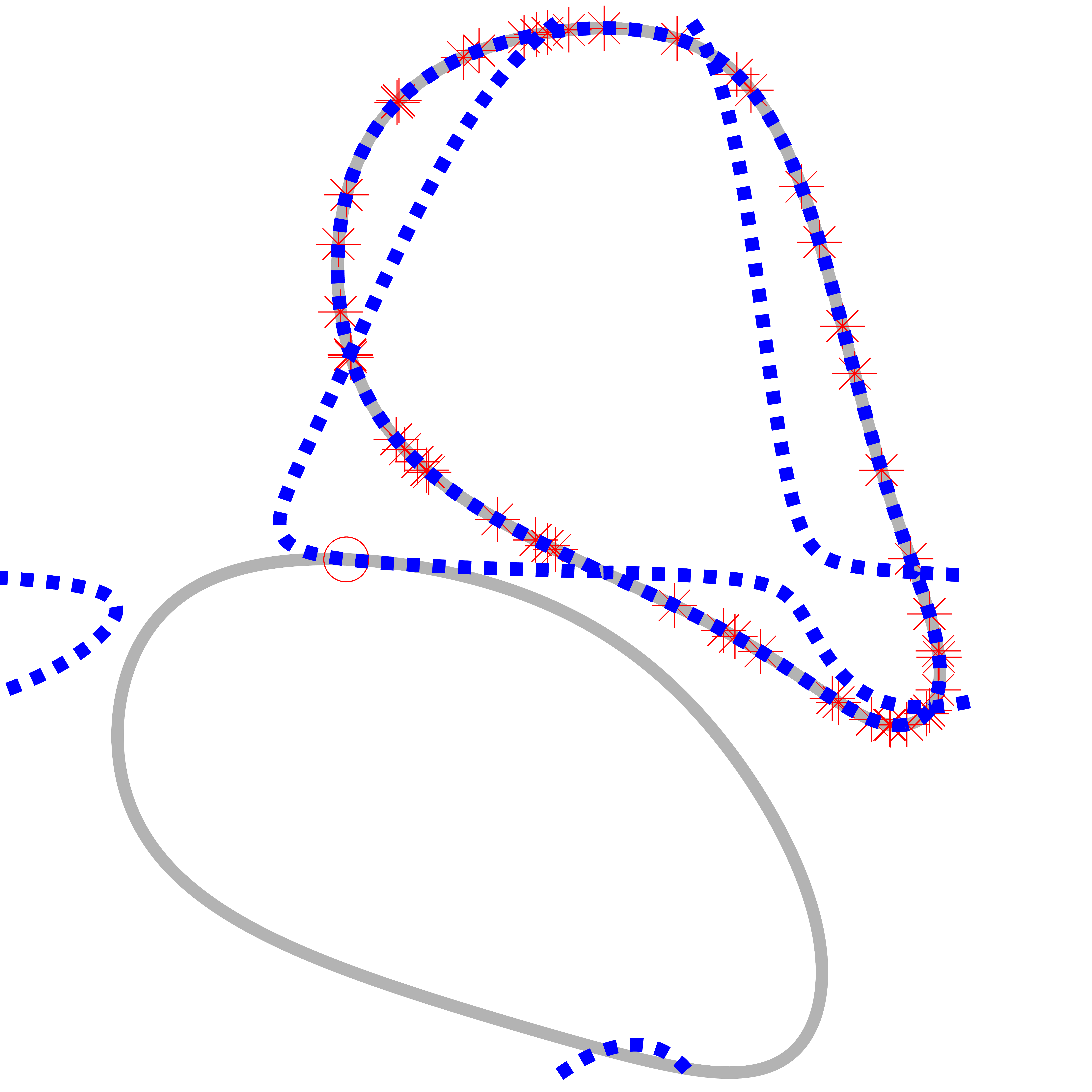}}
	\subfigure[$5+45$]{\includegraphics[width=0.135\textwidth]{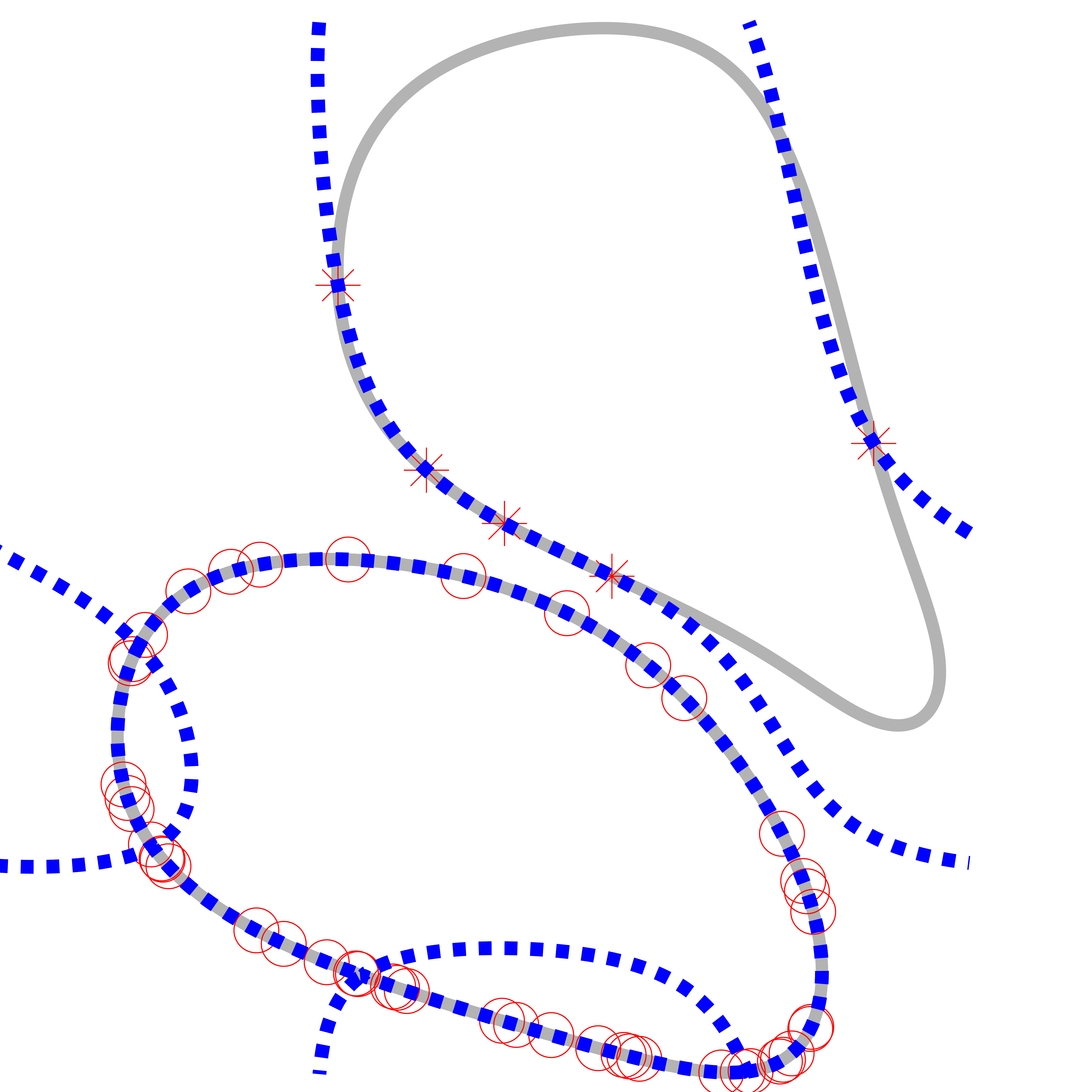}}
	\caption{Illustration of Proposition \ref{prop2D2}: We consider the same curve $C[\psi]$ as specified by Fig \ref{prop1Sim1} (b), which is given by the union of two irreducible curves with bandwidth $3\times 3$. So the bandwidth of $C[\psi]$ is $5\times 5$. According to Proposition \ref{prop2D2}, we will need to have around 100 samples to recover $C[\psi]$ and each of the two irreducible curves need to satisfy with the sampling condition. As we noted in Fig \ref{prop1Sim1}, our results are worst-case guarantees. We observe that when we have 50 points and those points are uniformly sampled on the two irreducible curves, we can successfully recover the whole curve, as shown in (a). Now, if we put most of the samples on one of the irreducible curves, we cannot fully recover the curve, as illustrated in (b) and (c). This implies that the sampling condition on each of the irreducible factors is necessary in Proposition \ref{prop2D2}.}
	\label{samplingdistribution}
\end{figure}

We demonstrate the above proposition in Fig. \ref{prop1Sim1}. We consider a curve $C[\psi]$, where  $\mathbf c_{\psi} \stackrel{\mathcal F}{\leftrightarrow} \psi$ is support limited to a $5\times5$ region. We note that there are three connected components in the above curve. We consider the recovery from different number of samples of $C[\psi]$ in the middle row, sampled randomly. The random strategy ensures that the samples are distributed to the factors, roughly satisfying the conditions in Proposition \ref{prop2D2}. Note that the theory guarantees recovery, when the number of samples exceeds around $(k_1+k_2)^2=100$ samples. We observe good recovery of the curve around 50 samples; note that our results are worst-case guarantees, and in practice fewer samples are sufficient for good recovery of most curves. We further study the distribution of the points in Fig. \ref{samplingdistribution}. The experiments demonstrate that each of the curves need to be sampled with a number proportional to the bandwidth of the curves as in (b). When the points are non-uniformly distributed as in (c) or (d), the recovery fails.

\begin{figure}[b!]
	\centering
	\includegraphics[width=0.4\textwidth]{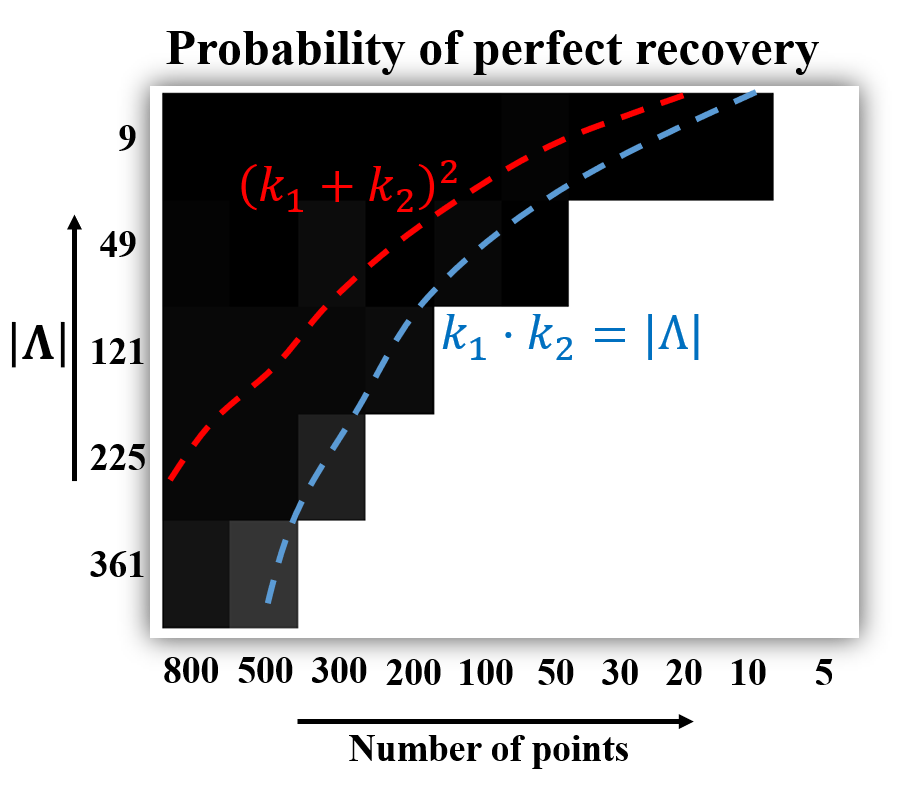}
	\caption{Effect of number of sampled points on perfect reconstruction. We randomly generated several curves with different bandwidth and number of sampled points, and recovered the curves from these samples. The success of reconstruction of the curves averaged over several trials are shown in the above phase transition plot, as a function of bandwidth and number of sampled entries. The color indicates the frequency of success; the color black indicates that the true curve cannot be recovered in any of the experiments, while the color white represents that the true curve is recovered in all the experiments. It is seen that perfect recovery occurs whenever we have $\geq (k_1+k_2)^2$ samples, as indicated by our worst-case guarantees. However, we note that good recovery is observed whenever the number of samples exceed the degrees of freedom $k_1\cdot k_2$}. 
	\label{prop1Sim2}
\end{figure}

We further studied the above proposition in Fig. \ref{prop1Sim2}. We considered several random curves, each with different bandwidth and considered their recovery from different number of samples. The sampling locations were picked at random. The colors indicate the average reconstruction error between the actual curve and the reconstructed curves. This reconstruction error is computed as the sum of distances between each point on one curve and the closest point to it on the other curve. We have also plotted the upper bound $(k_1+k_2)^2$ in red, while the number of unknowns in the curve representation $k_1\,k_2$ is plotted in blue. We note that the curve can be recovered accurately when the number of samples exceed the upper bound. We also note that in general, good recovery can be obtained for most curves, when the number of samples exceed $k_1\,k_2$.

\subsection{Curve recovery with unknown Fourier support}
\label{2Dunknown}
Propositions \ref{prop2D1} and \ref{prop2D2} assume that the true support of the Fourier coefficients of $\psi$, specified by $\Lambda$ is known, in addition to the points $\{\mathbf x_i\}_{i=1}^{N}$. However, typically only the points will be known and the filter support will be unknown. We now consider the case where the filter support is over-estimated as $\Gamma \supset \Lambda$. We focus on the recovery of the coefficients from the annihilation relation 
\begin{equation}
\label{newann}
\mathbf c^T\mathbf \Phi_{\Gamma} ~=0.
\end{equation}
The following result shows that the above matrix will have multiple linearly independent null-space vectors. However, if the curves are sampled as described below, the corresponding band-limited functions satisfy some desirable properties that facilitate the recovery of the curves.
\begin{prop}
\label{prop2D3}
Consider the zero level set of the band-limited polynomial $\psi(\mathbf x)$ with $J$ irreducible components, as described in Proposition \ref{prop2D2}. Let the assumed bandwidth of the curve be $\Gamma$ with $|\Gamma| = l_1\times l_2$ and $\Lambda \subset \Gamma$. Then, there exist multiple functions that satisfy $\mu(\mathbf x_i)=0; i=1,\cdots,N$. If the irreducible curves of the zero level set of $\psi$ are sampled with 
\begin{equation}
\label{overSamp}
N_j> (l_1+l_2)(k_{1,j}+k_{2,j}); ~~j=1,\ldots,J,
\end{equation}
all of the above functions, or equivalently the right nullspace vectors $\mathbf c_{\mu} \stackrel{\mathcal F}{\leftrightarrow} \mu$ of $\mathbf \Phi_{\Gamma}$, will be of the form: 
\begin{equation}
\label{annfunctions}
\mu(\mathbf x) = \psi(\mathbf x)\;\eta(\mathbf x)
\end{equation}
where $\eta(\mathbf x)$ is an arbitrary function such that ${\rm supp}(\mathbf c_{\mu}) = \Gamma$. 
\end{prop}

\begin{figure*}[t!] 
	\centering
          \subfigure[1st nullspace func.]{\includegraphics[width=0.145\textwidth]{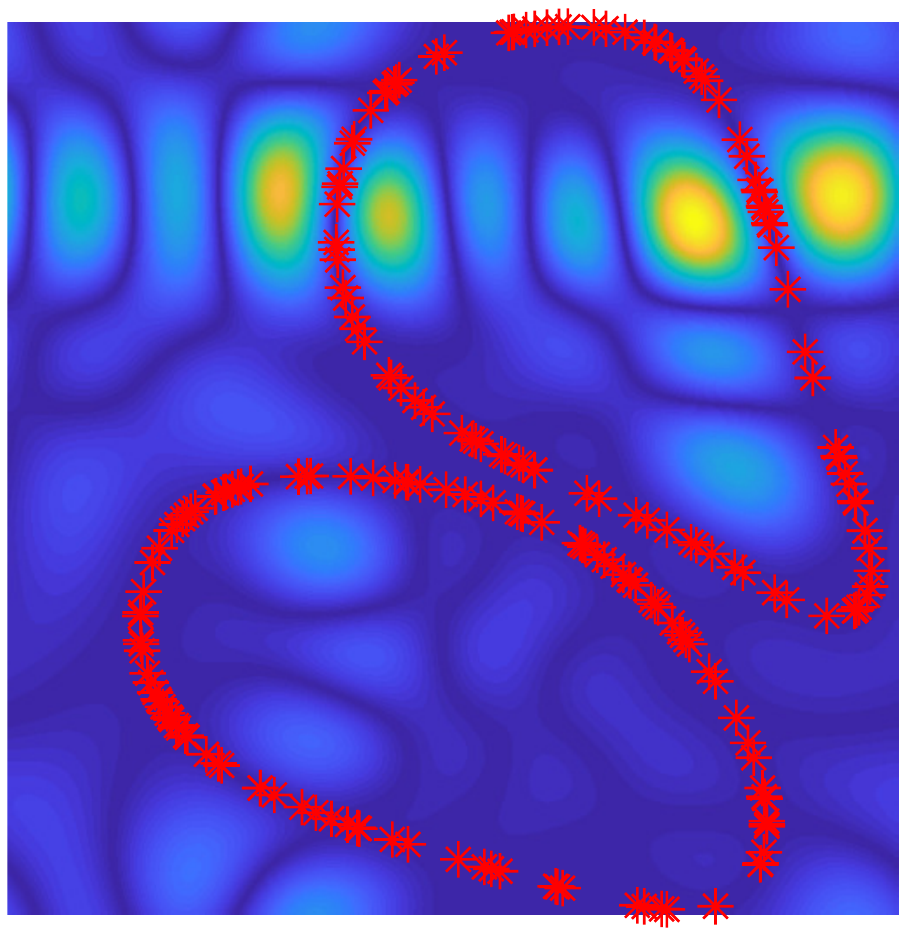}}\hspace{4ex}
          \subfigure[2nd nullspace func.]{\includegraphics[width=0.145\textwidth]{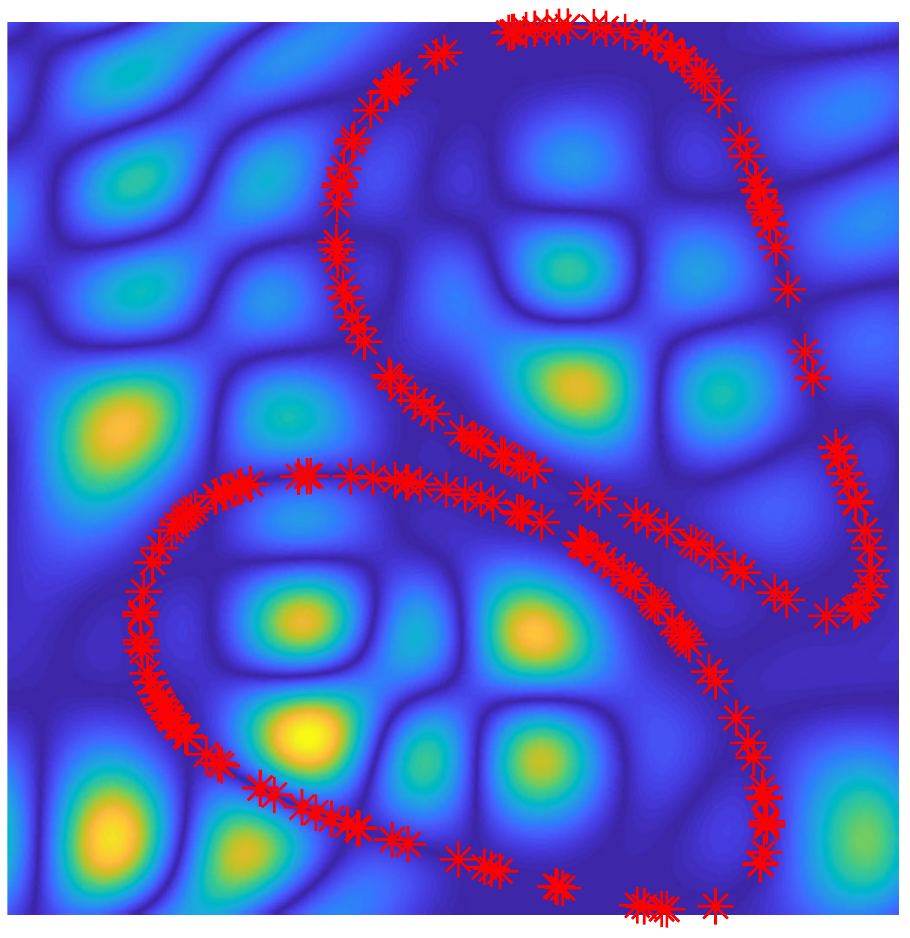}}\hspace{4ex}
          \subfigure[3rd nullspace func.]{\includegraphics[width=0.145\textwidth]{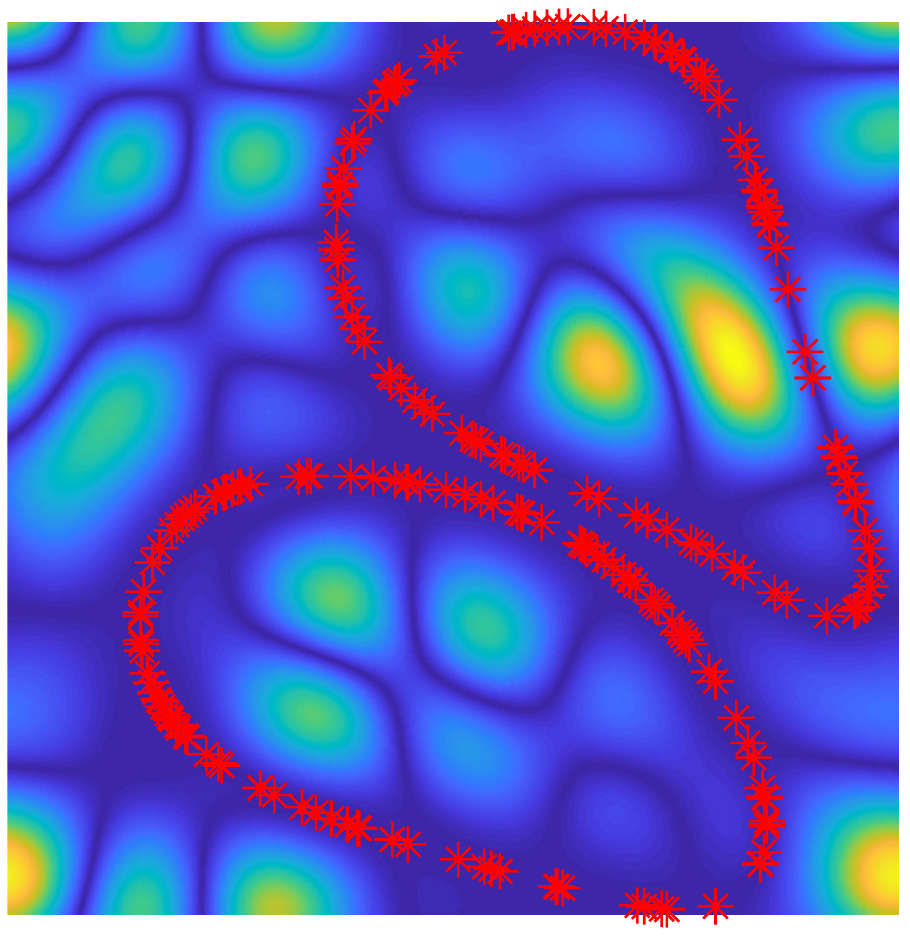}}\hspace{4ex}
          \subfigure[SOS polynomial]{\includegraphics[width=0.145\textwidth]{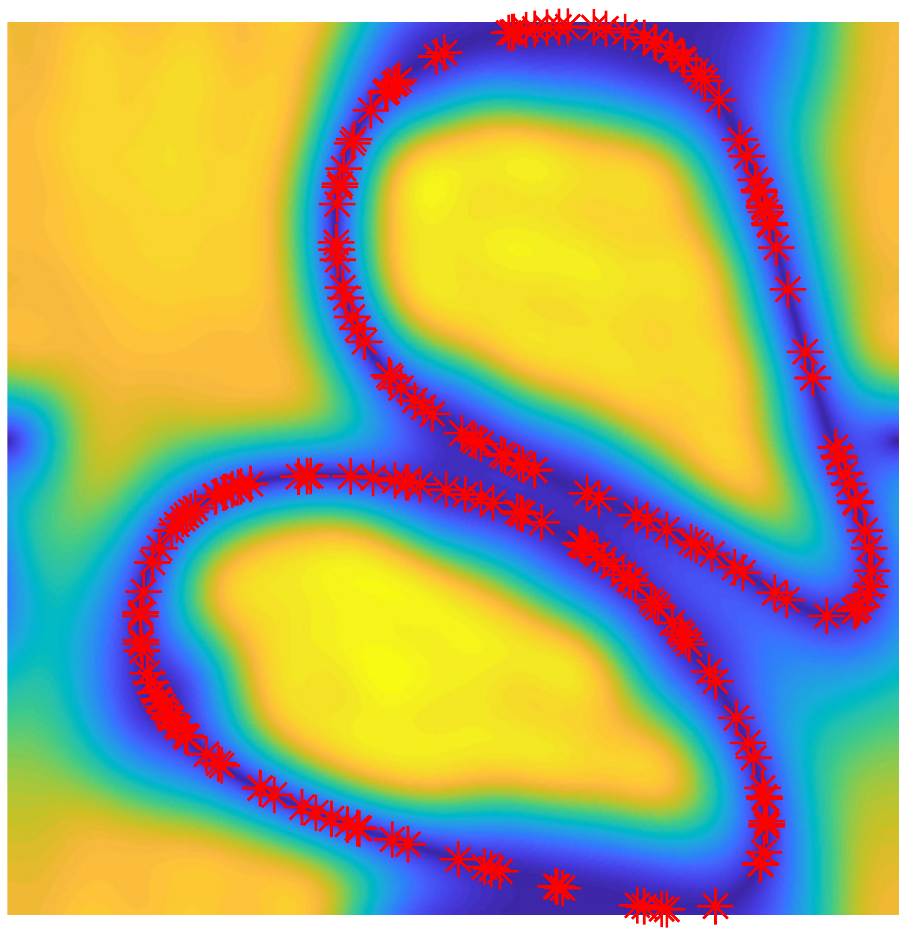}} \\
          \subfigure[1st nullspace func.]{\includegraphics[width=0.145\textwidth]{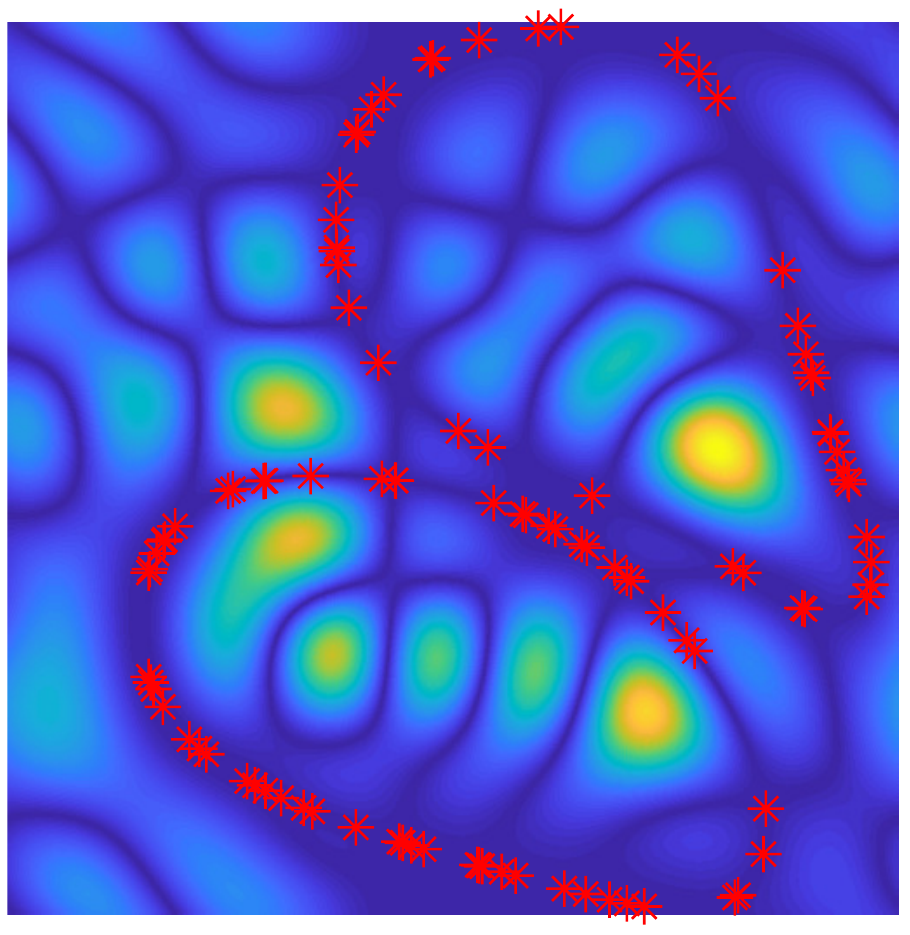}}\hspace{4ex}
          \subfigure[2nd nullspace func.]{\includegraphics[width=0.145\textwidth]{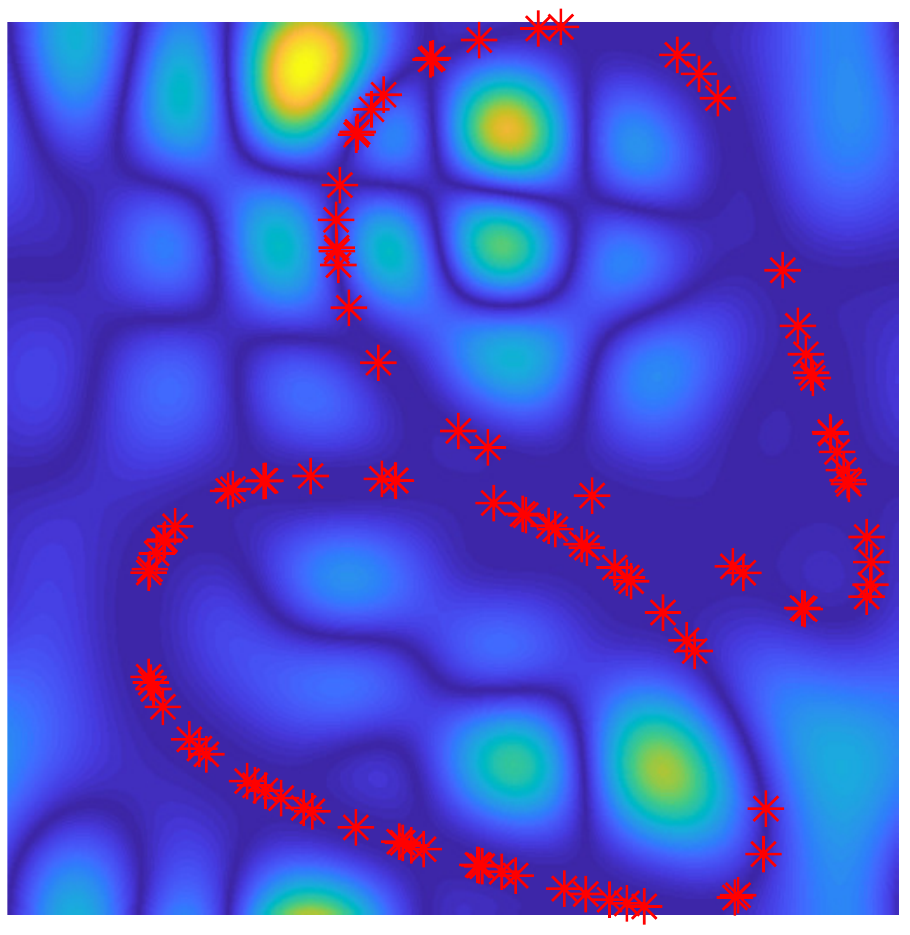}}\hspace{4ex}
          \subfigure[3rd nullspace func.]{\includegraphics[width=0.145\textwidth]{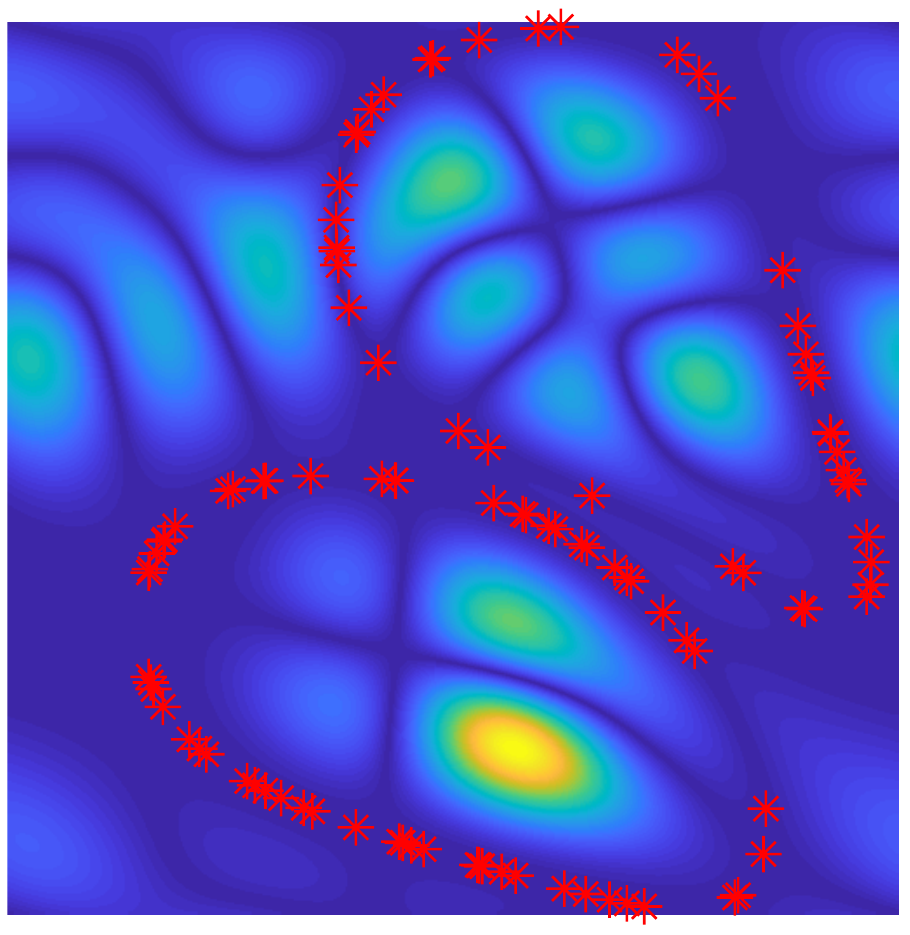}}\hspace{4ex}
          \subfigure[SOS polynomial]{\includegraphics[width=0.145\textwidth]{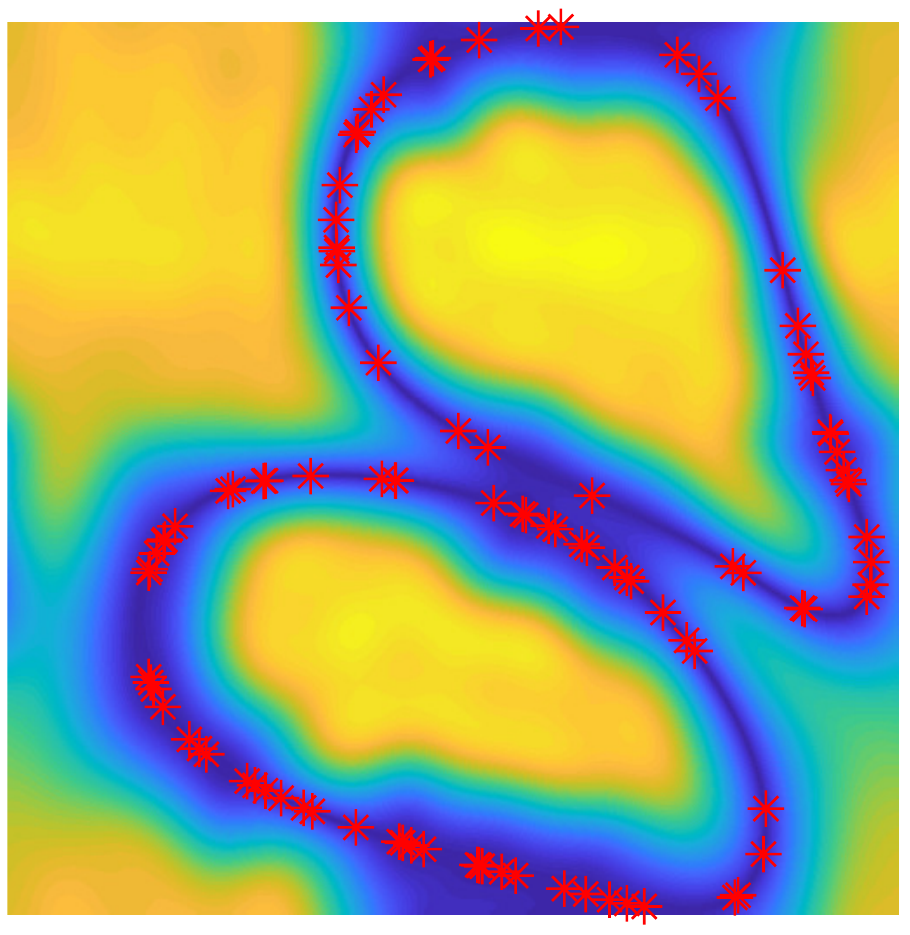}} 
          \caption{Illustration of Propositions \ref{prop2D3} \& \ref{propRank}: We consider the recovery of the curve $C[\psi]$ as specified by Fig \ref{prop1Sim1} (b), assuming unknown bandwidth. We over-estimate the support $\Gamma$ as 11x11, while the original support of $\mathbf c_{\psi} \stackrel{\mathcal F}{\leftrightarrow} \psi$ is $5\times5$. According to Propositions \ref{prop2D3}\& \ref{propRank}, when the number of samples exceed $(k_{1}+k_{2})(l_{1}+l_{2}) = 220$, the matrix is low-rank. The first row shows the results by using 220 samples. We display the Fourier transforms of the three null-space functions of $\Phi_{\Gamma}$ in (a), (b) and (c). This approach of visualizing the null-space functions is similar to the approaches in \cite{6678771,pan2014sampling}. All of these functions are zero on the  $\mathcal C[\psi]$, in addition to possessing several other zeros. The sum of squares function, denoted by \eqref{sos} is shown on the right column, captures the common zeros, which specifies the curve $\mathcal C[\psi]$. We use the SOS function as a surrogate for the greatest common divisor of the null-space functions. Note that the bound in Proposition \ref{prop2D3}  is also a worst-case guarantee. In the second row, the curve $C[\psi]$ was sampled on 100 random sampling locations, denoted by the red crosses. We see that the curve can be recovered well using just 100 samples. The computational time used to specify the curve using SOS function in this experiment is about 1.6 second }.
	\label{prop7Sim2}
\end{figure*}

Note that the minimal function $\psi(\mathbf x)$ is a special case of \eqref{annfunctions}, with $\eta=1$. The above result is proved in Appendix \ref{proof2D3}. Since $\psi(\mathbf x)$ is the common factor of all the annihilating functions, all of them will satisfy $\mu(\mathbf x)=0$, for any point on the original curve as well as the sampling locations. This also implies that $\psi(\mathbf x)$ is a common divisor of the above functions $\mu(\mathbf x)$. In fact, $\psi(\mathbf x)$ is the greatest common divisor as we will show it in the next paragraph. We now characterize the number of linearly independent annihilation functions, or equivalently the size of the right null space of $\Phi_{\Gamma}$.

\begin{prop}
\label{propRank}
We consider the trigonometric polynomial $\psi(\mathbf x)$ described in Proposition \ref{prop2D3}  and $\Lambda \subset \Gamma$. Then:
\begin{equation}
\label{rankbound}
{\rm rank}\left(\Phi_{\Gamma}(\mathbf X) \right) \leq \underbrace{ |\Gamma|-|\Gamma:\Lambda|}_r
\end{equation}
with equality if the sampling conditions of Proposition \ref{prop2D3} are satisfied.
\end{prop}
Here, \begin{equation}
\Gamma:\Lambda = \{\mathbf l \in \Gamma: \mathbf l -\mathbf k \in \Gamma, ~\forall ~\mathbf k \in \Lambda\}.
\end{equation}
This set is illustrated in Fig 5(a) from \cite{Ongie2016b} along with $\Gamma$ and $\Lambda$. The inequality of this result is same as the inequality of Proposition 5.1 in \cite{Ongie2016b}. Based on the inequality, we can then obtain the second part (the equality) of the result, which provides us a means to compute the original curve, even when the original bandwidth/support $\Lambda$ is unknown. Specifically, Proposition \ref{propRank} shows that $\Phi_{\Gamma}(\mathbf X)$ has $|\Gamma:\Lambda|$ null-space vectors, each of which satisfies \eqref{annfunctions}. Besides, from the proof of Proposition \ref{propRank}, we can see that any polynomial of the form \begin{equation}
\theta_{\mathbf l}=\exp(j2\pi \mathbf l^T \mathbf x)\;\psi(\mathbf x), ~~\forall \mathbf l \in \Gamma:\Lambda
\end{equation}
is a null-space vector of $\Phi_\Gamma(\mathbf X)$. Note that the exponentials $\exp(j2\pi \mathbf l^T \mathbf x), \forall \mathbf l \in \Gamma:\Lambda$ are linearly independent, and hence the set $\{\theta_{\mathbf l}; \mathbf l \in \Gamma:\Lambda\}$ spans the null space of $\Phi_\Gamma(\mathbf X)$. Since $\exp(j2\pi \mathbf l^T \mathbf x)$ does not vanish in the domain, the only common zeros of $\{\theta_{\mathbf l}; \mathbf l \in \Gamma:\Lambda\}$ will be the zeros of the minimal polynomial $\psi(\mathbf x)$, meaning that $\psi(\mathbf x)$ is the greatest common divisor of the functions that span the null-space of $\Phi_\Gamma(\mathbf X)$. Therefore, the common zeros of these functions, or equivalently the zeros of the greatest common divisor, will specify the curve. A cheaper alternative to evaluating the greatest common divisor is to evaluate the sum of squares polynomial, specified by:
\begin{equation}
\label{sos}
\gamma(\mathbf x) = \sum_{i=1}^{Q} \|\mu_i(\mathbf x)\|^2
\end{equation}
which will vanish only on points satisfying $\psi(\mathbf x)=0$. Here $Q = |\Gamma|-r$ is the dimension of the right null-space of $\Phi_{\Gamma}$. Since $\mathbf c_{\psi} \stackrel{\mathcal F}{\leftrightarrow}\psi$ is a valid right null-space vector of $\Phi_{\Gamma}$ that only vanishes on the true curve, the sum of squares function $\gamma$ specified in \eqref{sos} will only vanish on the true curve. Thus, if the total number of points sampled are $N = \sum_{j=1}^J N_j> (l_1+l_2)(k_1+k_2+2(J-1))$, and are arranged as \eqref{overSamp}, then the curve can be uniquely recovered.
 
We demonstrate the above result in Fig. \ref{prop7Sim2}. We considered the sampling of the same curve illustrated in Fig. \ref{prop1Sim1}, with the exception that we over-estimated the support to be $11\times 11$ as opposed to the true support of $5\times 5$. We considered 220 random samples, which satisfies the sampling conditions in Proposition \ref{prop2D3}. We show three of the annihilating functions in the first three columns of Fig. \ref{prop7Sim2}. We note that all of these functions are valid annihilating functions, but possess additional zeros. By contrast, the sum of square polynomial shown on the right uniquely specifies the curve.

\section{Applications}
The above section provided details on curve representation as well as provided guarantees on when the curve can be recovered from finite number of samples. We will now demonstrate the utility of the theory in some preliminary representative applications. We believe that the representation has the potential to improve upon the state of the art in several areas. However, these developments and rigorous validations are beyond the scope of this preliminary work. The experiments in this section are run on a laptop with Intel Core i7-9750H CPU.
\subsection{Application in segmentation}
\label{segmentation}
The Mumford Shah functional is a popular formulation for segmenting objects into piecewise constant regions. It approximates an image $f$ by a piecewise constant function
\begin{equation}
f = \sum_{k=1}^K a_k~ \chi_{\Omega_k},
\end{equation}
in the $\ell_2$ sense, where $\Omega_k, k=1,..,K$ are the regions and $a_k$ are the constants and $\chi$ represents the characteristic function on the set. The bounded curve is denoted by $\partial \Omega = \partial \Omega_1 \cup \partial \Omega_2\cup  \cdots\cup \partial \Omega_K$. Different penalties, including the length of $\partial \Omega$ or its smoothness are imposed to regularize the optimization problem. We propose to represent $\partial \Omega$ as the zero level set of a band-limited function $\psi$ specified by \eqref{model} as in \cite{Ongie2016b}. In this case, the piecewise constant function satisfies $
\widehat{\nabla f} * \mathbf {\tilde{c}} = 0$, which can be expressed in the matrix form as 
\begin{eqnarray}
\mathcal T\left(\widehat{\nabla f}\right) \mathbf c = 0,
\end{eqnarray}
where $\mathcal T$ is a block Toeplitz 2-D convolution matrix and $\mathbf {\tilde{c}}$ is the matrix version of vector $\mathbf c$. When the bandwidth is over-estimated, $\mathcal T\left(\widehat{\nabla f}\right) $ has multiple linearly independent null-space vectors and hence the matrix is low-rank. Note that the rank of the matrix can be considered as a surrogate for the complexity of the curve $\partial \Omega$. We hence formulate the segmentation task as the low-rank optimization problem, analogous to \cite{6678771}.
\begin{equation}
f^* = \arg \min_f \|f-h\|^2 + \lambda \sum_{i=r+1}^N \left\|\sigma_i\left[\mathcal T\left(\widehat{\nabla f}\right)\right]\right\|^2
\label{slr}
\end{equation}
where $h$ is the original image. Note that as $\lambda \rightarrow \infty$, $\mathcal T\left(\widehat{\nabla f}\right)$ approaches a rank $r$ matrix. Once $f^*$ is obtained, the sum of square function of the null space of $\mathcal T\left(\widehat{\nabla f}\right) $ will specify the curve and $f^*$ is the piecewise constant approximation. We use an alternating minimization strategy as reported in \cite{6678771} to solve the above optimization scheme.


\subsection{Recovery of noisy point clouds}
\label{pointclouds}
We now consider the case where we have noisy measurements of points lying on the curve $\psi(\mathbf x)=0$, where $\psi$ is represented as a linear combination of basis functions as in \eqref{curverep}. When the measurements are noisy, we propose to denoise them using the low-rank property of the features discussed in Proposition \ref{propRank}.

\subsubsection{Relation to kernel methods}

Proposition \ref{propRank} indicates that we can solve inverse problems by enforcing a low-rank constraint on the feature matrix. However, the size of the feature matrix grows with the dimensionality of the ambient space $n$ as well as the size of the over-estimated filter support $\Gamma$. Thus, in practice, it might be infeasible to form the feature matrix. However, the Gram matrix given by
\begin{equation}
\label{defKern}
\mathbf K_{\Gamma} = \Phi_{\Gamma}(\mathbf X)^T\Phi_{\Gamma}(\mathbf X)
\end{equation}
is of size $N\times N$, where $N$ is the number of points. Note that the complexity of an algorithm that depends on the Gram matrix is independent of the dimension of the ambient space and the chosen filter support $\Gamma$. We note that this approach is similar to the \emph{``kernel-trick''} used in various machine learning applications \cite{scholkopf}. Under our assumed model, the entries of this Gram matrix are given by
\begin{equation}
\label{kernel}
\left(\mathbf K_{\Gamma}\right)_{i,j} = \phi_{\Gamma}(\mathbf x_i)^T\phi_{\Gamma}(\mathbf x_j) =\underbrace{ \sum_{\mathbf k\in \Gamma}  \exp\left(j~2 \pi\mathbf k^T \left(\mathbf x_j-\mathbf x_i\right)\right)}_{\kappa_{\Gamma}(\mathbf x_j-\mathbf x_i)}
\end{equation}
When $\Gamma$ is a centered cube in $\mathbb{R}^n$, $\kappa_{\Gamma}(\mathbf r) $ is a Dirichlet function. Note that the entries of $\mathbf K_{\Gamma}$ can be evaluated without explicitly evaluating the feature matrix. The width of the Dirichlet function is dependent on the Fourier support $\Gamma$. The kernel matrix satisfies  ${\rm rank}(\mathbf K_{\Gamma}) \leq r$, where $r$ is given by \eqref{rankbound}. The relationship \eqref{defKern} implies that the kernel matrix is low-rank, which is a property that is widely used in kernel low-rank methods. 

A popular practical choice for kernel is Gaussian functions, or equivalently periodized Gaussian functions when the domain is restricted to $[0,1)^2$. Note that the Gaussian kernel is qualitatively similar to the Dirichlet kernel considered above, where the width of the Gaussian is a parameter similar to the size of the support $\Lambda$. We note that the Gaussian kernel function is less oscillatory and is isotropic, which makes it more attractive than Dirichlet \footnote{Our preliminary experiments show that Dirichlet kernels perform equally well in applications, but are computationally less efficient. Our future work is focused on making this approach computationally efficient. } kernels in applications. The Gaussian kernel correspond to feature maps of the form 
\begin{equation}
\left[\phi(\mathbf x)\right]_i = \exp\left(-\pi^2 \sigma^2\frac{\|\mathbf k_i\|^2}{2}\right)\cdot\exp(j2\pi \mathbf k_i^T\mathbf x)
\end{equation}
as $\Gamma\rightarrow \mathbb Z^n$. Note that this is the Fourier transform of a shifted Gaussian; this setting corresponds to the case where the level set function is being expressed as a shift invariant linear combination of Gaussian functions on a very fine grid. Since the Gaussian kernel matrix $\mathbf K_{\Gamma}$ is theoretically full rank, the theoretical analysis in the previous section is not directly applicablele to Gaussian kernels in the strict sense. However, we observe that the Fourier series coefficients of a Gaussian function may be approximated to be zero outside $|\mathbf k|< 3/\pi\sigma$, which translates to $|\Lambda| \approx \left(\frac{6}{\pi\sigma}\right)^n$. This implies that the Gaussian kernel matrix can be safely approximated to be low-rank, when  $\sigma$ is sufficiently high; this corresponds to a more localized kernel in space.

In the next subsection, we will describe how to use proposition \ref{propRank}, without explicitly forming the feature matrix $\Phi_{\Gamma}(\mathbf X)$, and computing only it's Gram matrix $\mathbf K_{\Gamma}$ instead.

\subsubsection{Denoising of point clouds using nuclear norm minimization}\label{pointdenoising}
With the addition of noise, the points deviate from the zero level set of $\psi$. A high bandwidth potential function is needed to represent the noisy curve. Let the noisy measurements of the matrix $\mathbf X$ be given by $\mathbf Y$. We propose to use the nuclear norm of the feature matrix as a regularizer in the recovery of the points from noisy measurements:
\begin{equation}
\label{opt}
\mathbf X^* = \arg\min_{\mathbf X} \|\mathbf X - \mathbf Y\|^2 + \lambda\|\mathbf \Phi(\mathbf X)\|_*
\end{equation}
We note that this approach has conceptual similarities to Cadzow denoising that is widely used in FRI methods \cite{pan2014sampling,dragotti2007sampling}. We use an IRLS approach \cite{fornasier2011low,mohan2012iterative} where the nuclear norm is approximated as:
\begin{equation}
\left\|\Phi(\mathbf X)\right\|_* = {\rm trace}\left[\left(\Phi(\mathbf X)^T \Phi(\mathbf X)\right)^{\frac{1}{2}}\right] \approx {\rm trace}\left[\mathcal K(\mathbf X)\mathbf P\right]
\end{equation}
where $\mathbf P = \left[\mathcal K(\mathbf X) + \gamma \mathbf I\right]^{-\frac{1}{2}}$ and $\mathcal K$ is the Gaussian kernel. Here, $\gamma$ is a small constant added to ensure that the inverse is well-defined. 

The IRLS algorithm alternates between the following two steps:
\begin{equation}
\label{kernelApprox}
\mathbf X^{(m)}= \arg\min_{\mathbf X} \underbrace{ \|\mathbf X - \mathbf Y\|_F^2 + \lambda~
{\rm trace}\left[\mathcal K(\mathbf X)\mathbf P^{(m-1)}\right]}_{\mathcal C(\mathbf X^{(m)})}
\end{equation}
where 
\begin{equation}
\mathbf P^{(m)} = \left[\mathcal K\left(\mathbf X^{(m)}\right) + \gamma^{(m)} \mathbf I\right]^{-\frac{1}{2}}.
\end{equation}
Here, $\gamma^{(m)} = \frac{\gamma^{(m-1)}}{\eta}$, and $\eta>1$ is a constant.

Note from \eqref{kernelApprox} that updating $\mathbf X$ involves the solution of a non-linear system of equations. We propose to linearize the gradient of the cost function in \eqref{kernelApprox} with respect to $\mathbf X$. \begin{equation}
 \nabla_{\mathbf x_i}\mathcal C = 2(\mathbf x_i - \mathbf y_i) + \lambda \sum_j \mathbf P^{(m-1)}_{ij}~\nabla_{\mathbf x_i}[\mathcal K(\mathbf X)]_{ij}
\end{equation}
Linearizing the gradient with respect to $\mathbf X_i$, we obtain:
\begin{equation}
\begin{split}
\nabla_{\mathbf X_i} \mathcal C& \approx 2(\mathbf X_i - \mathbf Y_i) +2\lambda \sum_j w_{ij}^{(m-1)}(\mathbf X_i- \mathbf X_j)
\end{split}
\end{equation}
where $w_{ij}^{(m-1)}$ is the $(i,j)^{th}$ entry of a matrix 
\begin{equation}
\mathbf W^{(m-1)} = -\frac{1}{\sigma^2}\mathcal K(\mathbf X^{(m-1)}) \odot \mathbf P^{(m-1)}.
\label{wt}
\end{equation}
In matrix form, the gradient can be rewritten as $\nabla_{\mathbf X} \mathcal C  = 2(\mathbf X - \mathbf Y) +  2\lambda \mathbf X \mathbf L^{(m-1)}$, 
where $\mathbf L^{(m)}$ is computed from the weight matrix $\mathbf W^{(m)}$ as 
\begin{equation}
\mathbf L^{(m)} = \mathbf D^{(m)} - \mathbf W^{(m)}
\label{laplacian} 
\end{equation}
Here, $\mathbf D^{(m)}$ is a diagonal matrix with elements defined as $\mathbf D_{ii}^{(m)} = \sum_j \mathbf W_{ij}^{(m)}$. 

This results in the following equivalent optimization problem for the estimation of $\mathbf X$ at the $n^{th}$ iteration, which can be solved analytically:
\begin{equation}
\label{Req}
\mathbf X^{(m)}  = \arg\min_{\mathbf X} \left\|\mathbf X - \mathbf Y\right\|_F^2 + \lambda~{\rm trace}\left(\mathbf X~ \mathbf L^{(m-1)} ~\mathbf X^{T} \right)
\end{equation}
Thus, we alternate between the estimation of $\mathbf X^{(m)}$ and $\mathbf L^{(m)}$ till convergence to solve \eqref{opt}. We note that this optimization algorithm is non-convex and does not come with any convergence guarantees. However, with reasonable initialization (e.g. $\mathbf X=\mathbf Y$), the algorithm yielded good results in practice.

\section{Comparisons with state of the art}

Now, we compare our planar curves recovery results with an algorithm that relies on level set evolution. Specifically, we re-engineer the level-set based method for curve recovery termed as ``distance regularized level set evolution'' (DRLSE), which was introduced in \cite{li2010distance} for image segmentation. DRLSE poses the image segmentation as the minimization of the cost function
\[\underbrace{\mathcal{E}(\phi)}_{\mbox{energy function}} = \lambda\underbrace{\mathcal{L}_g(\phi)}_{\mbox{ length}}+\alpha\underbrace{\mathcal{A}_g(\phi)}_{\mbox{area}}+ \mu\underbrace{\mathcal{R}(\phi)}_{\mbox{regularization}},\]
where $\phi$ is the level set function. Here, $\mathcal{R}_p(\phi)$ is a level-set regularization term which maintains the level-set function $\phi$ as a signed distance function. We choose the function to be (16) of \cite{li2010distance}. The first and second terms are the weighted length and area of the curve, respectively:
	\begin{eqnarray}\label{key}
	\mathcal L_g(\phi) &=& \int_{\Omega}g\delta(\phi)|\Delta\phi|d\mathbf{x}\\
	\mathcal L_g(\phi) &=& \int_{\Omega}gH(-\phi)d\mathbf{x}
	\end{eqnarray}
which are determined by the choice of edge indicator function $g$. Length and area minimizing flows are well-studied in the level-set literature, and correspond to curve velocities that are proportional to curvature and constant velocity along the curve normals \cite{siddiqi1997area}. The parameter $\mu$ for level set regularization term is determined by the time step. Once the time step is chosen, $\mu$ is almost determined because of the Courant-Friedrichs-Lewy (CFL) condition. We will re-engineer DRLSE to the curve recovery from samples by choosing the edge indicator function as the distance of the level-set function to points. Since DRLSE was originally designed for image segmentation, we use the edge-based edge indicator function discussed in \eqref{curvedistance}. We call the re-engineered DRLSE algorithm the level-set based algorithm.

\subsection{Curve recovery from samples}
\label{dlrsepts}
\begin{figure}[htbp!]
	\centering
	\subfigure[600 samples]{\includegraphics[width=0.15\textwidth]{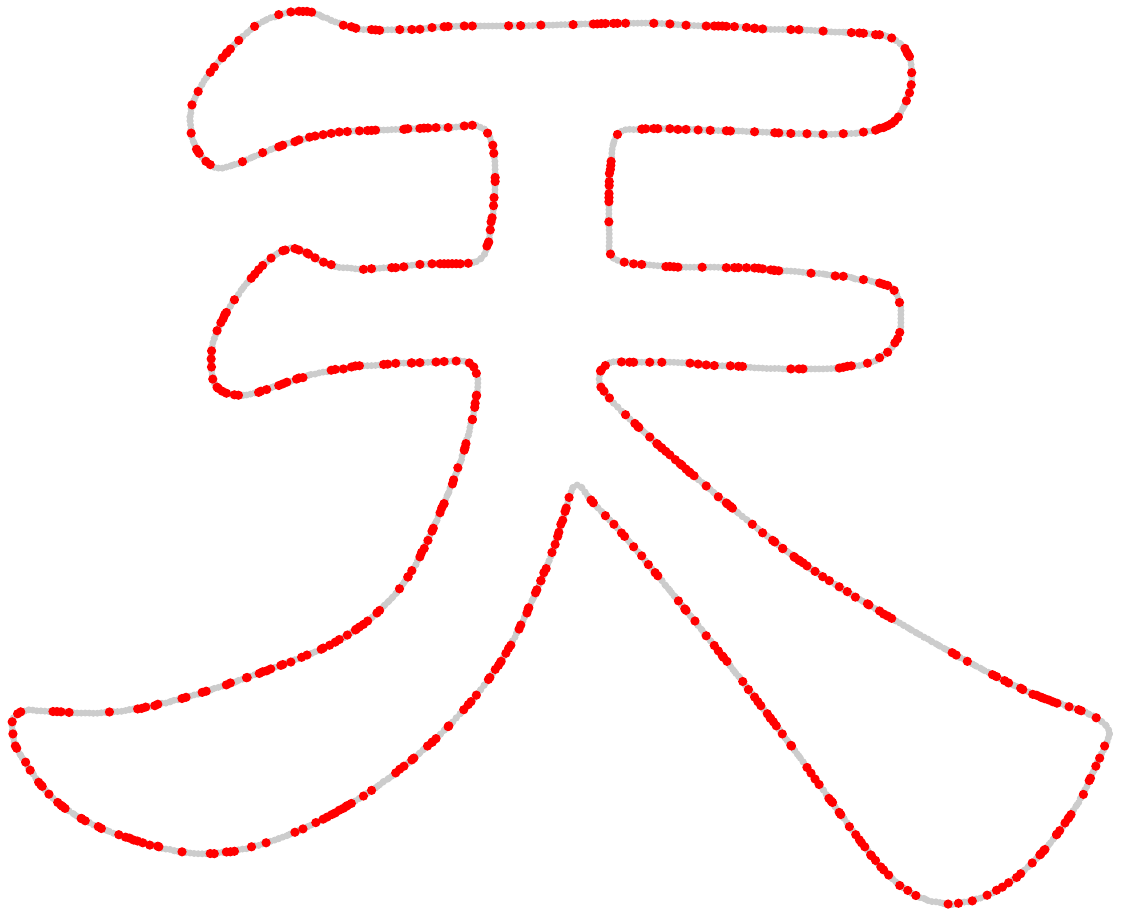}}
           \subfigure[level-set based: 600]{\includegraphics[width=0.15\textwidth]{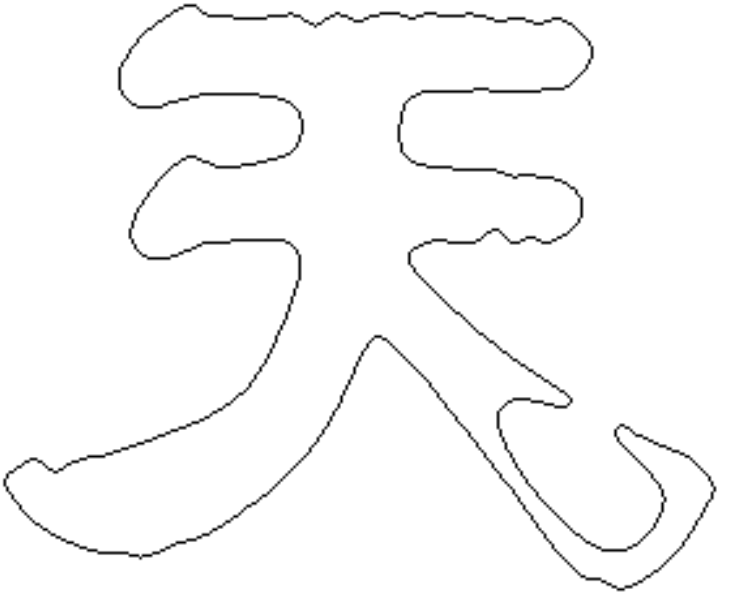}}
           \subfigure[Proposed: 600]{\includegraphics[width=0.15\textwidth]{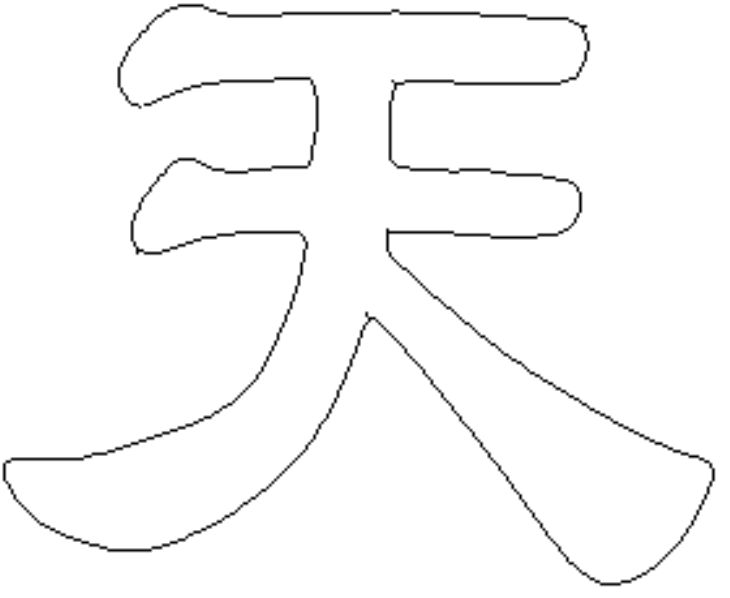}}\\
	\subfigure[1000 samples]{\includegraphics[width=0.15\textwidth]{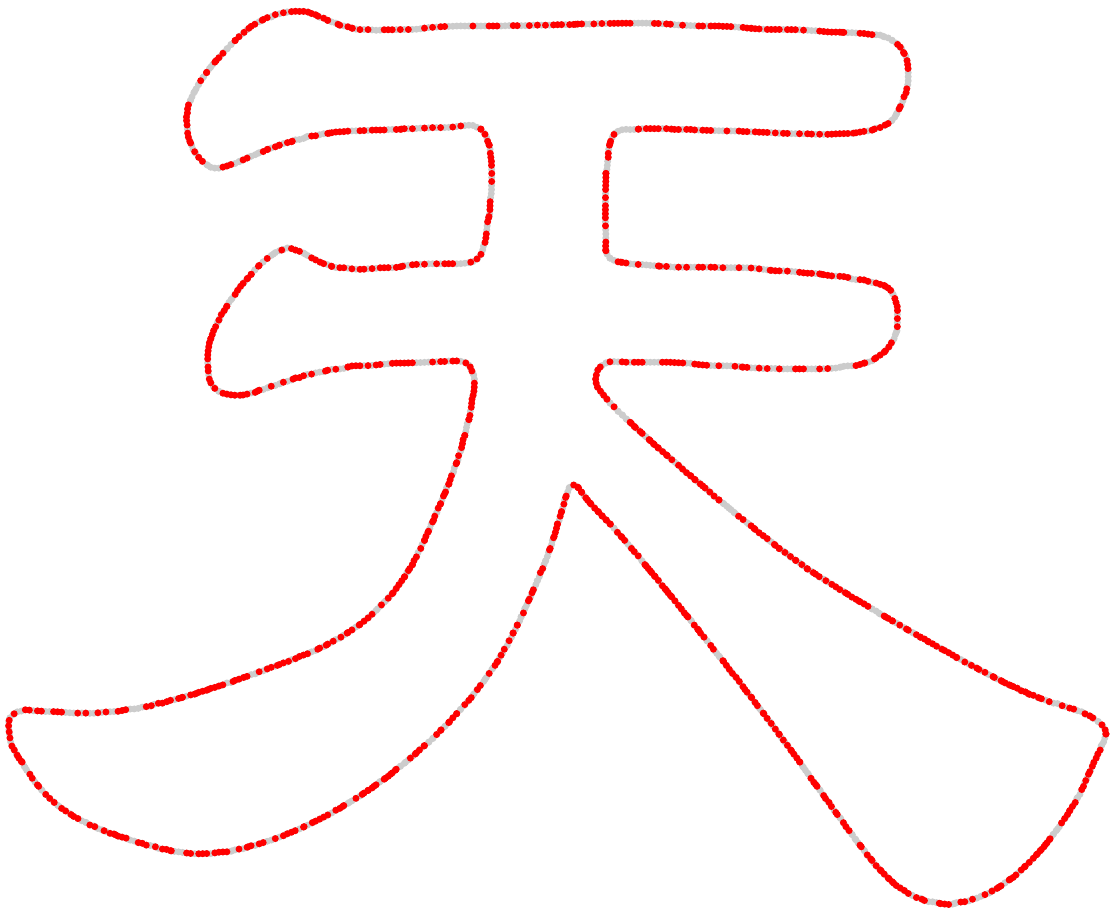}}	
	\subfigure[level-set based:1000]{\includegraphics[width=0.15\textwidth]{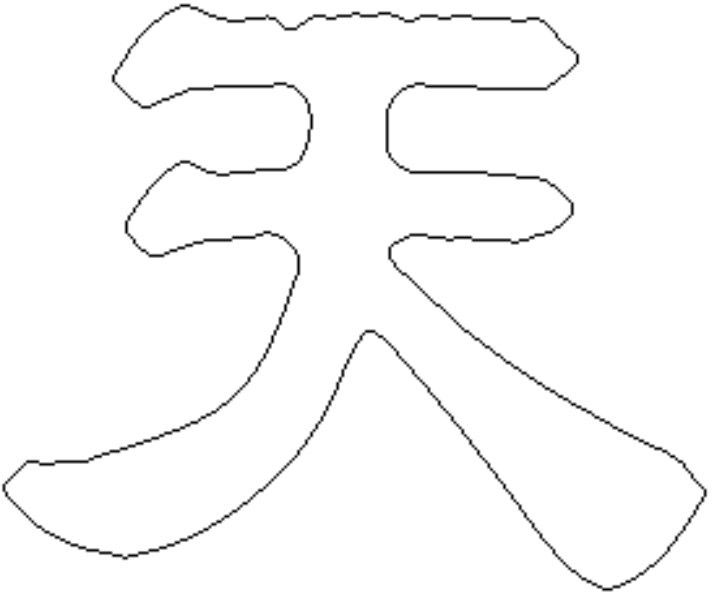}}
	\subfigure[Proposed: 1000]{\includegraphics[width=0.15\textwidth]{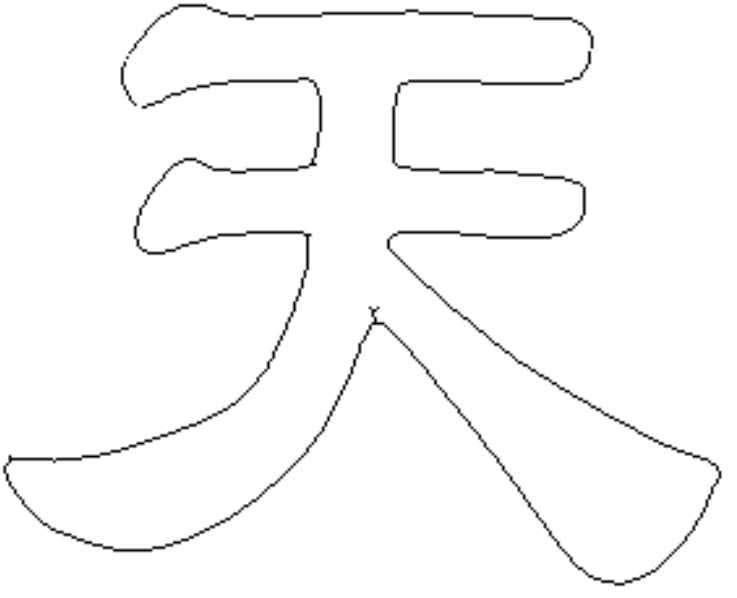}}\\
	\caption{Comparison of the proposed curve recovery scheme in Section \ref{curverecoveryalgorithm} with the adaptation of \cite{li2010distance} described in Section \ref{dlrsepts}. The shape is randomly sampled on the points shown in the first column. The second column consists of the curves recovered using the level-set based algorithm, while the last column shows the ones by the proposed scheme. The computational time required for the level-set based algorithm is about 66 seconds whereas the computational time required for the proposed algorithm is only about 6.4 seconds using 1000 samples.}
	\label{tian}
\end{figure}

In this section, we compare the proposed curve recovery scheme in Section \ref{sampling} with the level-set based algorithm. We choose the edge indicator function as the distance of the curve from the samples $\mathbf x_i; i=1,..,N$:
\begin{equation}\label{curvedistance}
g = \frac{1}{c}d(\mathbf x, \mathbf x_i)
\end{equation}	
where $d(\mathbf x, \mathbf x_i) = \min_{i=1,\cdots,N}\{c,\|\mathbf x-\mathbf x_i\|^2\}$ for all $\mathbf x$ in the image domain and $c$ is a large constant. We compare the two methods in the context of recovering the edge curve for the Chinese character ``Tian'' (meaning sky in English) in Fig. \ref{tian}.

In our method, we chose the bandwidth of the curve as $51\times 51$. For the level-set based algorithm, we choose the parameters as $\lambda=5$, $\alpha=10$ and the initialization curve as a square which includes the whole curve. Note that our method do not need any initialization. The two rows show the recovery results by the two different methods from 600 and 1000 samples respectively. The first column shows the samples we choose. The results obtained by using the level-set based algorithm are given in the second column. The numbers of iterations for obtaining (b) and (e) are 1510 and 2260. The third column shows the recovery results by using our method. By comparing (b) and (c), one can see that our method recover the curve successfully from 600 randomly chosen samples. For the level-set based algorithm, the curve is not successfully recovered from those 600 samples. Once we have 1000 samples, we can find that both the two methods succeed in recovering the curve, as shown in (e) and (f). However, the computational time required for our proposed algorithm is less than that of the level-set based algorithm. This example also demonstrates that both the two level-set based methods work well even though the curves have some sharp corners.


\begin{figure*}[htbp!]
	\centering
	\subfigure[DRLSE \#1]{\includegraphics[width=0.25\textwidth]{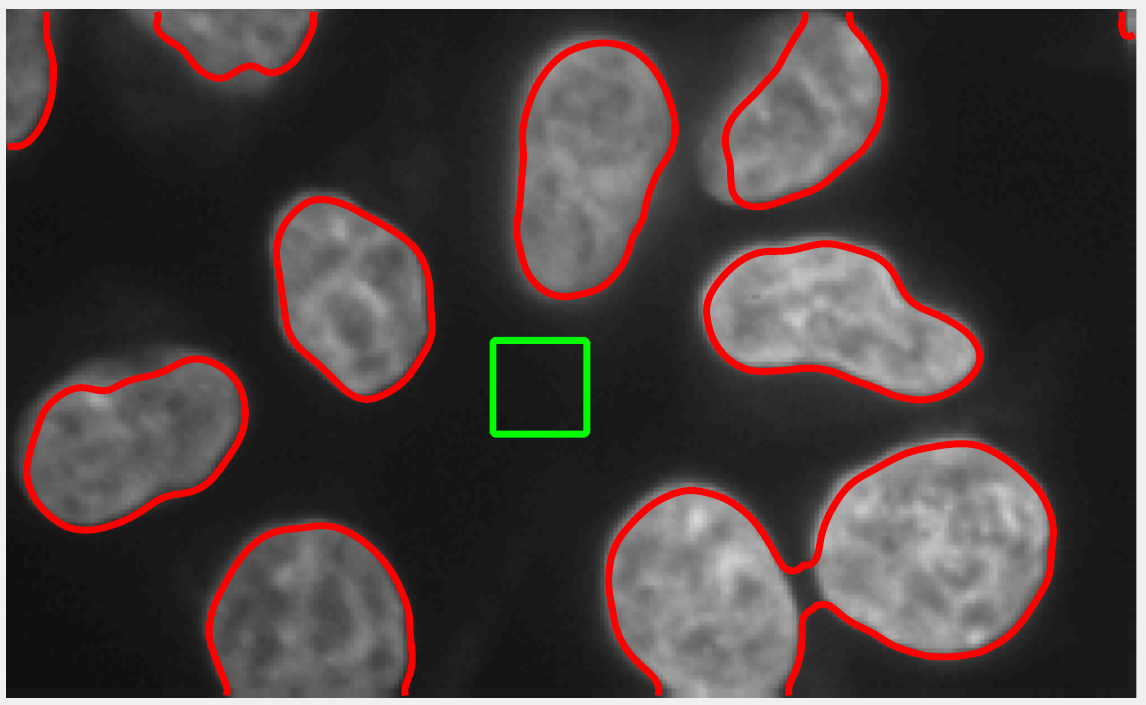}}\label{fig:1a}
	\subfigure[DRLSE \#2]{\includegraphics[width=0.25\textwidth]{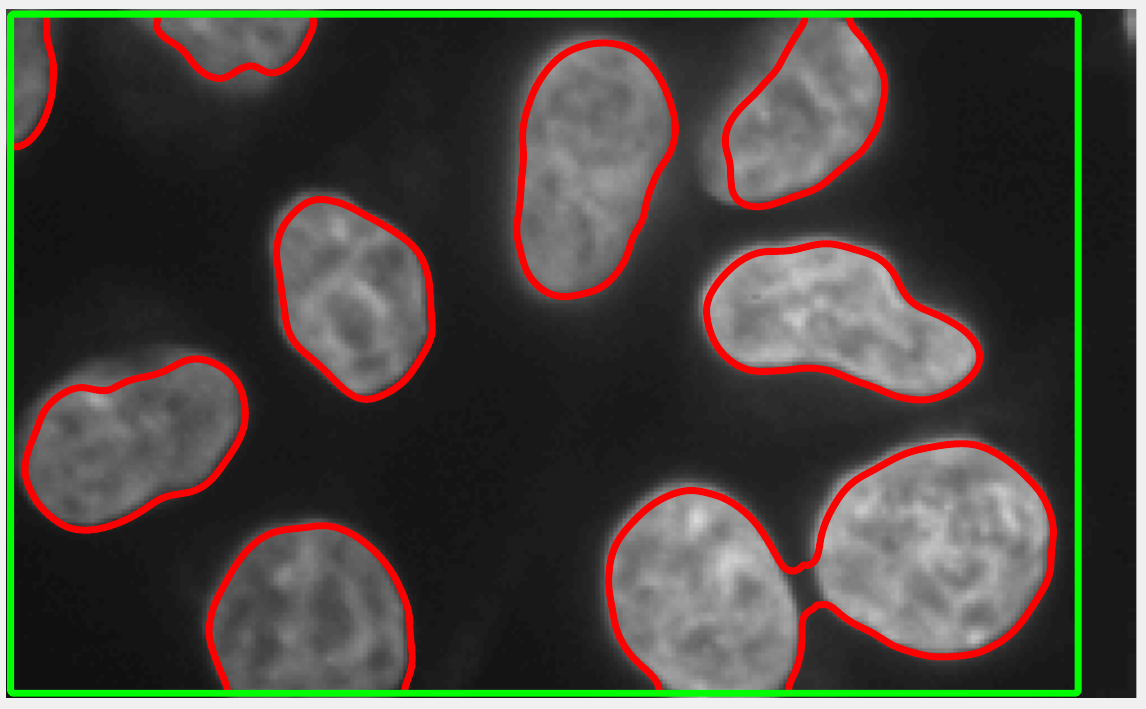}}\label{fig:1b}
	\subfigure[Proposed]{\includegraphics[width=0.25\textwidth]{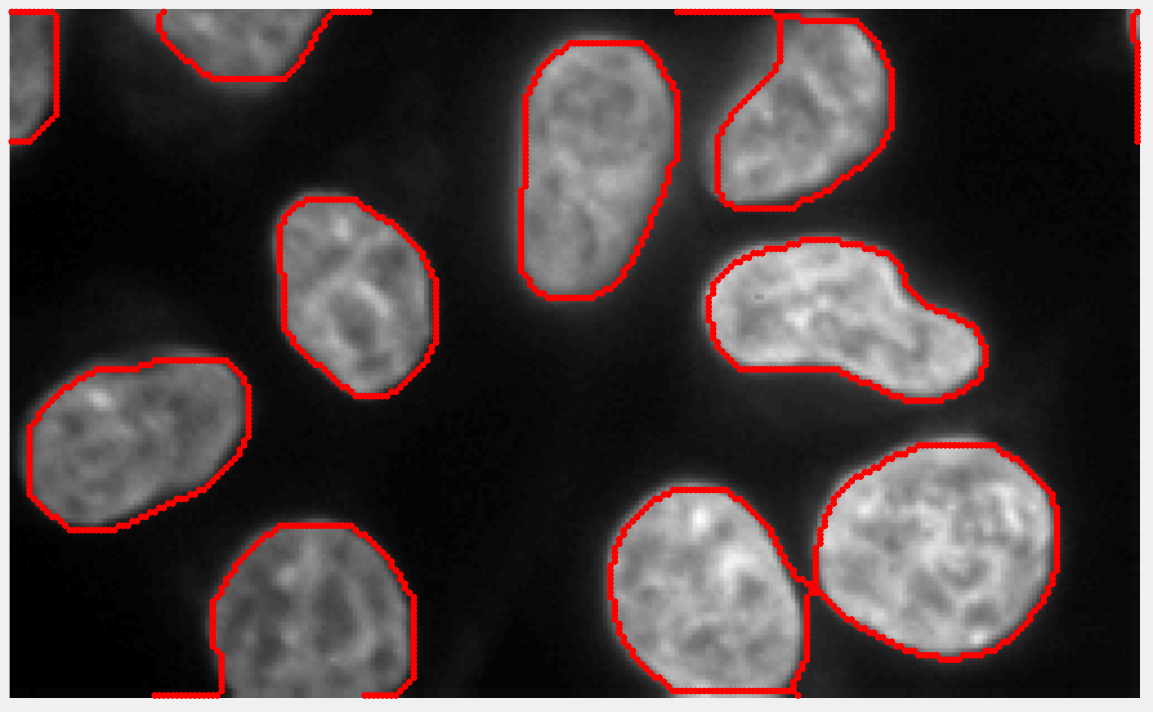}}\label{fig:1c}\\
	\subfigure[DRLSE \#1]{\includegraphics[width=0.25\textwidth]{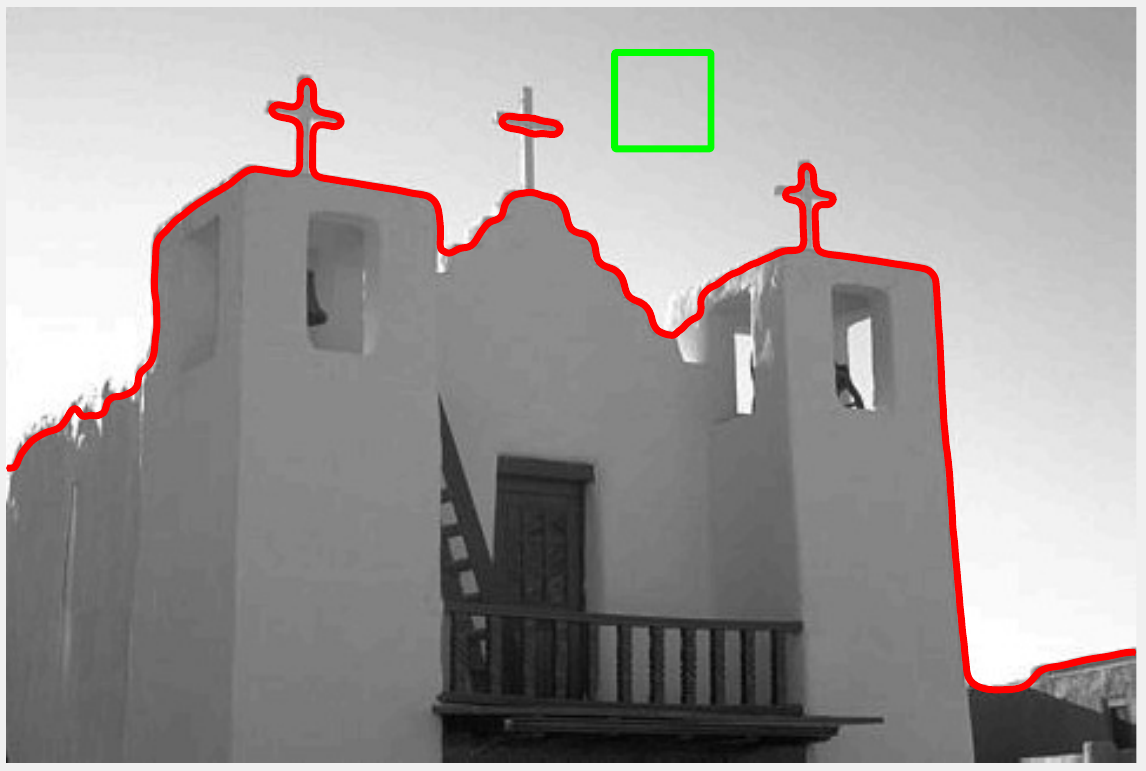}}\label{fig:2a}
	\subfigure[DRLSE \#2]{\includegraphics[width=0.25\textwidth]{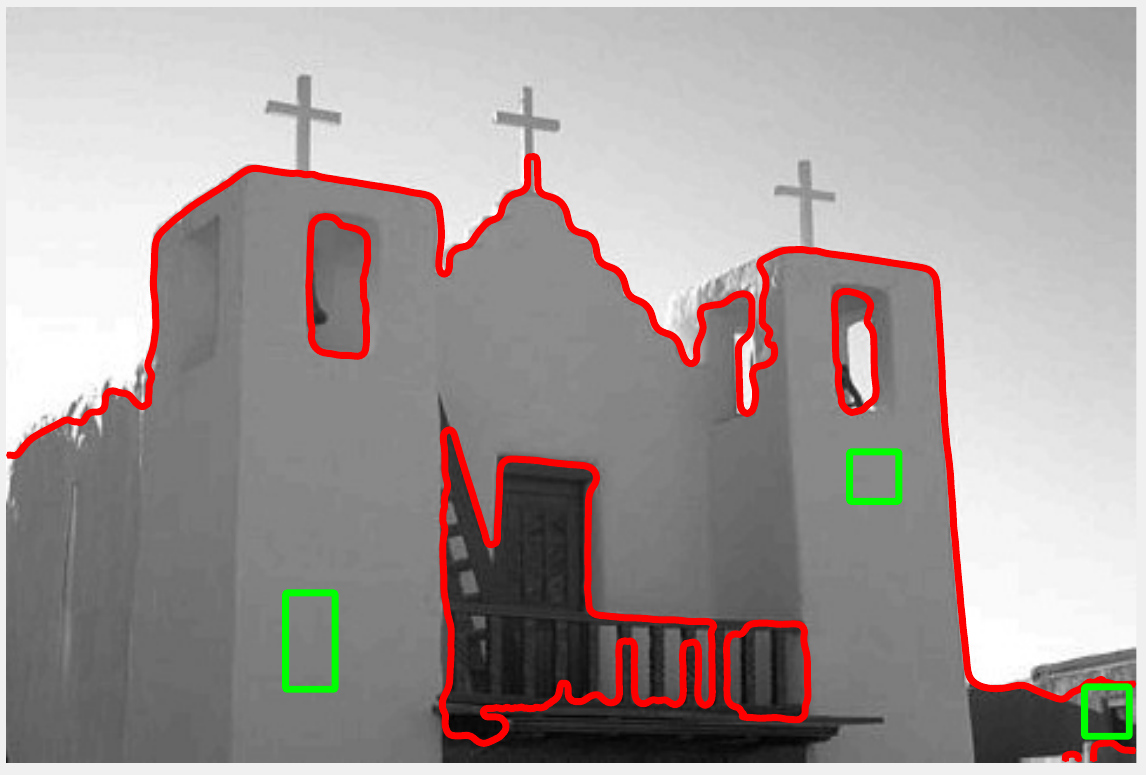}}\label{fig:2b}
	\subfigure[Proposed]{\includegraphics[width=0.25\textwidth]{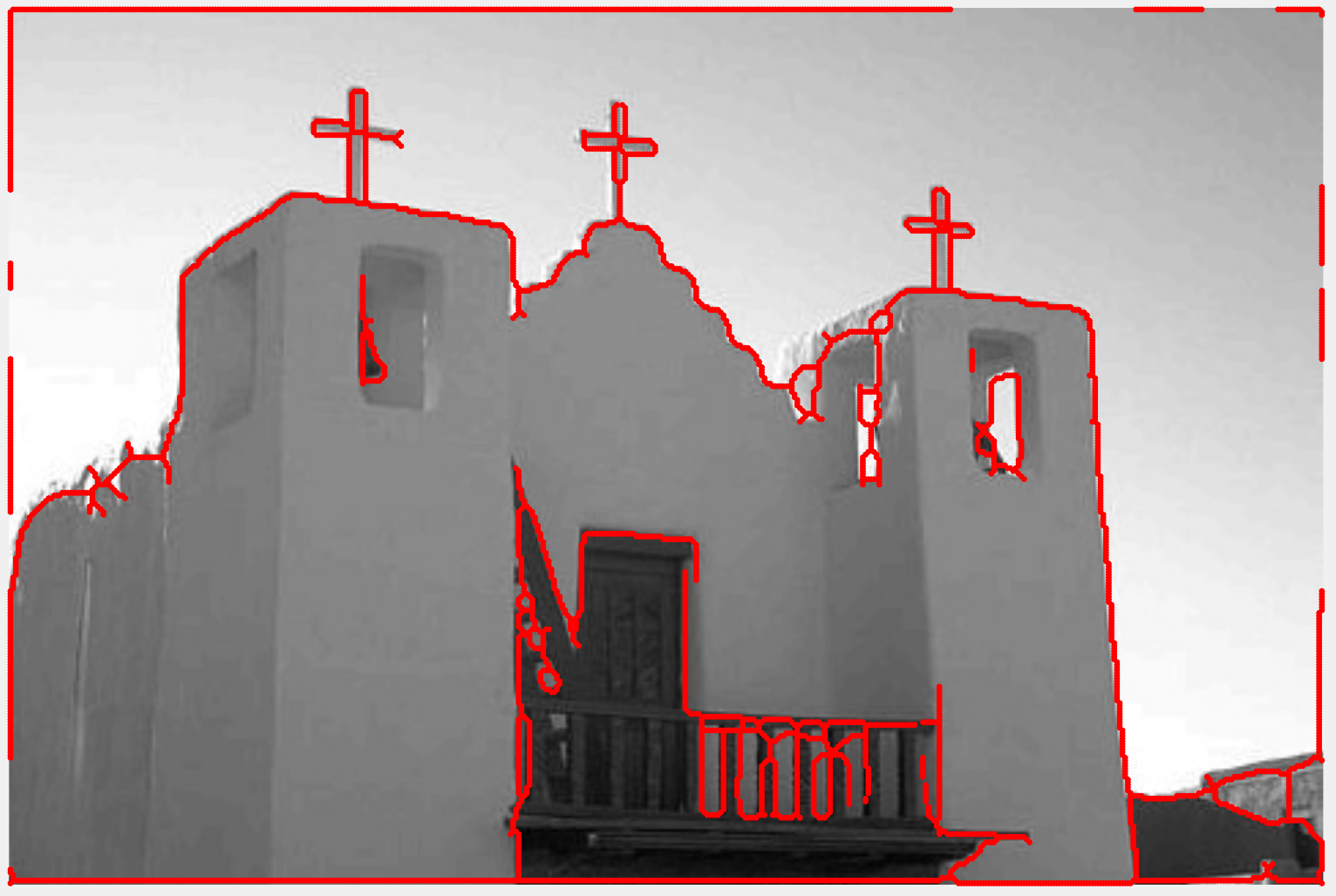}}\label{fig:2c}\\
	\caption{Illustration of edge based segmentation using the band-limited curve model using \eqref{slr} and the comparisons with the segmentation method DRLSE introduced in \cite{li2010distance}. The DLRSE scheme requires curve initialization, indicated by the green squares in the DLRSE results. The red curves in each case show the final curves. The parameters of the algorithms are optimized manually to yield the best results. The results show that the proposed scheme can provide similar segmentation as DLRSE, while it does not need initialization and is guaranteed to converge to global minimum. The ranks we choose here are 500 and 1200 for cells image and church image respectively.}
	\label{segment}
\end{figure*}

\begin{figure*}[htbp!]
	\centering
	\subfigure[rank = 200]{\includegraphics[width=0.25\textwidth]{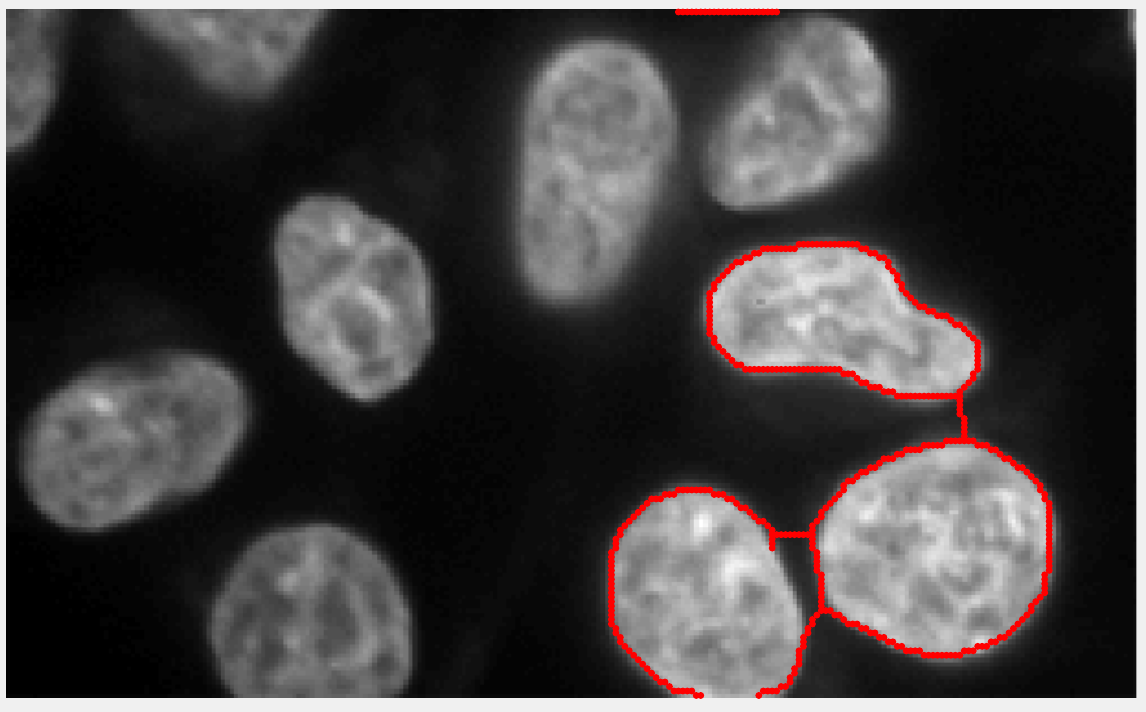}}\label{fig:1a}
	\subfigure[rank = 350]{\includegraphics[width=0.25\textwidth]{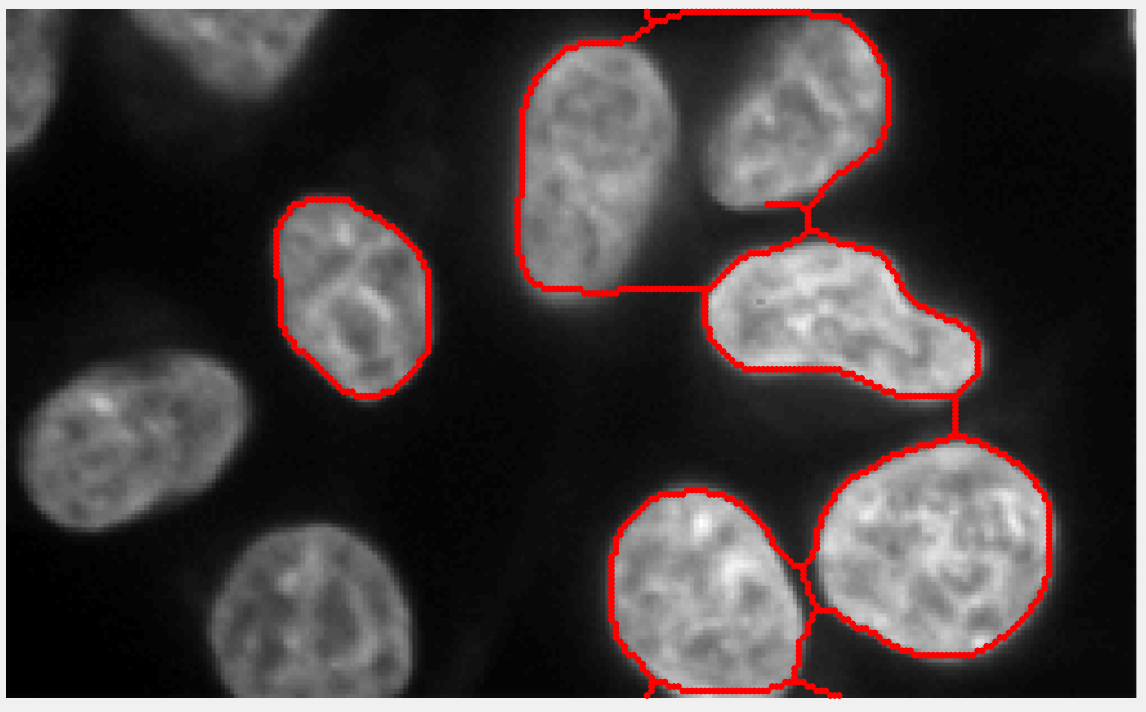}}\label{fig:1b}
	\subfigure[rank = 600]{\includegraphics[width=0.25\textwidth]{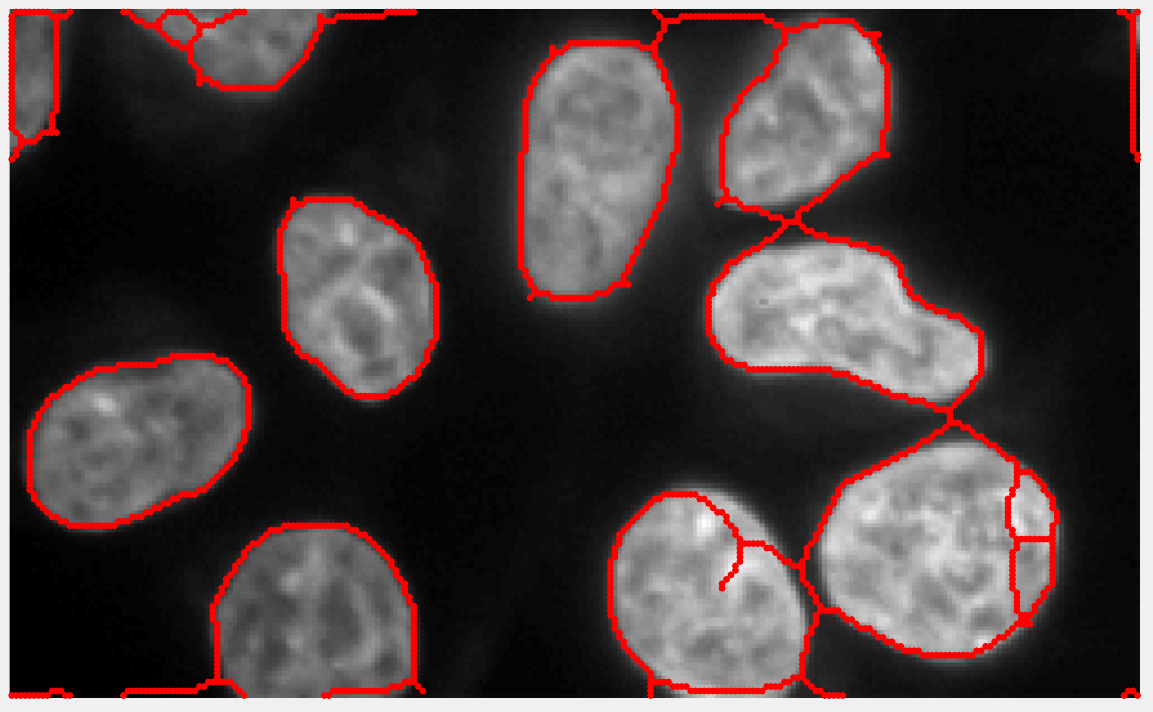}}\label{fig:1c}\\
	\subfigure[rank = 850]{\includegraphics[width=0.25\textwidth]{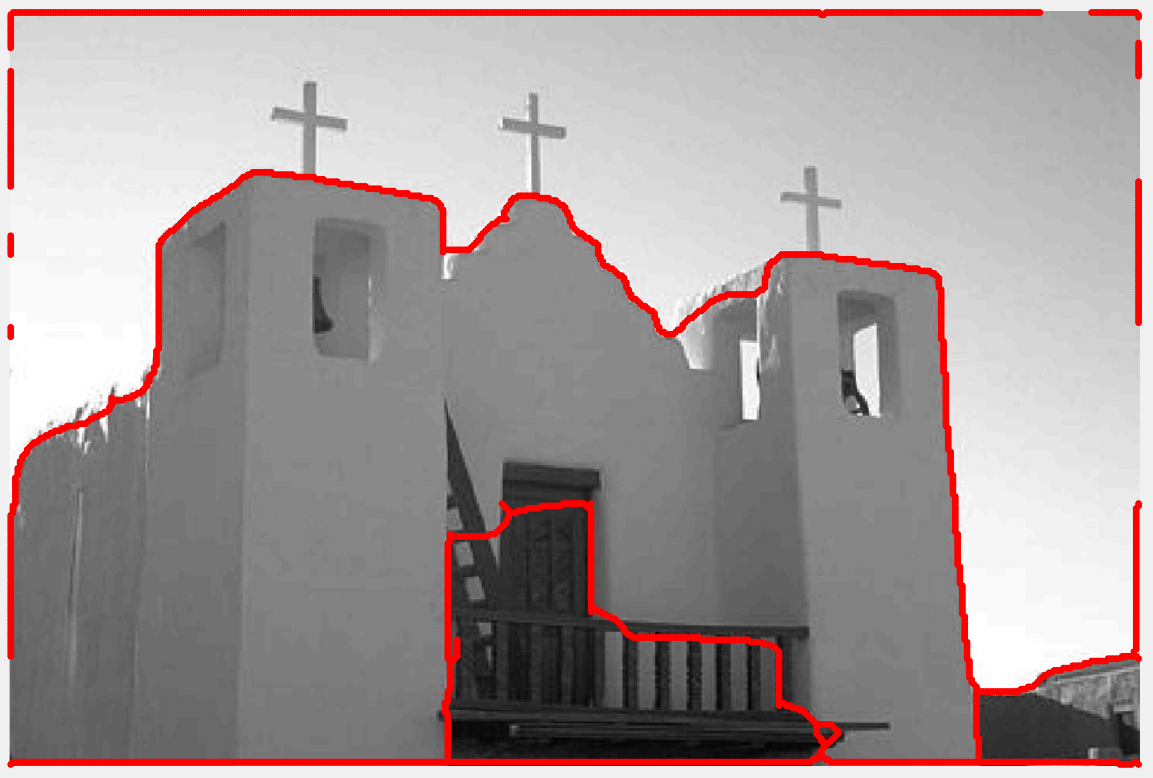}}\label{fig:2a}
	\subfigure[rank = 1000]{\includegraphics[width=0.25\textwidth]{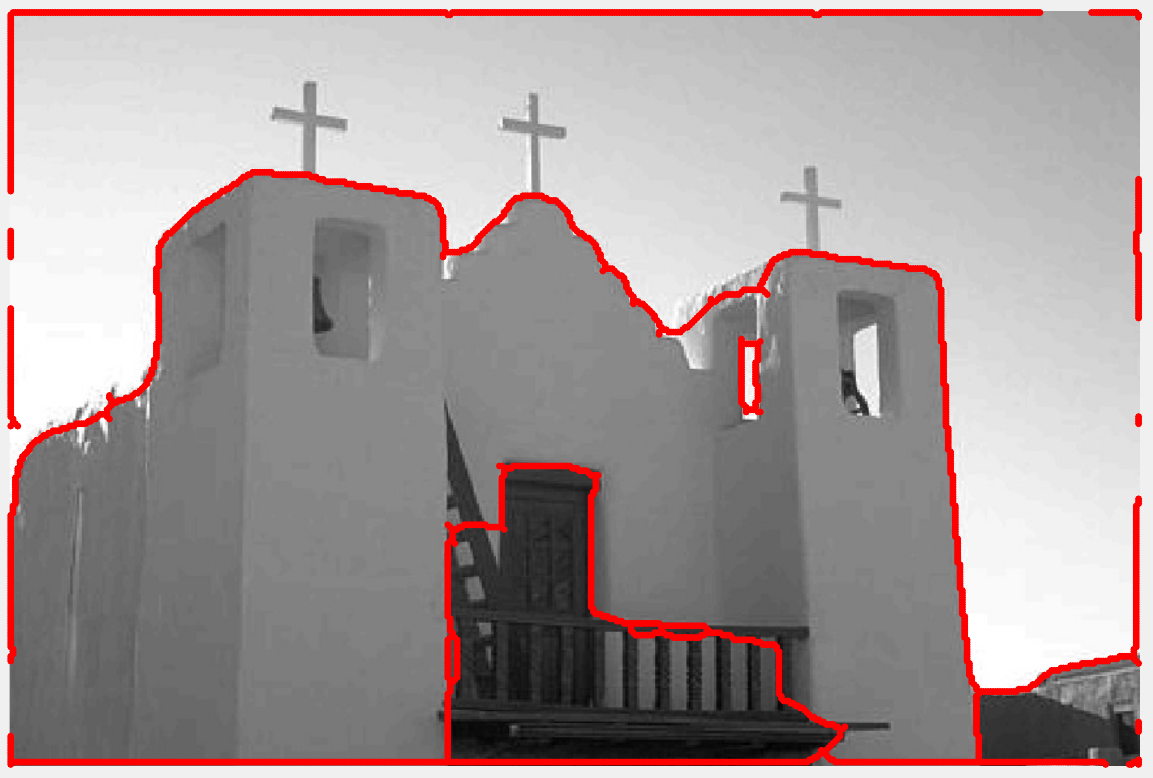}}\label{fig:2b}
	\subfigure[rank = 1350]{\includegraphics[width=0.25\textwidth]{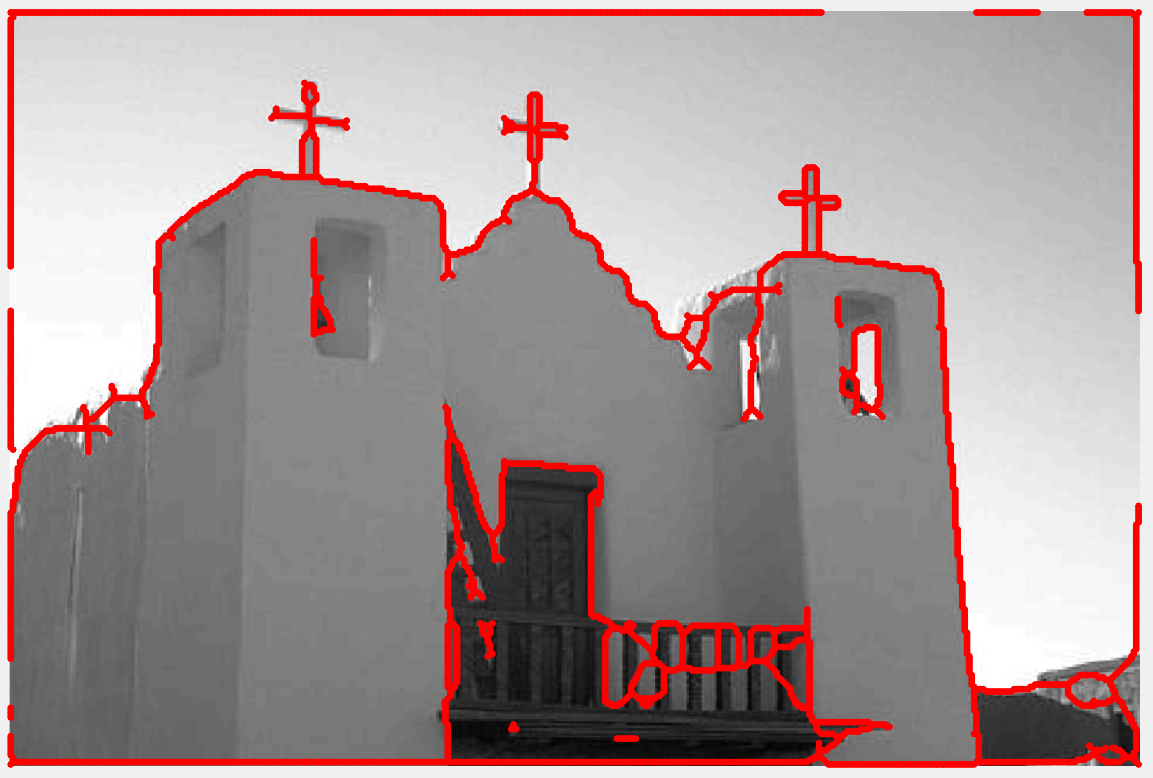}}\label{fig:2c}\\
	\caption{Illustration of sensitivity of our proposed image segmentation algorithm to the rank. From the segmentation results, we see that when the rank is small, simpler segmentation curves will be obtained. When the rank is chosen to be too high, we will obtain over-segmentation result. Thus, the rank is a good surrogate for the complexity of the curve.}
	\label{segment}
\end{figure*}

\subsection{Segmentation using structured low-rank methods}

We demonstrate the preliminary utility of the scheme in Section \ref{segmentation} in Fig. \ref{segment} on two images: the cells image and the church image. The proposed algorithm in \eqref{slr} is initialized with $f=h$ and iterated until convergence. The parameter $\lambda$ is set to a high value (e.g. $5\times 10^{9}$ in our experiments) to enforce the rank constraint. We note that the optimization scheme is capable of identifying the cells and the outline of the church, even though no curve initialization was provided. For the cells segmentation (c) and the church segmentation (f), we choose the rank to be 500 and 1200 respectively. The corresponding computational time for the two segmentations is 60 seconds and 226 seconds. We now compare our segmentation method with the level-set based segmentation method DLRSE, where the edge indicator function is chosen as
\begin{equation}\label{edgeenergy}
g= \frac{1}{1+|\nabla G_{\sigma}* I|^2},
\end{equation}	
where $G_{\sigma}$ is a Gaussian kernel with a standard deviation $\sigma$. We considered the initialization of DLRSE with two possible curves, indicated by the green squares. The parameters in DLRSE were chosen manually to yield the best results, which corresponded to $\lambda =6, \alpha=\pm 2, 1510$ iterations for the cells image. The parameters for the church image were $\lambda =4.8, \alpha=-2, 2710$ iterations. For the cells image segmentation, the time required for getting (a) and (b) is 47 seconds and 40 seconds. For chruch image segmentation, the computational time for getting (d) and (e) is 175 seconds and 249 seconds. These results show that the proposed scheme is comparable to DLRSE in segmentation performance and can capture sharp features. However, the main benefit is its insensitivity to initialization, compared to DLRSE seen from (d) and (e).

\subsection{Point cloud denoising: comparison with state-of-the-art}

\begin{figure*}[t!]
	\centering
	 \subfigure[Original \#1]{\includegraphics[width=0.215\textwidth]{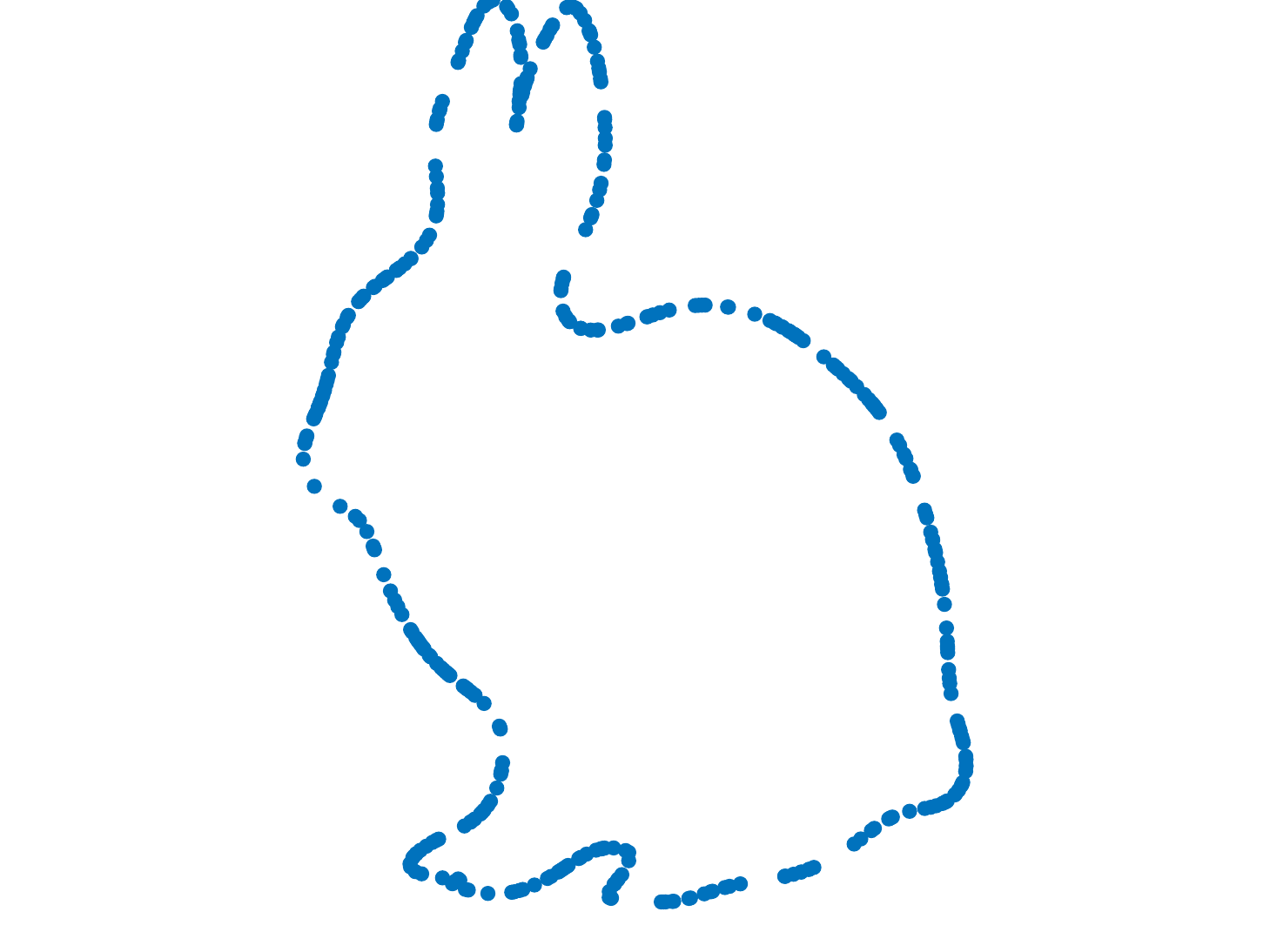}}\label{fig:1a}
	\subfigure[Noisy \#1, SNR = 31.85 dB]{\includegraphics[width=0.215\textwidth]{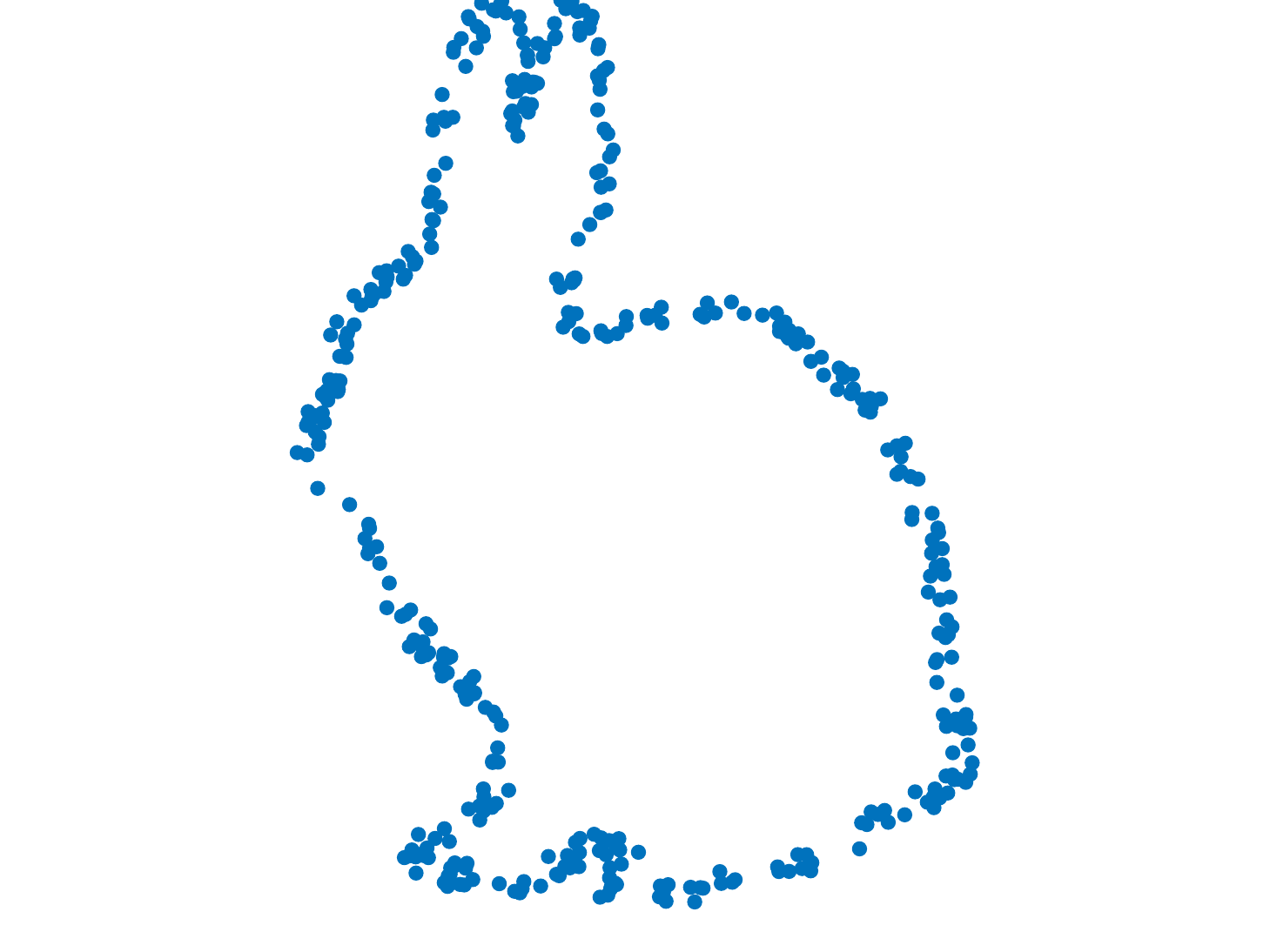}}\label{fig:1a}
	\subfigure[GLR, SNR = 32.09 dB]{\includegraphics[width=0.2125\textwidth]{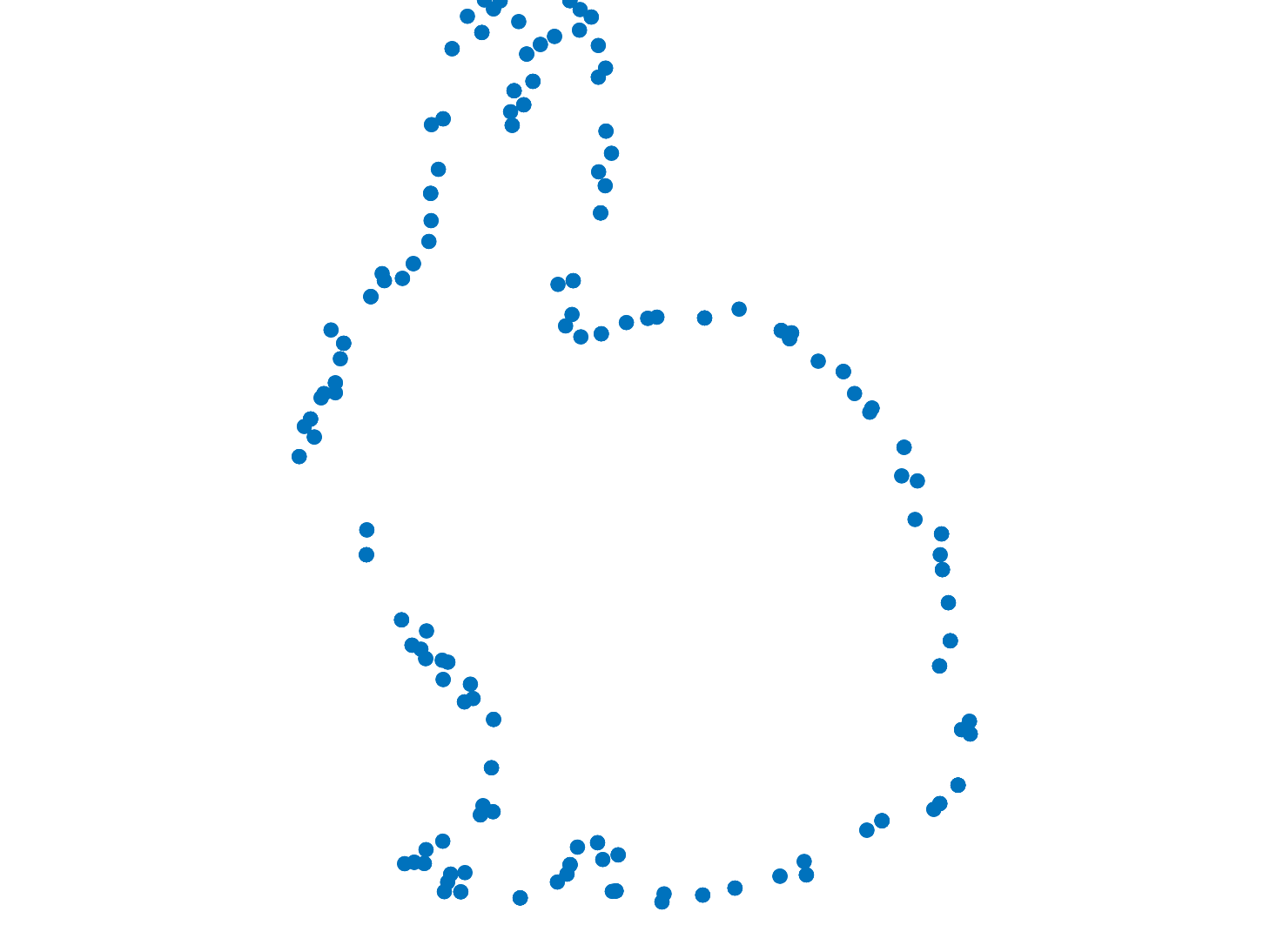}}\label{fig:1b}
	\subfigure[KLR, SNR = 35.21 dB]{\includegraphics[width=0.215\textwidth]{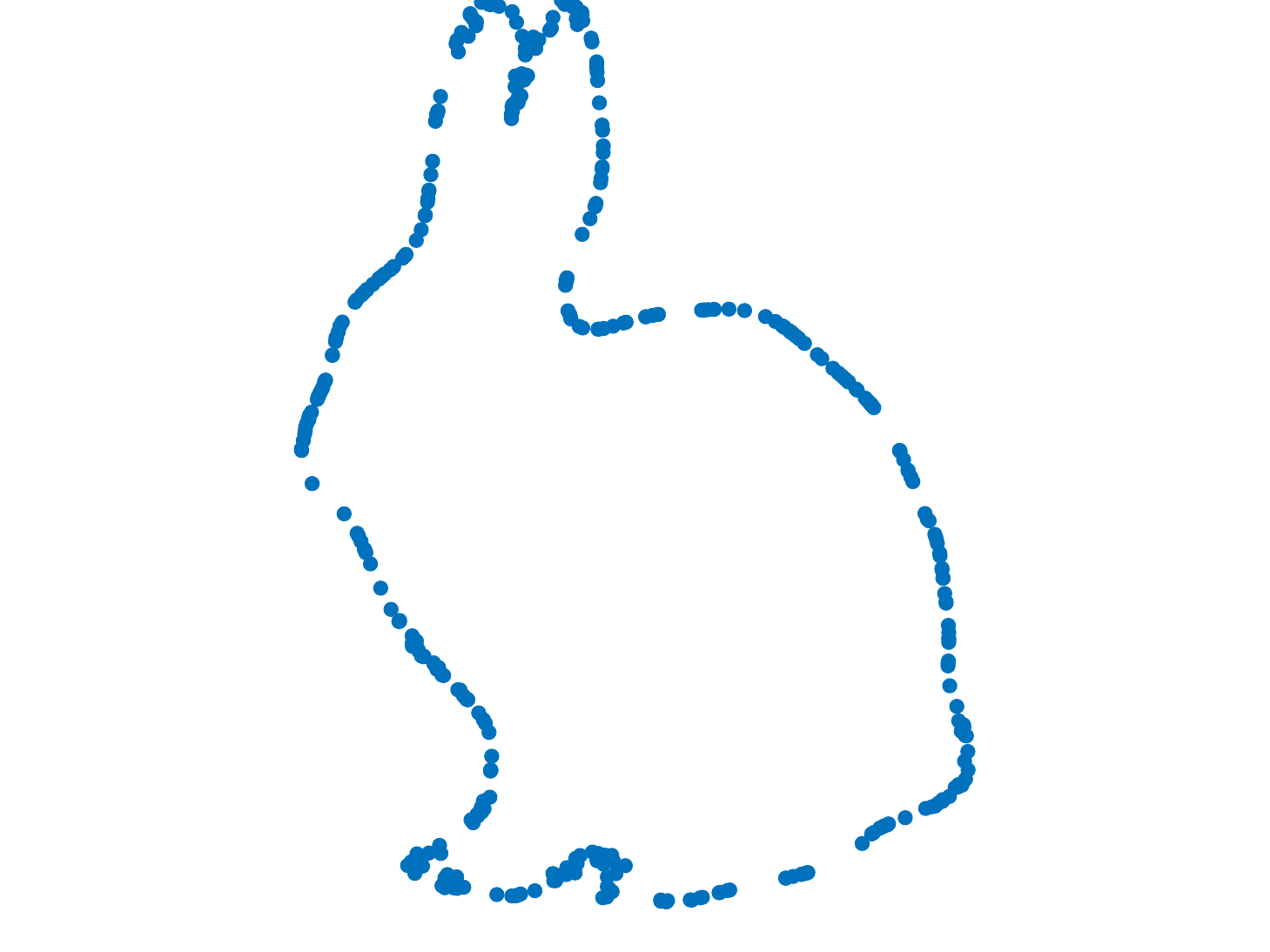}}\label{fig:1c}\\
	\subfigure[Original \#2]{\includegraphics[width=0.215\textwidth]{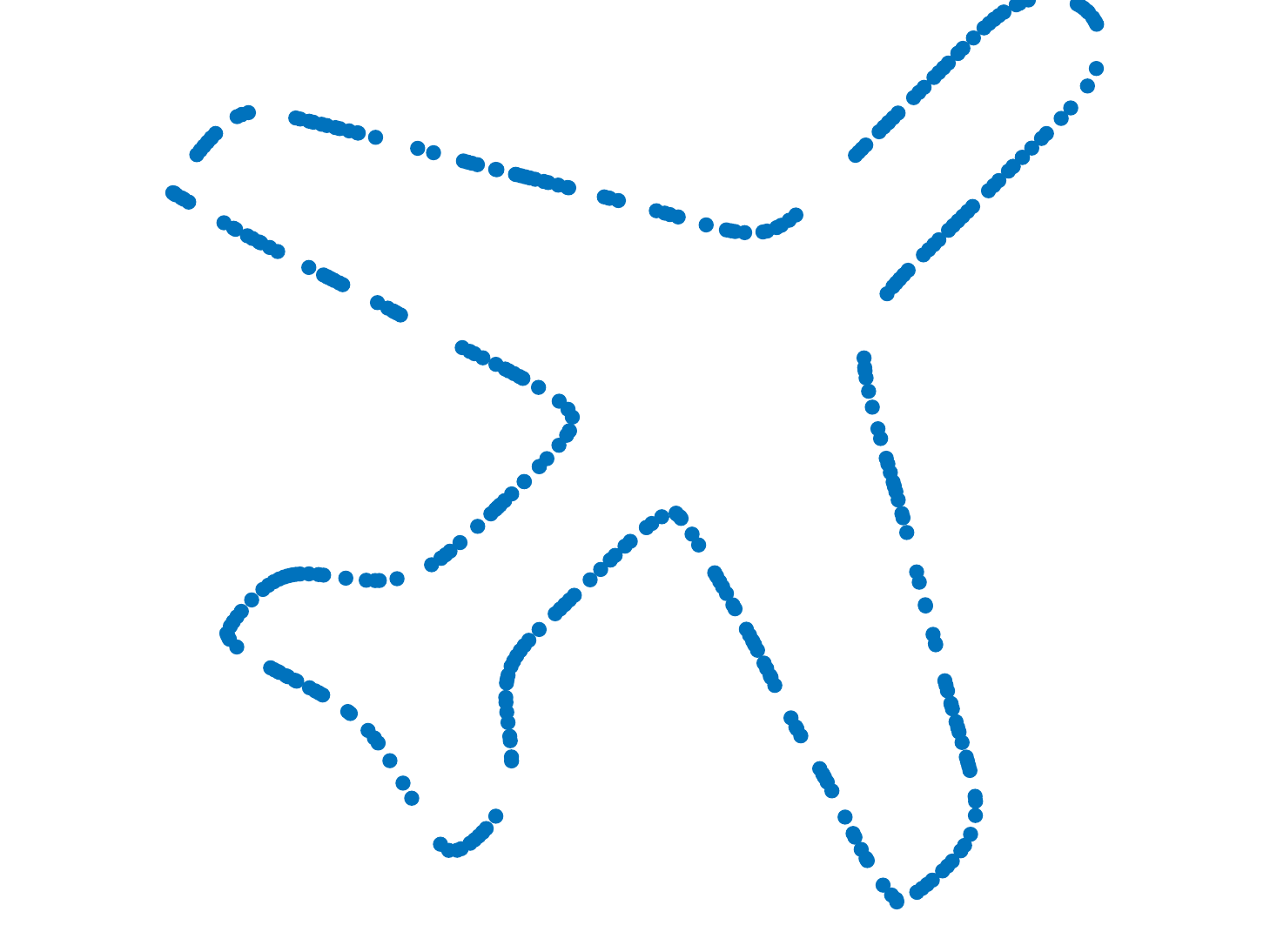}}\label{fig:1a}
	\subfigure[Noisy \#2, SNR = 29.94 dB]{\includegraphics[width=0.215\textwidth]{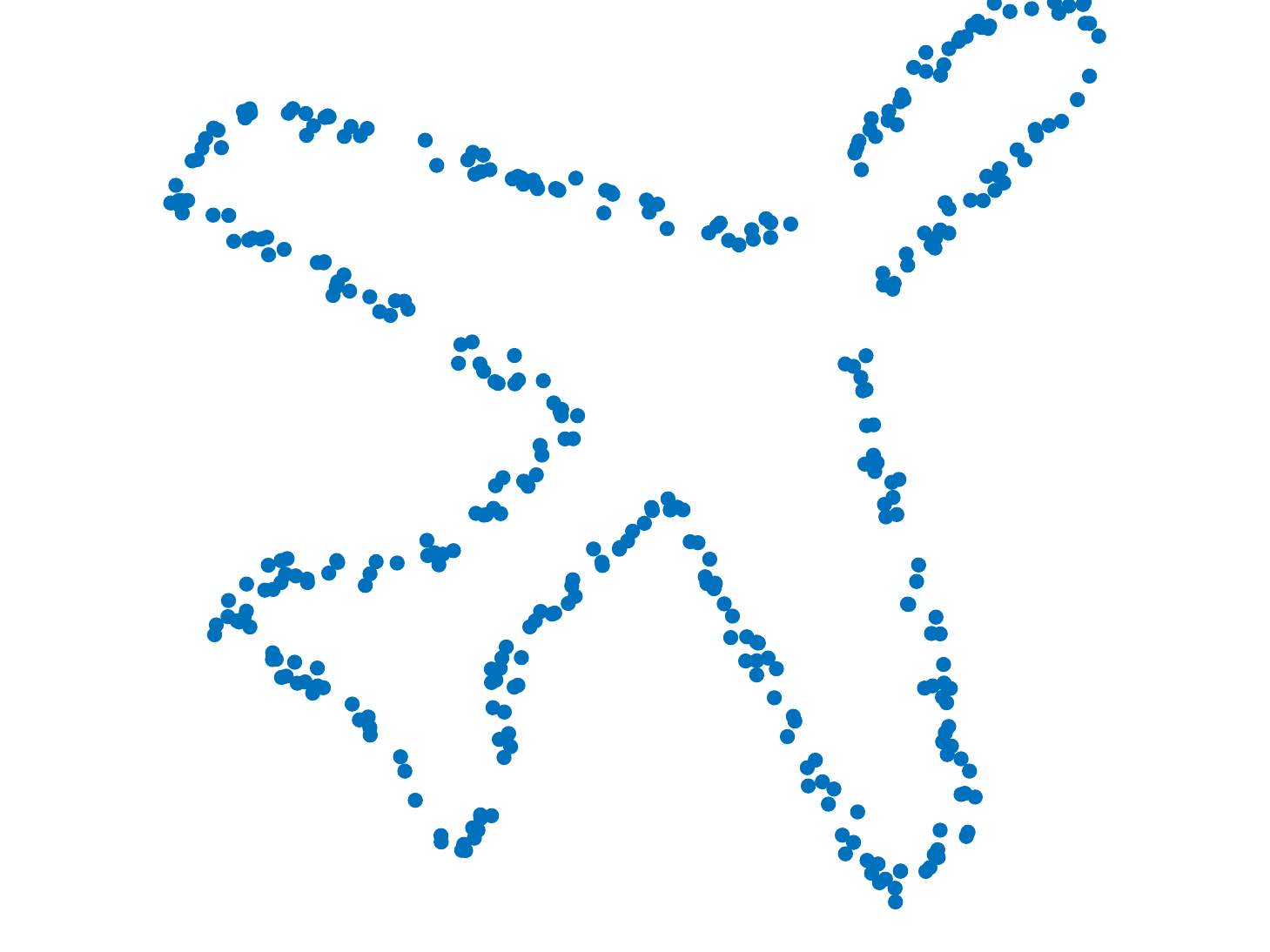}}\label{fig:1a}
	\subfigure[GLR, SNR = 30.01 dB]{\includegraphics[width=0.215\textwidth]{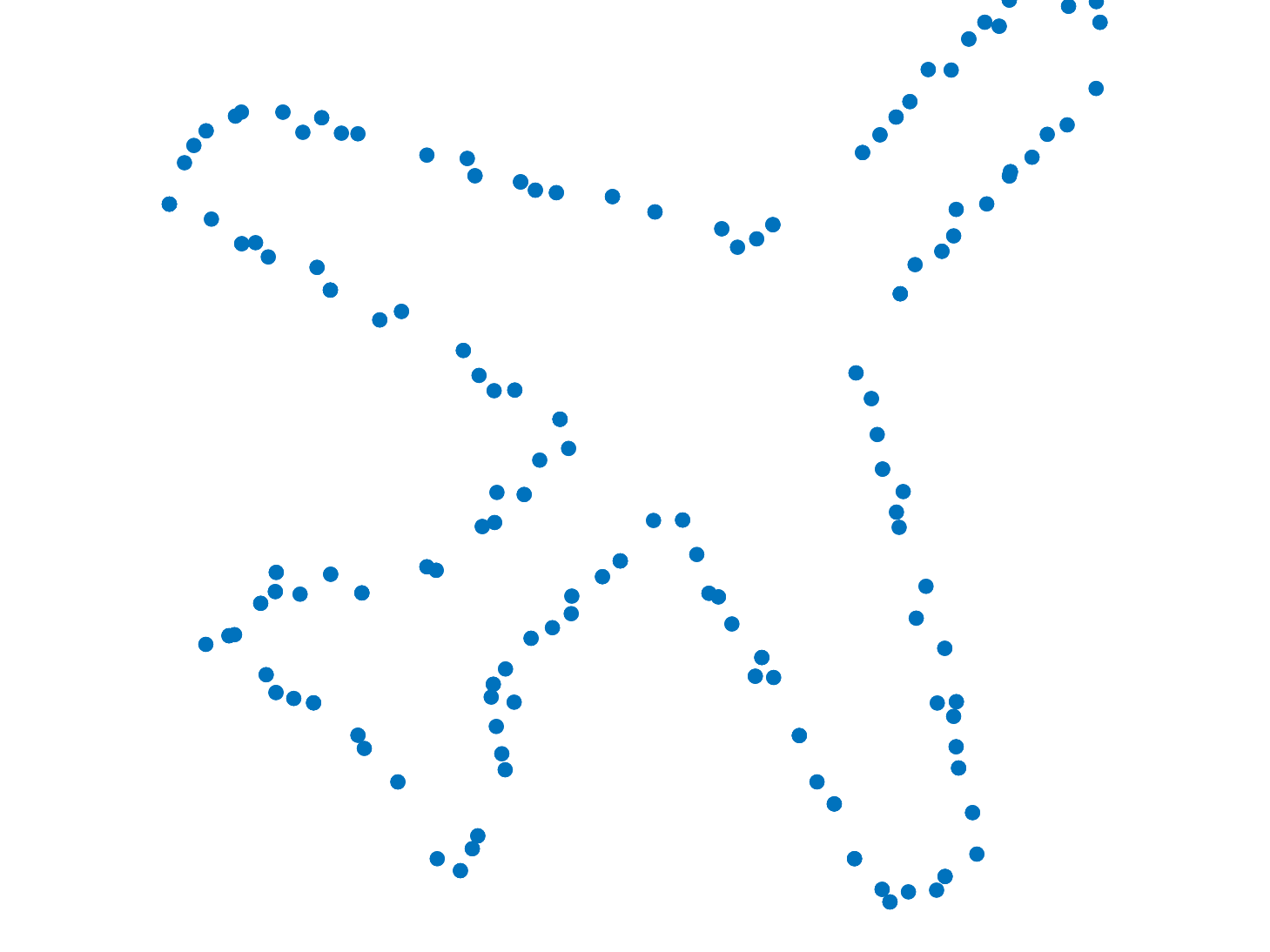}}\label{fig:1b}
	\subfigure[KLR, SNR = 33.01 dB]{\includegraphics[width=0.215\textwidth]{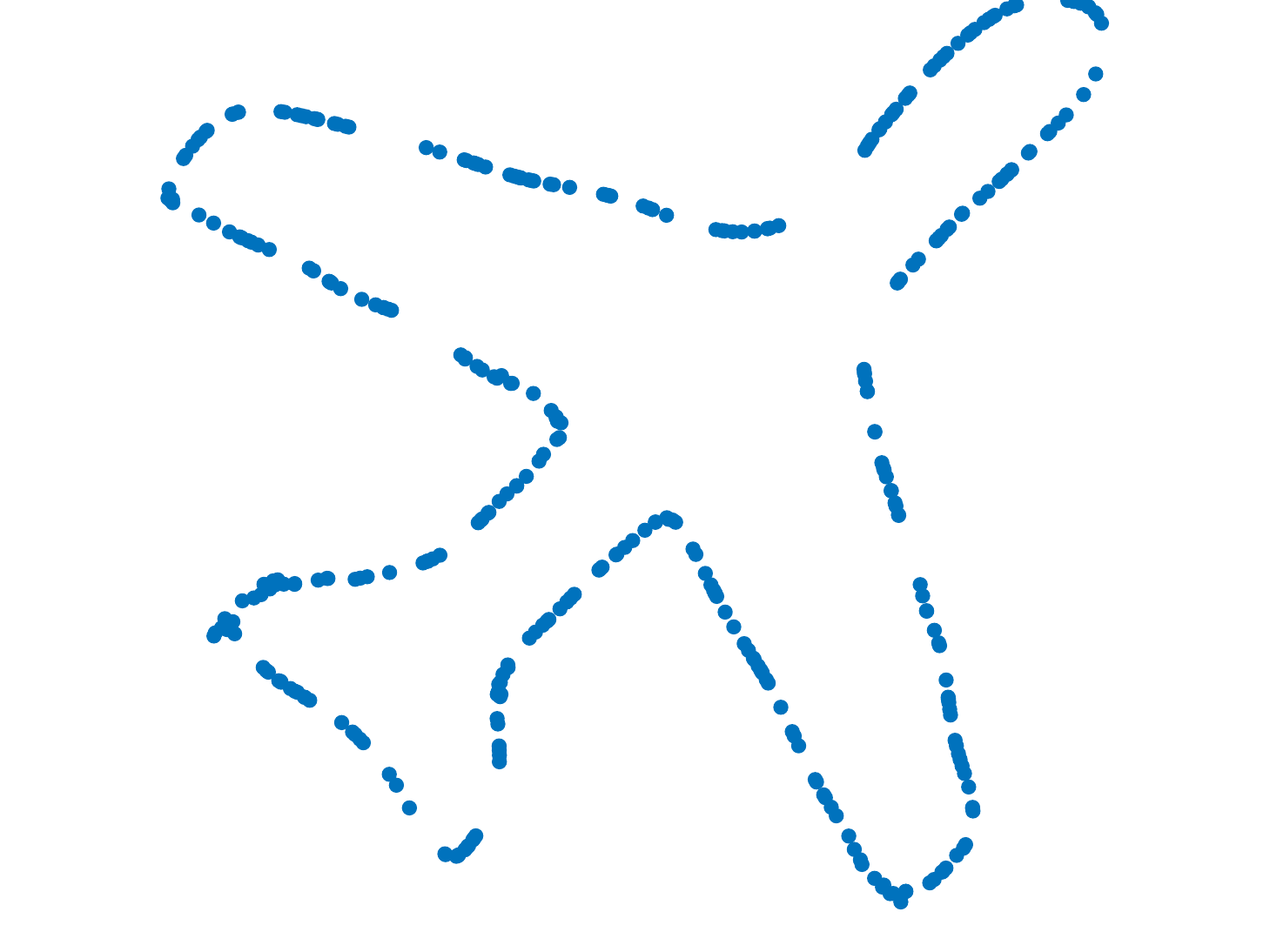}}\label{fig:1c}\\
	\subfigure[Original \#3]{\includegraphics[width=0.215\textwidth]{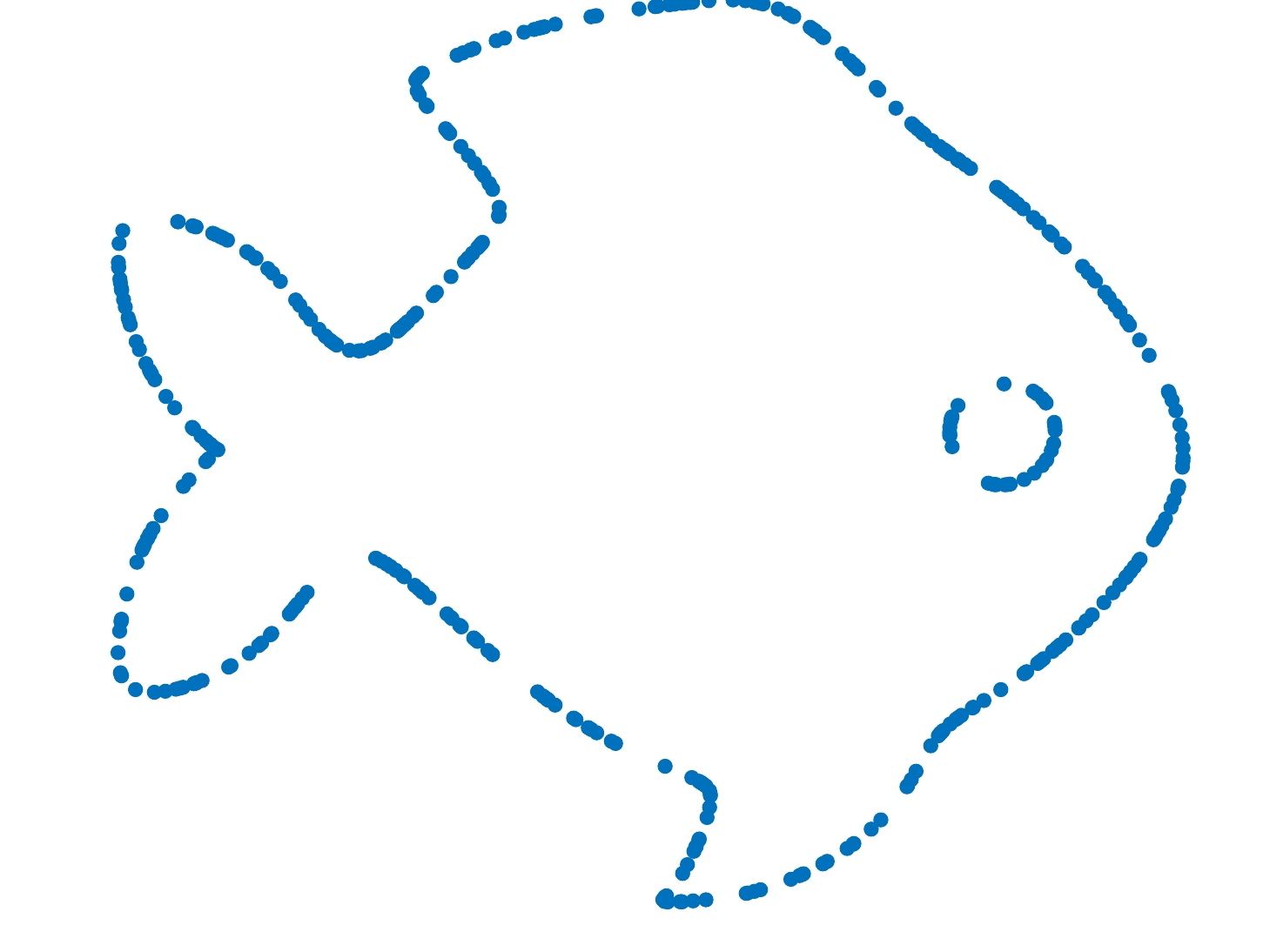}}\label{fig:1a}
	\subfigure[Noisy \#3, SNR= 28.33 dB]{\includegraphics[width=0.215\textwidth]{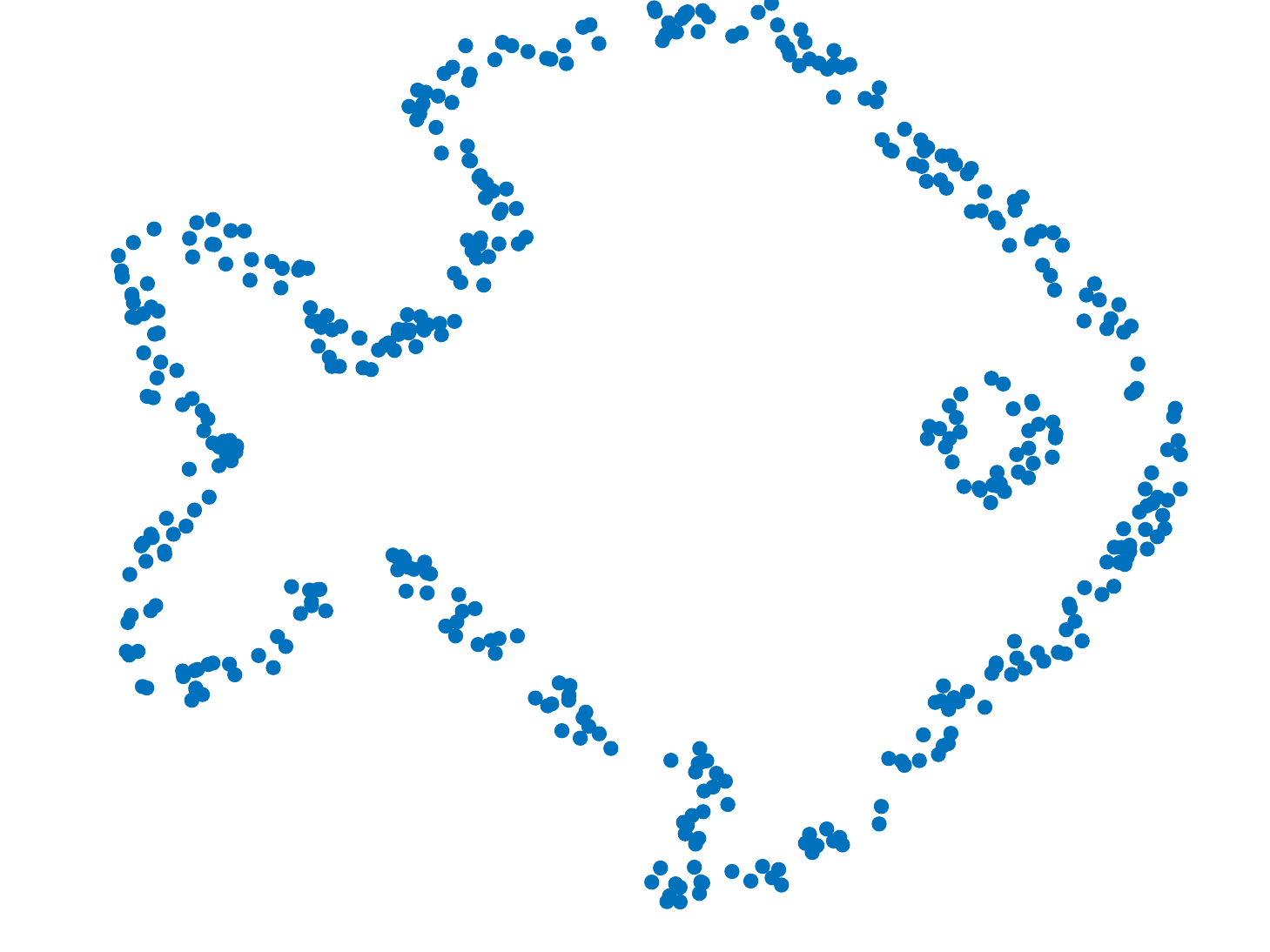}}\label{fig:1a}
	\subfigure[GLR, SNR = 28.95 dB]{\includegraphics[width=0.215\textwidth]{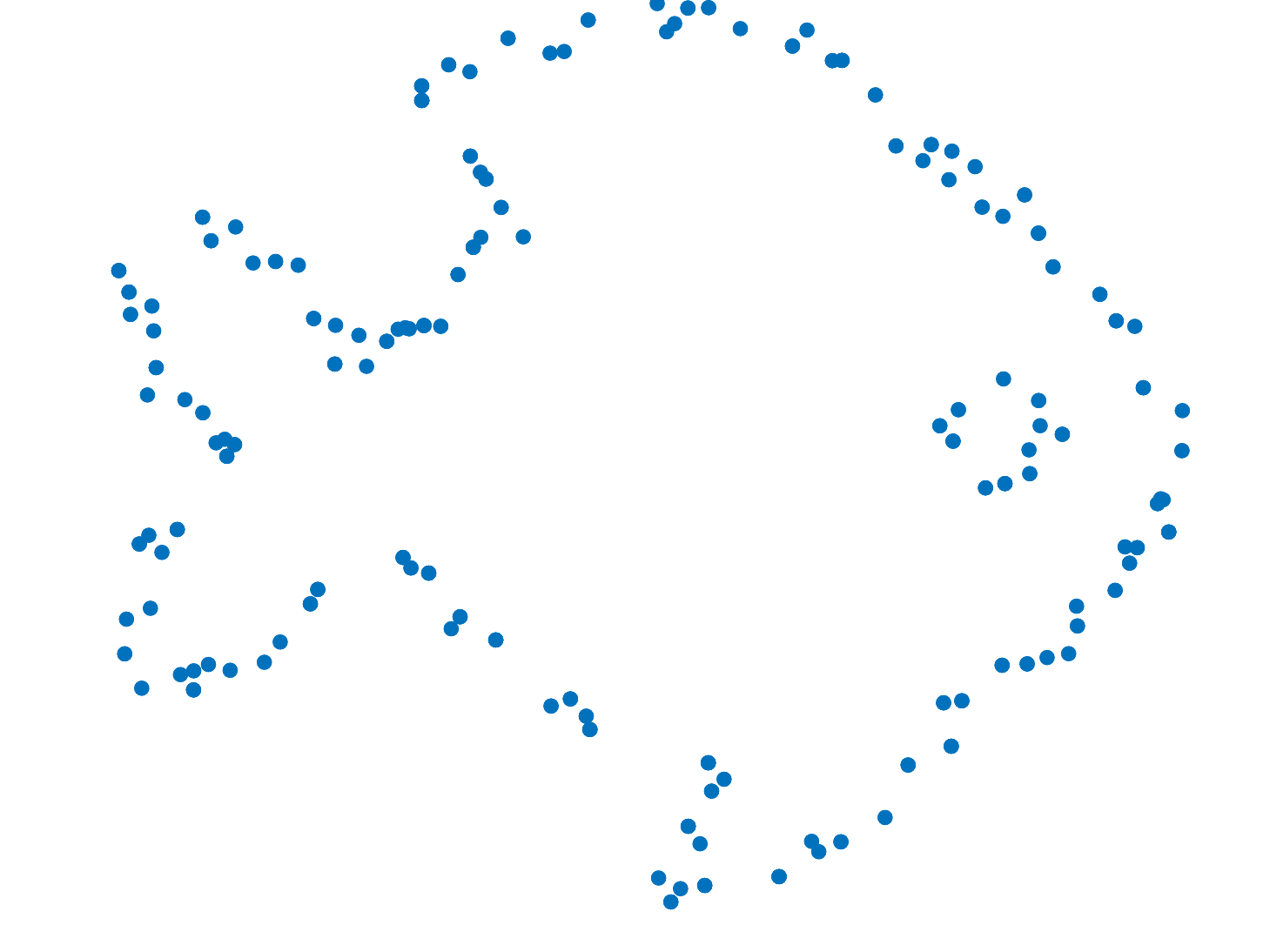}}\label{fig:1b}
	\subfigure[KLR, SNR = 31.84 dB]{\includegraphics[width=0.215\textwidth]{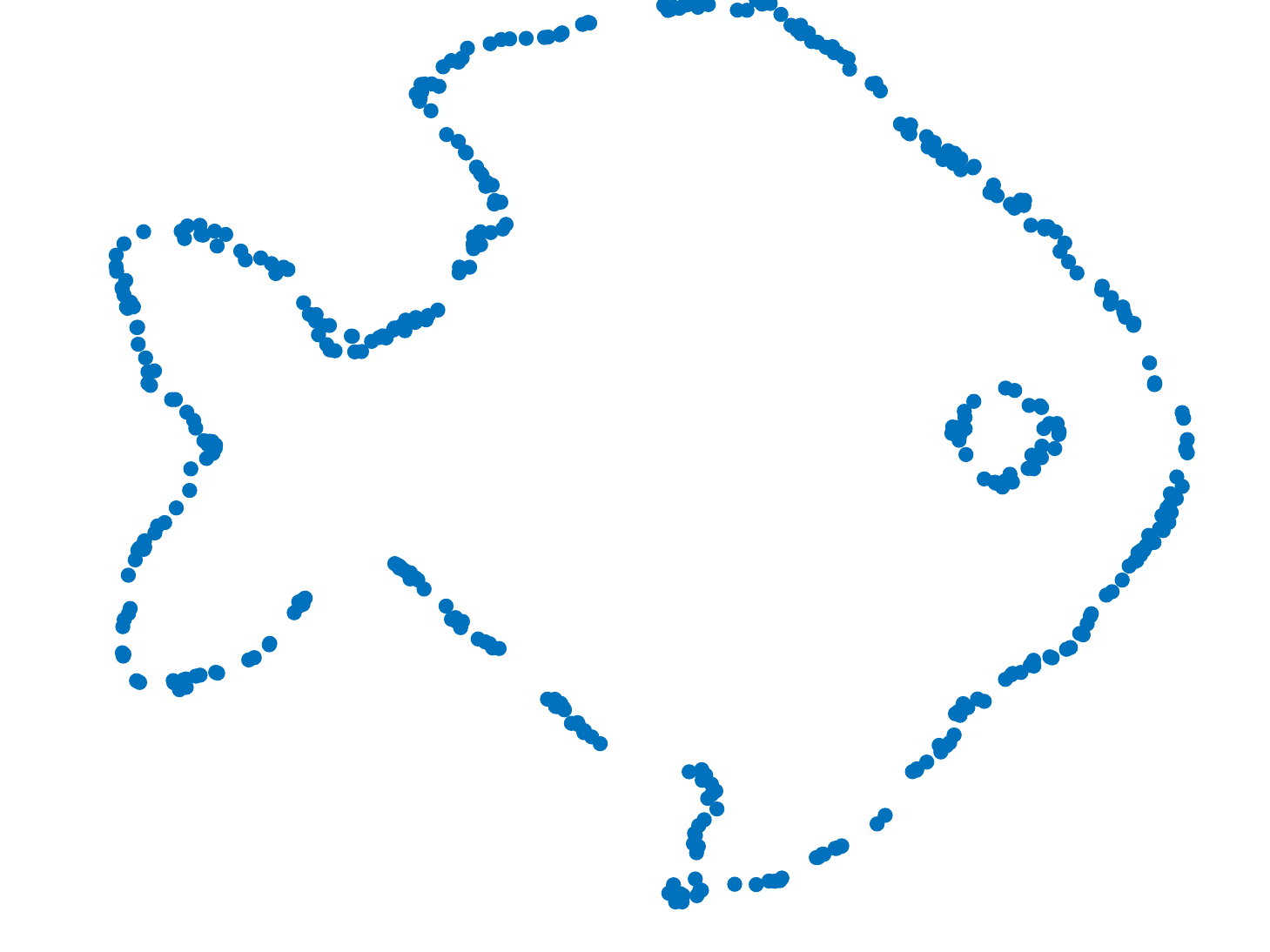}}\label{fig:1c}\\
	\caption{Comparison between proposed denoising algorithm (KLR) and Garph Laplacian Regularized denoising algorithm (GLR) introduced in \cite{zeng20183d}.}
	\label{Figherky}
\end{figure*}

We illustrate the utility of the kernel low-rank formulation (KLR) proposed in \ref{pointdenoising} to denoise 2D points in Fig. \ref{Figherky}. In Fig. \ref{Figherky}, we also compared our method with another point-set denoising method called “Graph Laplacian regularized point cloud denoising (GLR)”, which was introduced recently in \cite{zeng20183d}. We choose 3 examples to perform the point sets denoising algorithms and in each example, we add Gaussian noise to the point sets. In the first example, we randomly choose 409 points on the edge set of a rabbit as shown in (a). We use GLR and KLR to denoise the noisy points respectively. For GLR, we set the parameter $\mu$ to be 1000 and after 34 iterations, we obtained the denoising result (c). In KLR, we get the denoising result (d) after 80 iterations using about 4.4 seconds. In the second examples, we choose 385 points on the edge set of a plane. We again set the parameter $\mu$ to be 1000 and after 31 iterations, we have the denoising result (g). For KLR, we get the result (h) by iterating 80 times using about 4.0 seconds. In the third example, we choose 451 points on the shape of a fish. For GLR, after 34 iterations by setting the parameter $\mu$ in the algorithm to be 1000, we obtain the denosing result (k). For our proposed denoising algorithm, we raised the number of iterations to 450 and it takes about 32.7 seconds to obtain the denoising result (l). By comparing the denosing results (c) and (d), (g) and (h), (k) and (l), we can see that both the two methods work for denosing the noisy points. While for GLR, we can see that the some points will get closer along the right curve. To compare the experimental performance mathematically, we introduce an evaluation metric, signal-to-noise ratio (SNR), for point cloud denoising. Suppose the ground-truth and predict point clouds are $\{\mathbf{x}_i\}_{i=1}^{N_1}$ and $\{\mathbf{y}_i\}_{i=1}^{N_2}$. We define the SNR, which is measured in dB by
\[{\rm{SNR}} = 10\log\frac{1/N_2 \sum_{\mathbf{y}_i}||\mathbf{y}_i||_2^2}{{\rm{MSE}}},\]
where MSE is the mean-square-error defined as
\[{\rm{MSE}} = \frac{1}{2N_1}\sum_{\mathbf{x}_i}\min_{\mathbf{y}_j}||\mathbf{x}_i-\mathbf{y}_j||_2^2 +\frac{1}{2N_2}\sum_{\mathbf{y}_i}\min_{\mathbf{x}_j}||\mathbf{y}_i-\mathbf{x}_j||_2^2.\]

\section{Discussion and Conclusion}

We introduced a continuous domain framework for the recovery of points on a band-limited curve. The proposed bandlimited representation have several desirable geometric properties, which make it an attractive tool in a variety of shape estimation problems. We have introduced novel algorithms with sampling guarantees for the recovery of both irreducible and union of irreducible bandlimited curves from few of their samples; our experiments show that the bounds are sharp. We can see from Fig. \ref{prop1Sim2} that with less samples, perfect recovery will also happen. We will discuss this phenomenon in our feature work. Furthermore, we do not use the property that the coefficients of the polynomial are Hermitian symmetric \cite{pan2014sampling} in the proofs and estimation. We note that the use of the Hermitian symmetry property may result in improved sampling bounds. However, this extension is beyond the scope of the present work.

We also demonstrated the utility of the representation in practical applications including image segmentation and denoising of a point cloud, which can be modeled by a curve. The main benefit of the curve recovery from points as well as image segmentation over the state of the art is the convex formulation, which makes the algorithm insensitive to local minima errors as well as initialization. The segmentation and point cloud denoising experiments show that the proposed scheme can exploit the global structure of the points better than competing methods that rely on local curve properties such as smoothness and curvature, which makes the algorithms less sensitive to non-uniformity of sampling. In our future work, we will extend our results to high dimensional spaces. We note that the proofs based on B\'ezout's inequality cannot be readily extended to the higher dimensional setting and thus a new method need to be introduced for the extension.

 \section*{Acknowledgement}

The authors gratefully acknowledge helpful comments from Dr. Greg Ongie.


\section{Appendix}

\subsection{Proof of Lemma \ref{prop0}}
\label{proofBez}

We first state the well-known result for complex polynomials, which we extend to the band-limited setting.
 \begin{lem}\cite{igor}  \label{lem1}
Let $p_1$ and $p_2$ be two nonconstant polynomials in $\mathbb{C}[z_1,z_2]$ of degrees $d_1$ and $d_2$ respectively. If $p_1$ and $p_2$ have no common component, then the system of equations
\begin{equation}
\label{2poly}
p_1=p_2=0
\end{equation}
has at most $d_1d_2$ solutions.
 \end{lem}



Lemma \ref{prop0} can be proved by simply substituting $p_1=\mathcal P[\mu]$ and $p_2=\mathcal P[\eta]$ in Lemma \ref{lem1}. Specifically, the degree of $\mathcal P[\mu]$ and $\mathcal P[\eta]$ are $(k_1+k_2)$ and $(l_1+l_2)$ respectively. Hence, the maximum number of solutions to \eqref{syseq} is given by $(k_1+k_2)(l_1+l_2)$.

\subsection{Proof of Proposition \ref{prop2D1}}
\label{proof2D1}
\begin{proof}
The Fourier coefficients of $\psi(\mathbf x)$ is support limited within $\Lambda$, which is the minimal support. Let $\eta(\mathbf x)$ be another band-limited polynomial, whose Fourier coefficients are support limited within $\Lambda$ and satisfies $\eta(\mathbf x_i)=0$, for $i=1,\ldots,N$. When the number of samples satisfy \eqref{nsamples1}, this is only possible if $\eta$ is a factor of $\psi$, according to B\'ezout's inequality. Thus, $\psi(\mathbf x)$ must be a factor of $\eta(\mathbf x)$. Since $\psi$ is irreducible, this implies that it is the unique band-limited irreducible polynomial satisfying $\psi(\mathbf x_i)=0$.
\end{proof}

\subsection{Proof of Proposition \ref{prop2D2}}
\label{proof2D2}
\begin{proof}
The polynomial $\psi(\mathbf x)$ is represented in terms of its irreducible factors as:
\begin{equation}
\psi(\mathbf x) = \psi_1(\mathbf x)\psi_2(\mathbf x)\ldots \psi_J(\mathbf x)
\end{equation}
where the bandwidth of $\psi_j(\mathbf x)$ is $k_{1,j}\times k_{2,j}$. 

Let $\eta(\mathbf x)$ be another polynomial with bandwidth $k_1 \times k_2$ satisfying $\eta(\mathbf x_i)=0$, for $i=1,\ldots,N$. Consider one of the irreducible sub-curves $\{\psi_j(\mathbf x)=0\}$, that is sampled on $N_j$ points satisfying \eqref{nsamples2}. According to Lemma \ref{prop0}, both $\psi_j$ and $\eta$ can be simultaneously zero at these sampling locations only if $\psi_j$ and $\eta$ have a common factor. Since $\psi_j$ is irreducible, this implies that $\psi_j$ is a factor of $\eta$. Repeating this line of reasoning for all factors $\{\psi_j\}$, we conclude that $\psi(\mathbf x)$ divides $\eta(\mathbf x)$. Since both $\psi(\mathbf x)$ and $\eta(\mathbf x)$ have the same bandwidth, the only possibility is that $\eta(\mathbf x)$ is a scalar multiple of $\psi(\mathbf x)$. This implies that the curve $\psi(\mathbf x) = 0$ can be uniquely recovered in \eqref{nsamples2} is satisfied. 

The total number of points to be sampled is $N = \sum_{j=1}^{J}N_j > (k_1+k_2)\sum_{j=1}^{J}(k_{1,j}+k_{2,j})$. 

The support of the Fourier coefficients of $\psi$ can be expressed in terms of the supports of $\{\psi_j\}$. Using convolution properties, we get: $k_1 = 1 + \sum_{j=1}^J (k_{1,j} - 1)$ and $k_2 = 1 + \sum_{j=1}^J (k_{2,j} - 1)$. Thus, $\sum_{j=1}^{J}(k_{1,j}+k_{2,j}) = k_1+k_2+2(J-1)$ and it can be concluded that $N >(k_1+k_2)(k_1+k_2+2(J-1))$.
\end{proof}

\subsection{Proof of Proposition \ref{prop2D3}}
\label{proof2D3}
\begin{proof}
Following the steps of the proof for Proposition \ref{prop2D2}, we can conclude that $\psi(\mathbf x)$ is a factor of $\mu(\mathbf x)$. Since $\Lambda \subset \Gamma$, it follows that $\mu(\mathbf x) = \psi(\mathbf x)\;\eta(\mathbf x)$, where $\eta(\mathbf x)$ is some arbitrary function such that $\mu(\mathbf x)$ is band-limited to $\Gamma$.
\end{proof}

\subsection{Proof of Proposition \ref{propRank}}
\label{proofRank}
\begin{proof}
Let $\mathbf c$ be the minimal filter of bandwidth $|\Lambda|$, associated with the polynomial $\psi(\mathbf x)$. We define the following filters supported in $\Gamma$ for all $\mathbf l \in \Gamma:\Lambda$.
\begin{equation}
\mathbf c_{\mathbf l}[\mathbf k] = \begin{cases}
\mathbf c[\mathbf k - \mathbf l], & \text{if $\mathbf k - \mathbf l \in \Lambda$}.\\
0, & \text{otherwise}.
\end{cases}
\end{equation}
$\mathbf c_{\mathbf l}$ are the Fourier coefficients of $\exp(j2\pi \mathbf l^T \mathbf x)\psi(\mathbf x)$, and are all null-space vectors of the feature matrix $\Phi_{\Gamma}(\mathbf X)$. The number of such filters is $|\Gamma:\Lambda|$. Hence, we get the rank bound: ${\rm rank}\left(\Phi_{\Gamma}(\mathbf X) \right) \leq |\Gamma|-|\Gamma:\Lambda|$. 

If the sampling conditions of Proposition \ref{prop2D3} are satisfied, then all the polynomials corresponding to null-space vectors of $\Phi_{\Gamma}$ are of the form: $\mu(\mathbf x) = \psi(\mathbf x)\;\eta(\mathbf x)$. Alternatively, in the Fourier domain, the filters are of the form:
\begin{equation}
\mathbf c_{\mu}[\mathbf k] = \sum_{\mathbf l \in \Gamma :\Lambda} \mathbf d_{\mathbf l}\mathbf c_{\mathbf l}[\mathbf k]
\end{equation}
where $\mathbf d_{\mathbf l}$ are the Fourier coefficients of the arbitrary polynomial $\eta(\mathbf x)$. Thus, all the null-space filters can be represented in terms of the basis set $\{\mathbf c_{\mathbf l}\}$. This leads to the relation: ${\rm rank}\left(\Phi_{\Gamma}(\mathbf X) \right) = |\Gamma|-|\Gamma:\Lambda|$.
\end{proof}
\bibliographystyle{IEEEbib}
\bibliography{curves}

\end{document}